\documentclass[floatfix,twocolumn,showpacs,aps,tightenlines,superscriptaddress]{revtex4-1}
\usepackage{graphicx}
\usepackage{color}
\usepackage{psfrag}
\usepackage{dcolumn}
\usepackage{bm}

\usepackage{amssymb}
\usepackage{amsmath}
\usepackage{amsfonts}
\usepackage{longtable}
\usepackage{xspace}

\newcommand{\beq}{\begin{equation}}
\newcommand{\eeq}{\end{equation}}

\newcommand{\be}{\begin{equation}}
\newcommand{\ee}{\end{equation}}
\newcommand{\bea}{\begin{eqnarray}}
\newcommand{\eea}{\end{eqnarray}}
\newcommand{\bes}{\begin{subequations}}
\newcommand{\ees}{\end{subequations}}

\usepackage{bm}

\begin{document}

\title{Spin flips in generic black hole binaries}

\author{Carlos O. Lousto}
\affiliation{Center for Computational Relativity and Gravitation,
School of Mathematical Sciences,
Rochester Institute of Technology, 85 Lomb Memorial Drive, Rochester,
New York 14623}

\author{James Healy}
\affiliation{Center for Computational Relativity and Gravitation,
School of Mathematical Sciences,
Rochester Institute of Technology, 85 Lomb Memorial Drive, Rochester,
New York 14623}

\author{Hiroyuki Nakano}
\affiliation{Center for Computational Relativity and Gravitation,
School of Mathematical Sciences,
Rochester Institute of Technology, 85 Lomb Memorial Drive, Rochester,
New York 14623}
\affiliation{Department of Physics, Kyoto University, Kyoto 606-8502, Japan}

\date{\today}

\begin{abstract}
We study the spin dynamics of individual black holes in a binary system.
In particular we focus on the polar precession of spins and the possibility
of a complete flip of spins with respect to the orbital plane. We perform
a full numerical simulation that displays these characteristics. We evolve
equal mass binary spinning black holes for $t=20,000M$ from an initial
proper separation of $d=25M$ down to merger after 48.5 orbits. We
compute the gravitational radiation from this system and compare it to
3.5 post-Newtonian generated waveforms finding close agreement. We then 
further use 3.5 post-Newtonian evolutions to show the extension of 
this spin {\it flip-flop} phenomenon to unequal mass binaries. 
We also provide analytic expressions
to approximate the maximum {\it flip-flop} angle and frequency in terms
of the binary spins and mass ratio parameters at a given orbital radius.
Finally we discuss the effect this spin {\it flip-flop} would have on 
accreting matter and other potential observational effects.
\end{abstract}

 \maketitle

\section{Introduction}\label{sec:intro}

In this first decade since the
breakthroughs~\cite{Pretorius:2005gq, Campanelli:2005dd,Baker:2005vv}
that allowed to numerically solve General Relativity's 
field equations for the evolution of black hole binaries (BHB), 
we have gained many new insights on these systems. 
The computation of theoretical gravitational waveforms
from these BHB systems is very important for first detection and estimation
of binary's
parameters~\cite{Aylott:2009ya, Aylott:2009tn, Ajith:2012az, Aasi:2014tra}.
It is also of astrophysical 
interest to study the BHB orbital and spin dynamics in precessing 
systems and how accreting matter interacts with them.

The spin of each individual black hole (BH) can notably affect its orbital
motion as displayed, for instance, by the {\it hangup}
mechanism~\cite{Campanelli:2006uy}, which delays or prompts the merger of the
binary according to the sign of the spin-orbit coupling, thus ensuring
the formation of a {\it horizon} (cosmic censorship hypothesis) at merger
before the final hole settles to a Kerr
BH~\cite{Campanelli:2008dv,Owen:2010vw} (no hair theorem).

One of the most notable predictions of numerical relativity is
that the remnant of the merger of two highly spinning BHs
may receive a recoil of thousands of 
km/s~\cite{Campanelli:2007ew,Campanelli:2007cga,Lousto:2011kp}
due to asymmetrical emission of gravitational radiation induced by the BH
spins~\cite{Lousto:2012su,Lousto:2012gt}.
This developed intense astronomical searches for such fast moving
BHs that may completely escape from their host galaxies
or produce observable disturbances to the velocity field of stars
in their cores~\cite{Komossa:2012cy}.

New developments in numerical relativity allowed the study of
relatively far separated binaries, up to distances of $100M$ in Ref.~\cite{Lousto:2013oza},
and mass ratios of $100:1$ in Refs.~\cite{Lousto:2010ut,Sperhake:2011ik}. 
Very long term evolutions are now
possible~\cite{Lousto:2014ida,Szilagyi:2015rwa} 
as well as the study of near maximally spinning
BHBs~\cite{Lovelace:2014twa,Scheel:2014ina,Ruchlin:2014zva}.

In Ref.~\cite{Lousto:2014ida}, we have found that the spin of BHs
can completely reverse sign during its orbital stage. This {\it flip-flop} 
of spins is due to a spin-spin coupling effect and may have important
observational consequences when the binary system is accreting gas from a
galactic environment~\cite{Bogdanovic:2014cua}. 
In this paper we extend the analysis to unequal
mass binaries using 3.5 post-Newtonian (PN) evolutions.
Those proved a reliable description for this scenario
when compared to the long term full numerical simulation of
an equal mass binary~\cite{Lousto:2014ida}.

This paper is organized as follows. In Sec.~\ref{sec:Equal}
we revisit the equal mass binary scenario providing further detail
and analysis of our full numerical simulation. We also provide
a simple vector analysis to understand the mechanism behind the
{\it flip-flop} and then a 2PN analytic study to describe its spin dynamics.
In Sec.~\ref{sec:Unequal} we study the case of unequal mass binaries.
We use 3.5PN evolutions of the equations of motion coupled with
the 2PN spin dynamics to find the configurations that maximize
the {\it flip-flop} angle and the likelihood of a {\it flip-flop} angle
to occur given plausible astrophysical distribution of binary parameters.
We also provide 2PN analytic expressions to approximate the {\it flip-flop}
angle and frequency as measured in the orbital frame and asymptotic frame
described in terms of projections with respect to the orbital angular
momentum $\vec{L}$ and total angular momentum $\vec{J}$.
We end the paper with Sec.~\ref{sec:Discussion} where we
make some estimates of the effect this {\it flip-flop}
phenomena could have in realistic astrophysical scenarios and suggest
that detailed simulations involving the magnetohydrodynamical description
of accreting matter onto BHs are needed to find the 
characteristic electromagnetic signatures of the {\it flip-flop}.
The appendix~\ref{app:PNSpin} provides details on the PN computation of the
flip-flop frequencies and angles.

\section{Equal Masses Binaries}\label{sec:Equal}

We first study the case of equal mass binaries since they
cleanly display the {flip-flop} effect on the spins of
the holes and allow for a simple approximate analytic
model of the process. In Sec.~\ref{sec:Unequal} we will
study in detail its mass ratio dependence.

\subsection{Full Numerical Evolution}\label{subsec:FN}

\begin{figure}
\includegraphics[width=0.8\columnwidth]{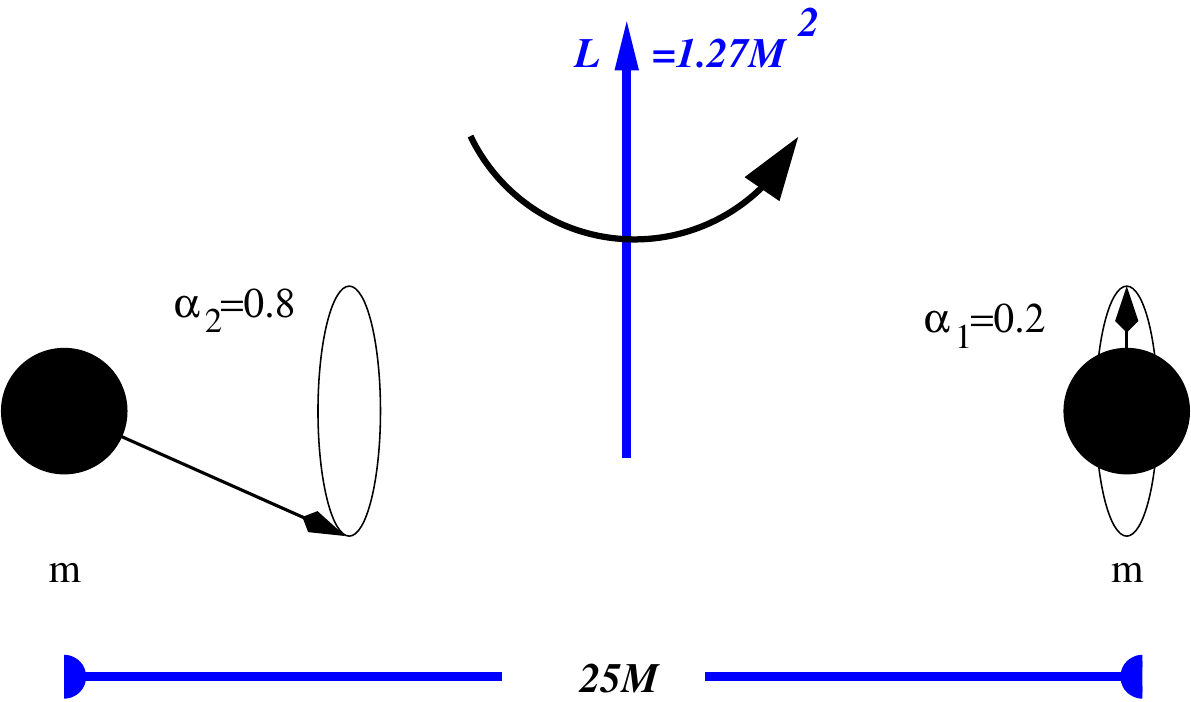}
\caption{Initial configuration of the simulated binary.
\label{fig:config}}
\end{figure}

In order to verify the realization of the spin 
{flip-flops} in the presence of gravitational radiation and
the full non-linearities of General Relativity during
the final stages of the inspiral,
we consider a long-term full numerical simulation with
initial configuration as described in Table~\ref{tab:ID} and
depicted in Fig.~\ref{fig:config}.

\begin{table}
\caption{Initial data parameters and system details.  The punctures are located
at $\vec r_1 = (x_1,\,0,\,z)$ and $\vec r_2 = (x_2,0,z)$, with momenta
$\vec P=\pm (0,\,P,\,0)$, spins $\vec S_1 = (0,\,0,\,S_{1z})$
and $\vec S_2 = (S_{2x},\,0,\,S_{2z} )$, 
mass parameters $m^p$, horizon (Christodoulou) masses $m^H$, total ADM mass
$M_{\rm ADM}$, and dimensionless spins $\alpha = a/m_H = S/m_H^2$. The horizon
masses and spins are given after the gauge settles, and the
errors in those quantities, denoted as $\delta m^H$ and 
$\delta\alpha$, are determined by the drift in the quantity 
during the inspiral.  Also provided are the simple proper distance $d$,
eccentricity at the start of the inspiral $e_i$, 
and eccentricity $e_f$ and the number of orbits $N$ just before merger.
}
\label{tab:ID}
\begin{ruledtabular}
\begin{tabular}{ccccc}
$x_1/m$ & $x_2/m$  & $z/m$ & $P/m$ & $d/m$\\
10.73983 & -10.76016 & -0.01968 & 0.05909 & 25.37 \\
\hline
$m^p_1/m$ & $m^p_2/m$ & $S_{1z}/m^2$ & $S_{2x}/m^2$ & $S_{2z}/m^2$\\
0.48543 & 0.30697 & 0.05 & 0.19365 & -0.05 \\
\hline
$M_{\rm ADM}/m$ & $J_{\rm ADM}/m^2$ & $e_i$ & $e_f$ & $N$ \\
0.99472 & 1.2704344 & 0.0322 & 0.0006 & 48.5 \\
\hline
$m^H_1/m$ & $\delta m^H_1/m$ & $m^H_2/m$ & $\delta m^H_2/m$ & \\
0.50000 & 0.00002 & 0.49974 & 0.00001 & \\
\hline
$\alpha_1$ & $\delta \alpha_1 $ & $\alpha_2$ & $\delta \alpha_2$& \\
0.20003 & 0.00056 & 0.80088 & 0.00066 & \\
\end{tabular}
\end{ruledtabular}
\end{table}

We evolve the following BHB data sets using the {\sc
LazEv}~\cite{Zlochower:2005bj} implementation of the moving puncture
approach~\cite{Campanelli:2005dd,Baker:2005vv} with the conformal
function $W=\sqrt{\chi}=\exp(-2\phi)$ suggested by
Ref.~\cite{Marronetti:2007wz}.  For the run presented here, we use
centered, eighth-order finite differencing in
space~\cite{Lousto:2007rj} and a fourth-order Runge Kutta time
integrator (note that we do not upwind the advection terms).

Our code uses the {\sc EinsteinToolkit}~\cite{Loffler:2011ay,
einsteintoolkit} / {\sc Cactus}~\cite{cactus_web} /
{\sc Carpet}~\cite{Schnetter-etal-03b}
infrastructure.  The {\sc
Carpet} mesh refinement driver provides a
``moving boxes'' style of mesh refinement. In this approach, refined
grids of fixed size are arranged about the coordinate centers of both
holes.  The {\sc Carpet} code then moves these fine grids about the
computational domain by following the trajectories of the two BHs.

We use {\sc AHFinderDirect}~\cite{Thornburg2003:AH-finding} to locate
apparent horizons.  We measure the magnitude of the horizon spin using
the {\it isolated horizon} (IH) algorithm detailed in
Ref.~\cite{Dreyer02a} and as implemented in Ref.~\cite{Campanelli:2006fy}.
Note that once we have the horizon spin, we can calculate the horizon
mass via the Christodoulou formula
\begin{equation}
{m_H} = \sqrt{m_{\rm irr}^2 + S_H^2/(4 m_{\rm irr}^2)} \,,
\end{equation}
where $m_{\rm irr} = \sqrt{A/(16 \pi)}$, $A$ is the surface area of
the horizon, and $S_H$ is the spin angular momentum of the BH (in
units of $M^2$).  In the tables below, we use the variation in the
measured horizon irreducible mass and spin during the simulation as a
measure of the error in computing these quantities.  
We measure radiated energy,
linear momentum, and angular momentum, in terms of the radiative Weyl
Scalar $\psi_4$, using the formulas provided in
Refs.~\cite{Campanelli:1998jv,Lousto:2007mh}. However, rather than
using the full $\psi_4$, we decompose it into $\ell$ and $m$ modes and
solve for the radiated linear momentum, dropping terms with $\ell >
6$.  The formulas in Refs.~\cite{Campanelli:1998jv,Lousto:2007mh} are
valid at $r=\infty$.  We extract the radiated energy-momentum at
finite radius and extrapolate to $r=\infty$, and find that the new
perturbative extrapolation described in Ref.~\cite{Nakano:2015pta} provides the
most accurate waveforms. While the difference of fitting both linear and
quadratic extrapolations provides an independent measure of the error.

\begin{figure}
\includegraphics[width=\columnwidth]{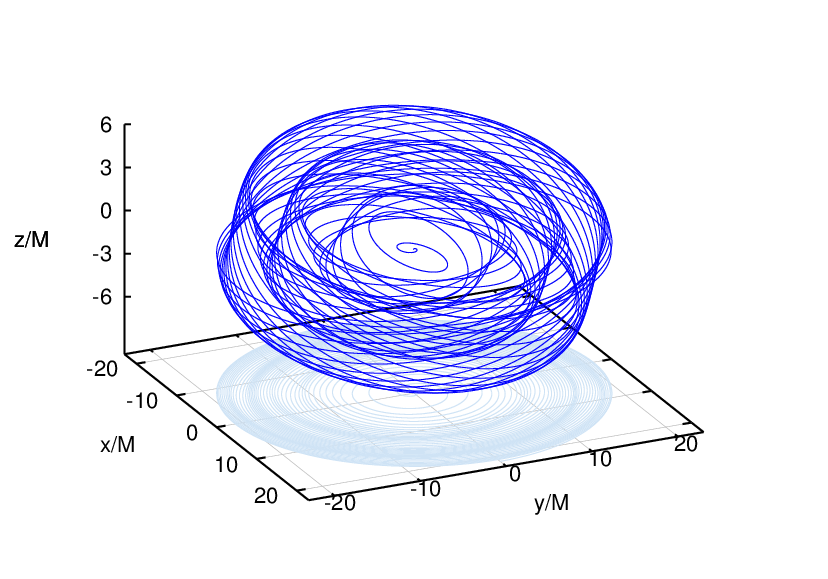}
\caption{Precession of the orbital plane as displayed by the distance
vector $\vec{d}=\vec{x}_1(t)-\vec{x}_2(t)$.
\label{fig:precession}}
\end{figure}

Figure~\ref{fig:precession} highlights the important effects of 
precession of the orbital plane during the whole evolution and
the three precession cycles occurring over the nearly 50 orbits.

\begin{figure}
\includegraphics[angle=270,width=0.49\columnwidth]{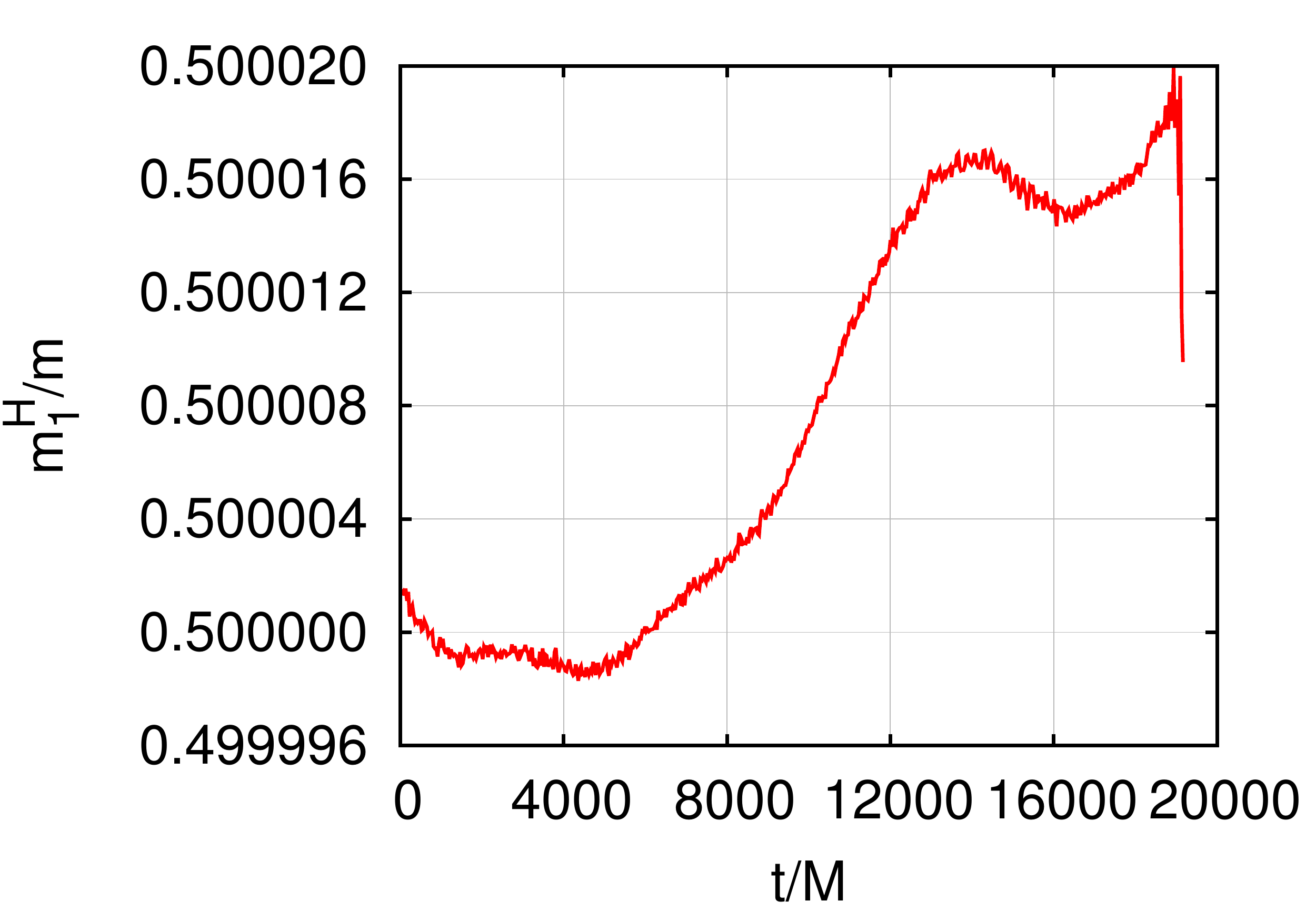}
\includegraphics[angle=270,width=0.49\columnwidth]{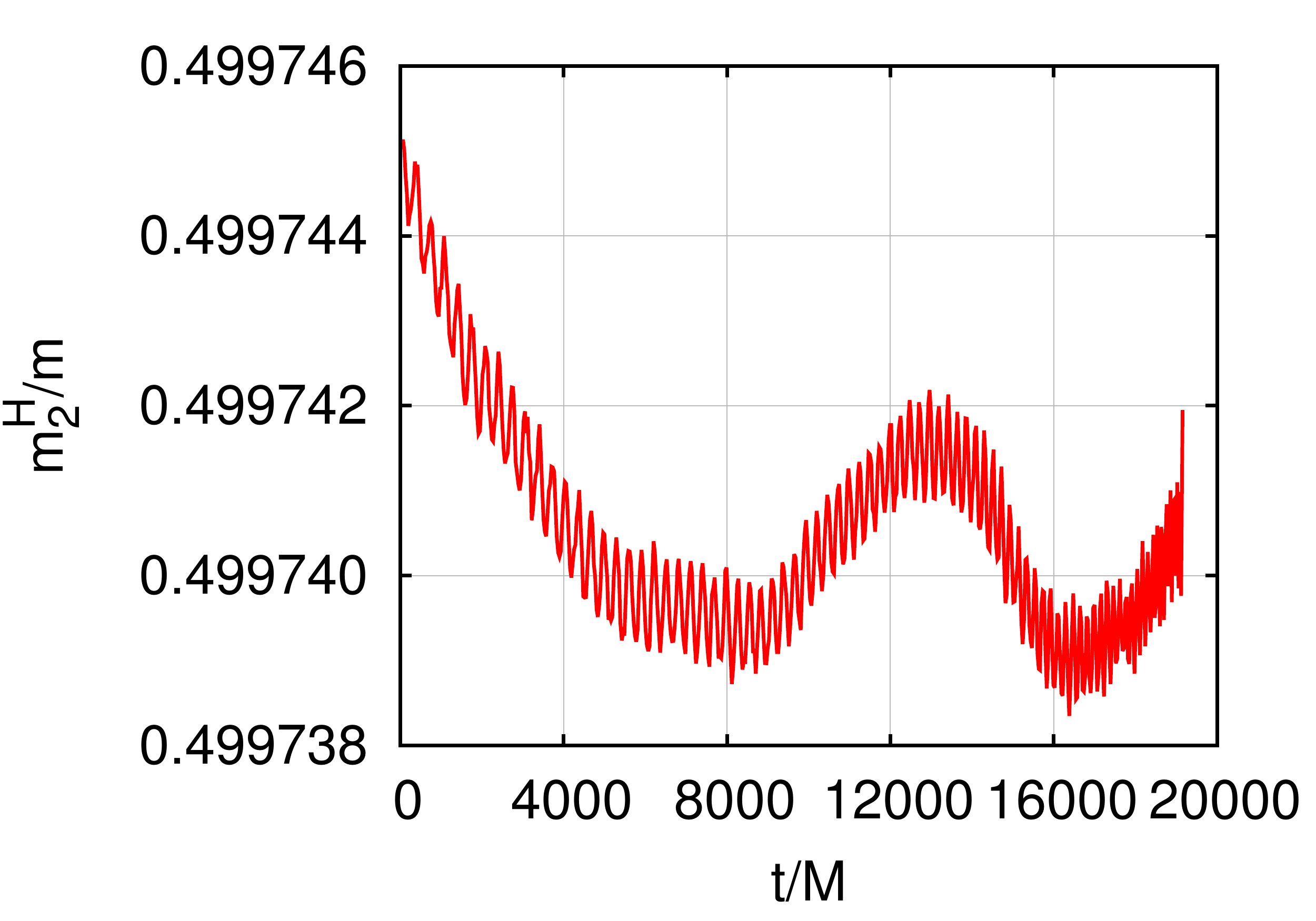}\\
\includegraphics[angle=270,width=0.49\columnwidth]{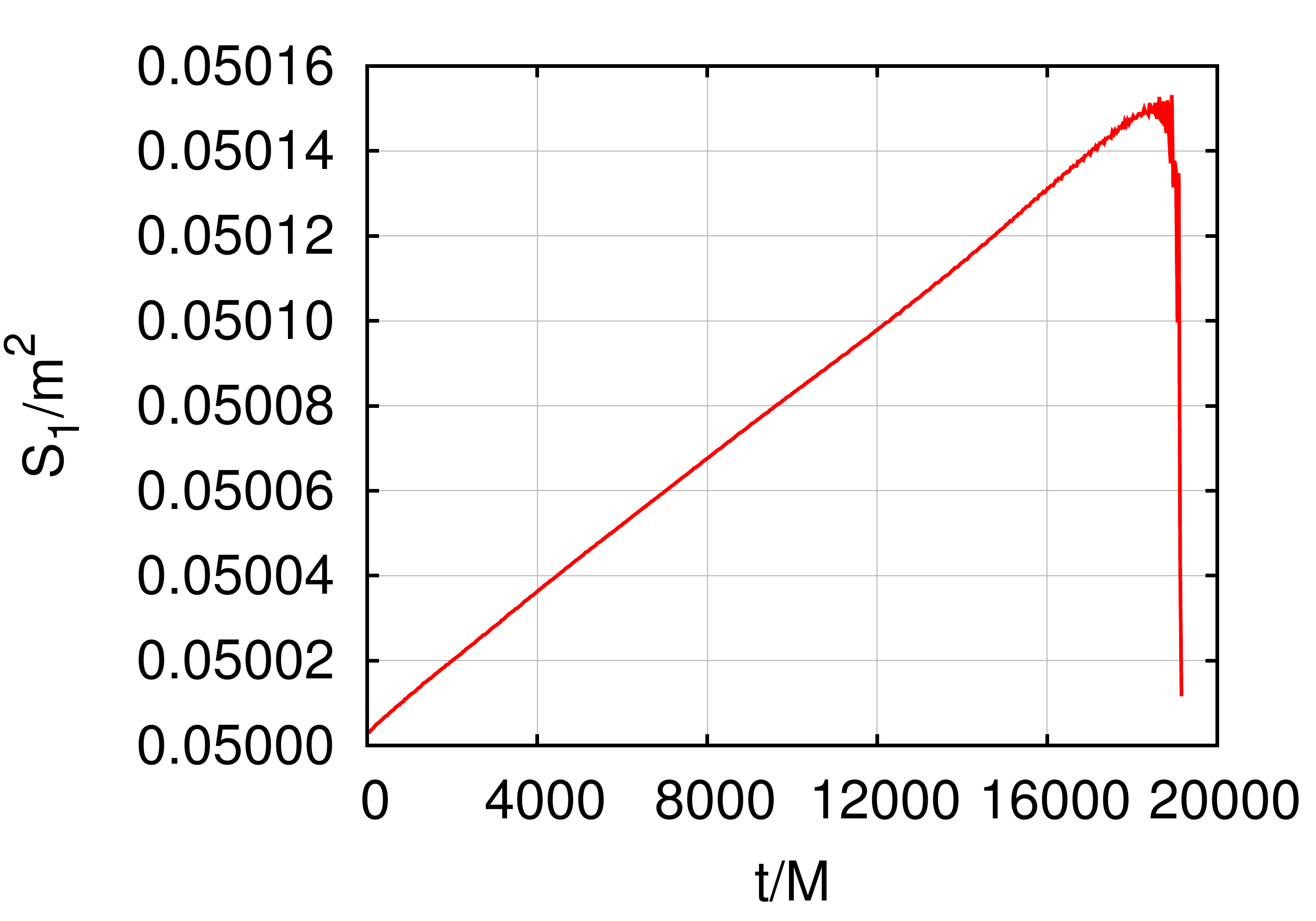}
\includegraphics[angle=270,width=0.49\columnwidth]{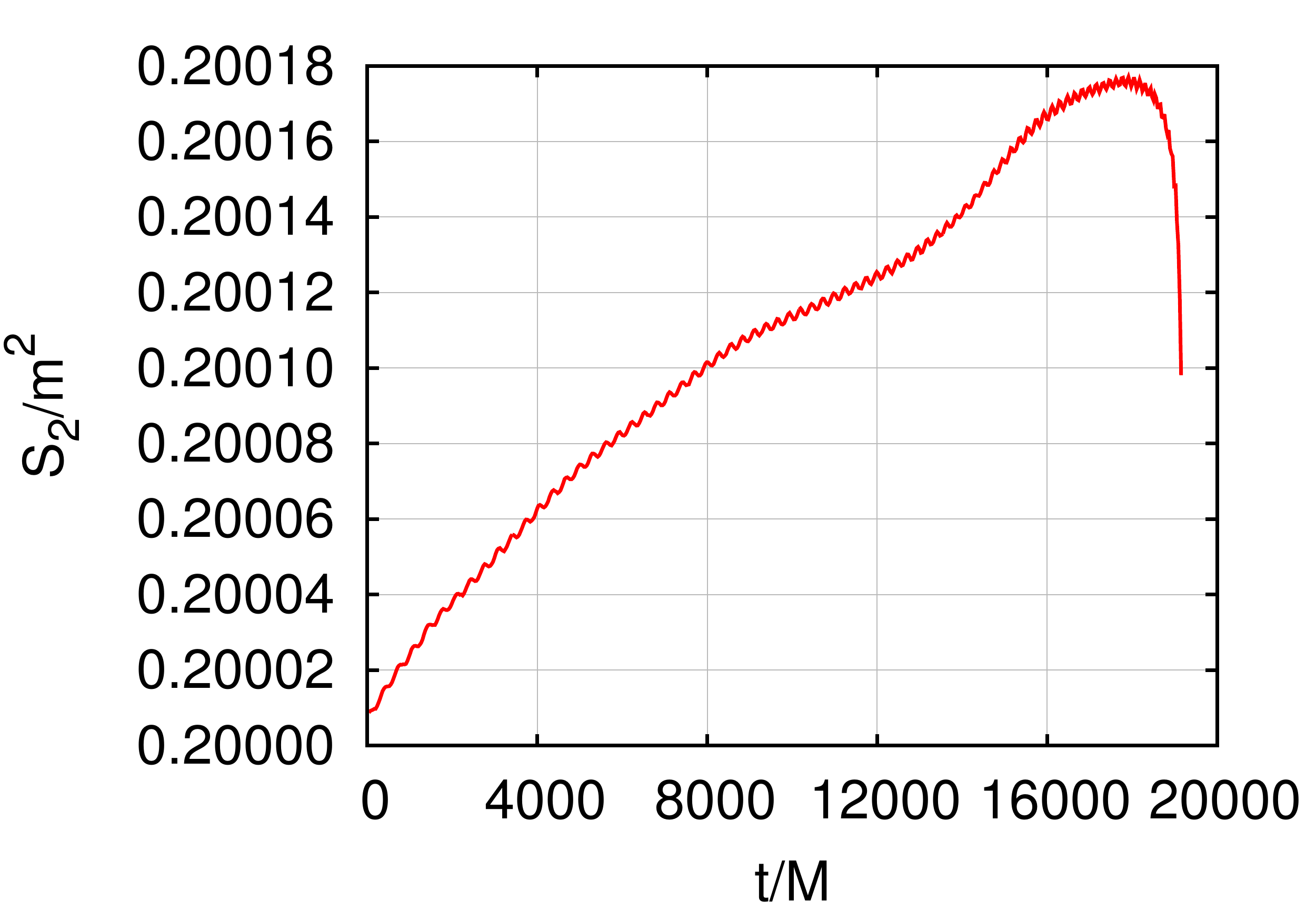}\\
\caption{The conservation of the individual masses and spin 
magnitudes of the
holes during the binary evolution.
\label{fig:massspinbinary}}
\end{figure}

Through evolution we track the horizon mass and magnitude and components 
of the individual spins of the BHs. Figure~\ref{fig:massspinbinary}
displays the levels at which these quantities are conserved during
the $t=20,000M$ of full numerical evolution and this provides a 
measure of the accuracy of the simulation, i.e., the variations
in $\Delta m_1^H/m_1^H\approx4\times10^{-5}$, 
$\Delta S_1^H/S_1^H\approx3\times10^{-3}$, 
$\Delta m_2^H/m_2^H\approx1.2\times10^{-5}$, and 
$\Delta S_2^H/S_2^H\approx9\times10^{-4}$.

\begin{figure}
\includegraphics[angle=270,width=\columnwidth]{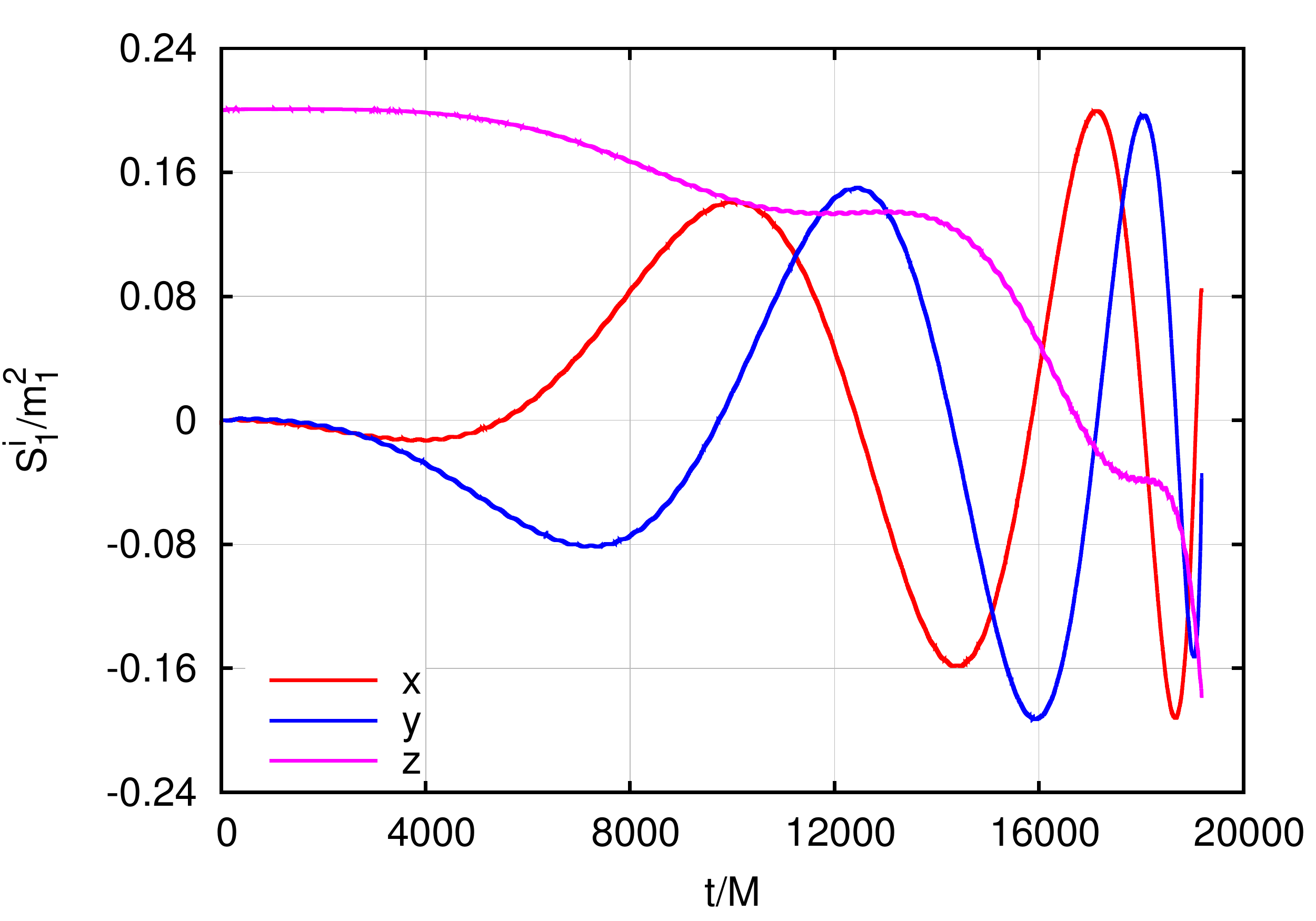}
\includegraphics[angle=270,width=\columnwidth]{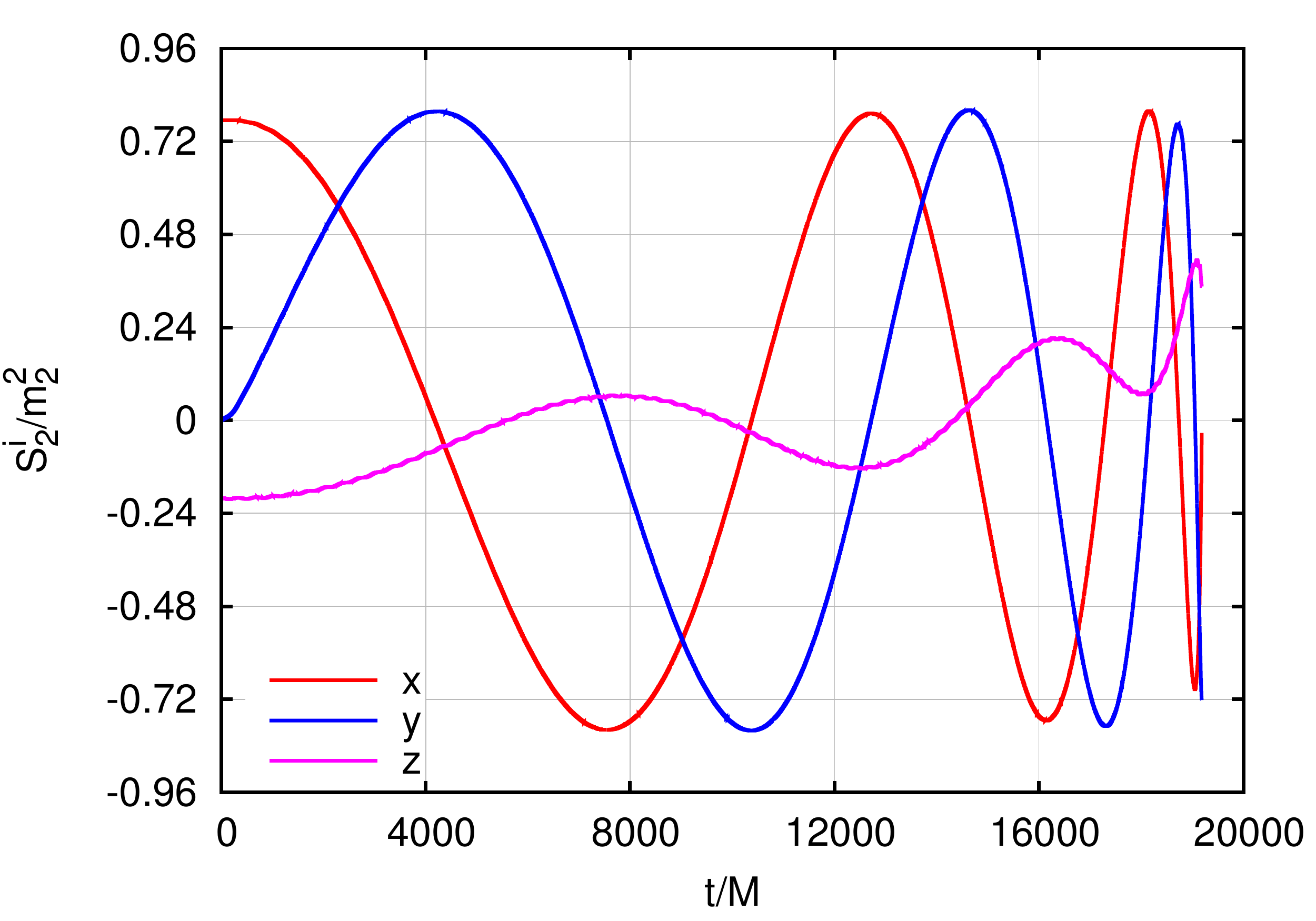}
\caption{BH (1,2) spin components in the coordinate frame. The top panel is BH1
and the bottom panel is BH2.
\label{fig:spindirections}}
\end{figure}

The spin components in the simulation coordinate system
are displayed in Fig.~\ref{fig:spindirections}. They show
the polar component of $\vec{S}_1$ from alignment at $t=0$ to almost
misalignment at merger and the complementary behavior for
$\vec{S}_2$. From the variations with time of the $(\hat{x},\,\hat{y})$ 
components of the spins one can also read-off the precessional
effects. While these are coordinate based periods, one can
verify, for instance from the oscillations in the amplitude
of the $(\ell=2,\,m=1)$ mode of the gravitational waveform in a gauge
invariant way that those periods correspond to the physical
dynamics of spins (see Fig.~\ref{fig:waveforms} in this paper 
and also Ref.~\cite{Campanelli:2008nk}).

\begin{figure}
\includegraphics[width=0.49\columnwidth]{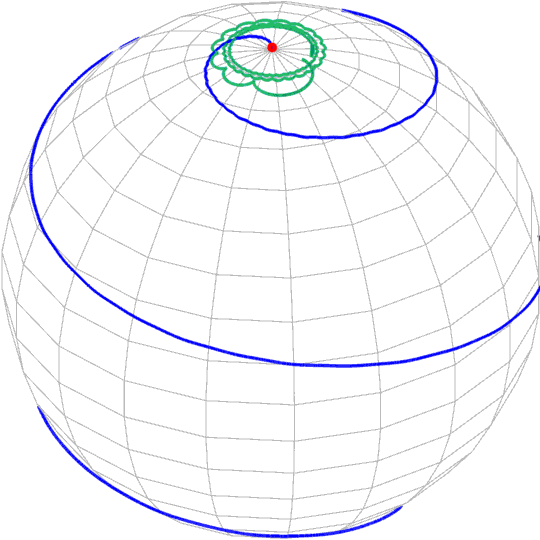}
\includegraphics[width=0.49\columnwidth]{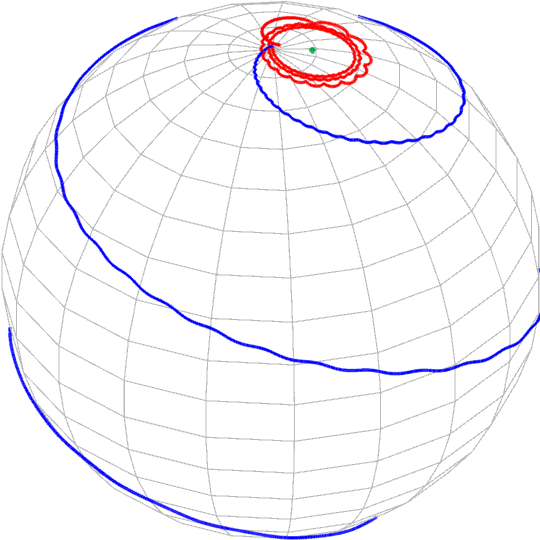}
\caption{Change in the spin direction (in blue) of the BH1
(with the smaller spin magnitude) in
the orbital frame, $\hat{L}$ (left),
and the coordinate frame $\hat{z}$ (right). Plotted also the directions of
$\vec{L}$ in red and $\vec{J}$ in green.
\label{fig:spinsphere}}
\end{figure}

\begin{widetext}

\begin{figure}
\includegraphics[angle=270,width=\columnwidth]{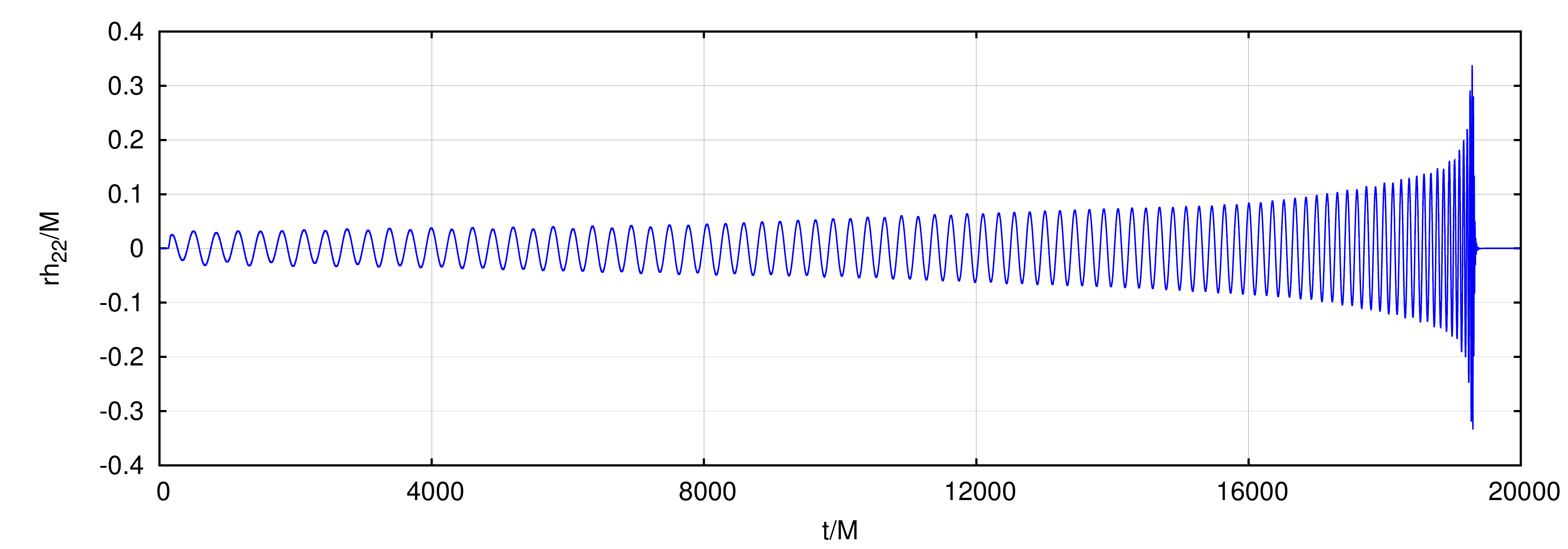}\\
\includegraphics[angle=270,width=\columnwidth]{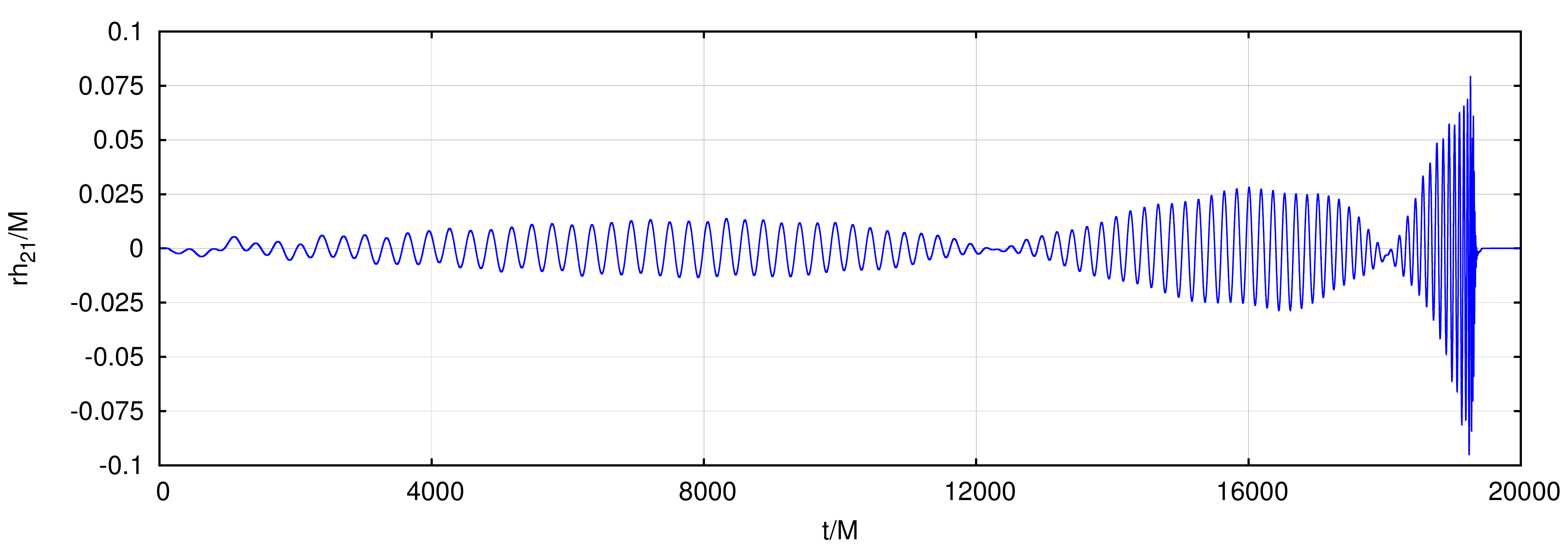}
\caption{The full numerical 
$(\ell=2,\,m=2)$ waveform (above) and
$(\ell=2,\,m=1)$ waveform (below) as extracted by observers
at $r=175M$.
\label{fig:waveforms}}
\end{figure}

\end{widetext}

Another useful representation of the spins~\cite{Lousto:2014ida} is given in
Fig.~\ref{fig:spinsphere} where the flip-flopping spin's 
($\hat{S}_1$) trajectory can be described as a 
composition of the polar (flip-flop) and azimuthal (precession)
dynamics which resembles that of an orange peeling.
For the sake of future interpretations
and applications we also provide the spin trajectories as seen
in an (approximately) inertial frame where the direction of
the total angular momentum $\hat{J}$ is conserved and coincident
(approximately) with the coordinate direction $\hat{z}$.

The gravitational waveforms from spinning, precessing BHB
systems are of great interest for detection and
parameter estimation in the forthcoming observation of
gravitational waves by Advanced LIGO~\cite{TheLIGOScientific:2014jea}
and other laser interferometric gravitational wave observatories.
Our simulation produced one of the longest waveforms studied to date,
starting from an initial configuration with proper separation, $d=25M$.
This is well in the post-Newtonian regime (see also the recent
simulation~\cite{Szilagyi:2015rwa}
for nonspinning, unequal mass binaries starting at an initial
separation $d=27M$). In Fig.~\ref{fig:waveforms} the leading
modes $(\ell=2,\,m=2)$ and $(2,\,1)$ are displayed~\cite{Lousto:2014ida} 
and the first shows
the well known chirp behavior with essentially monotonic
increase of amplitude and frequency until merger.
The second waveform, corresponding to the $(\ell=2,\,m=1)$ mode,
displays a modulation in amplitude that corresponds to
the period of precession of the orbital plane (and in
this equal mass binary case this is coincident with the
precession of the spins). Thus this waveform provides
an independent and gauge invariant measure of the precession
period.

Of particular interest is to determine the precision of this
waveform. Due to the high computational demand of this simulation,
$2.5$ million service units on 25 
to 30 nodes of our local cluster ``Blue Sky'' with dual Intel Xeon E5-2680 
processors nearing $100M$ of evolution per day, we have not
produced detailed convergence studies involving several runs at
different resolutions. Instead we compare our full numerical
waveforms with those produced with 3.5PN evolutions and found
excellent agreement, particularly at early times, when 3.5PN is
expected to be very accurate, given the large initial separation
of the holes. 

\begin{figure}
\includegraphics[angle=270,width=\columnwidth]{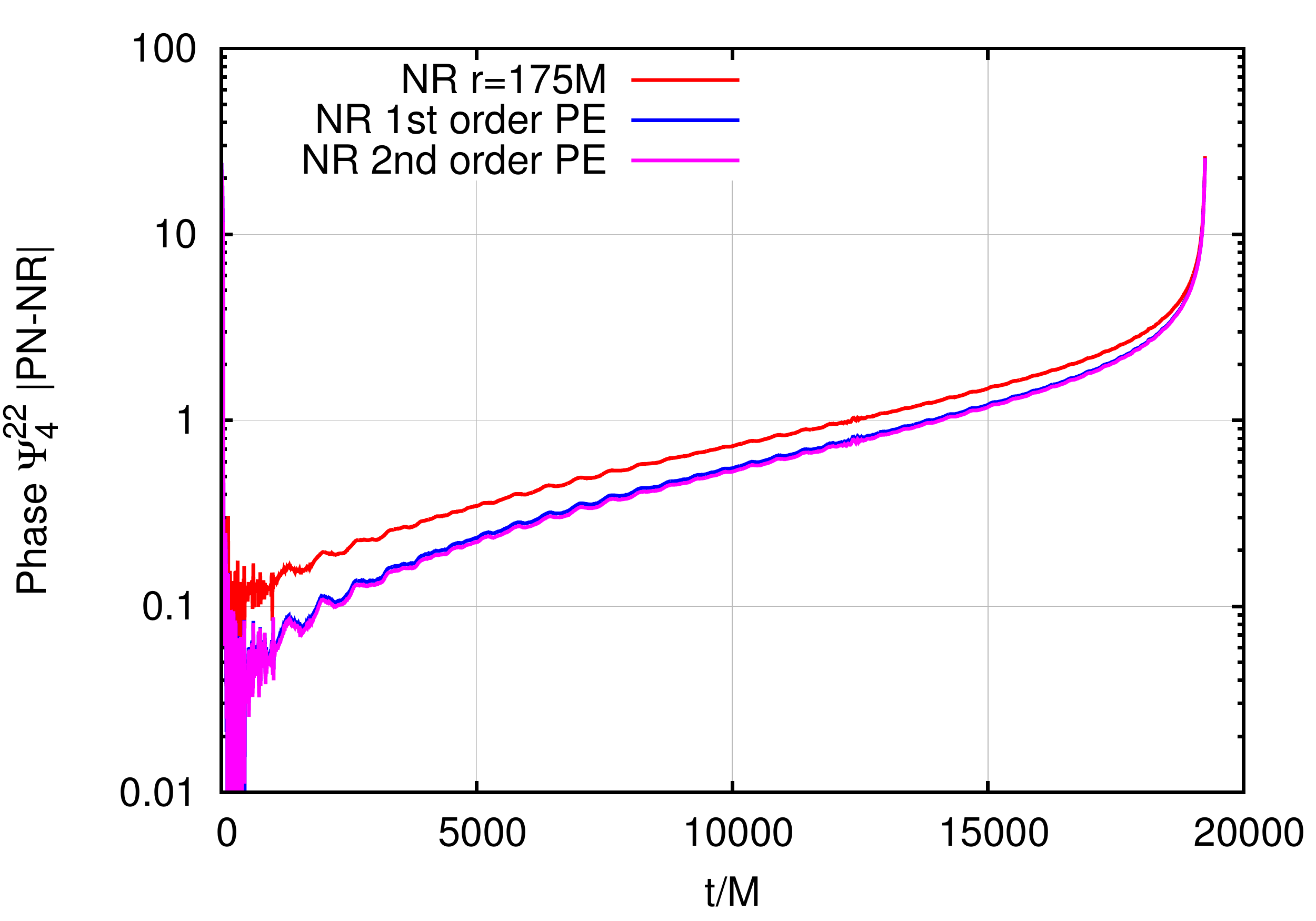}
\caption{
The phase difference between the full numerically generated
$(\ell=2,\,m=2)$ waveform (NR) and 
the 3.5PN waveform. Differences are notably reduced
when the NR-waveform, extracted at $r=175M$, is extrapolated to
an infinite observer location by the perturbative formulae (PE) of
1st order, accurate to $(1/r)$ terms and second order, accurate to $(1/r^2)$.
\label{fig:phase}}
\end{figure}

\begin{figure}
\includegraphics[angle=270,width=\columnwidth]{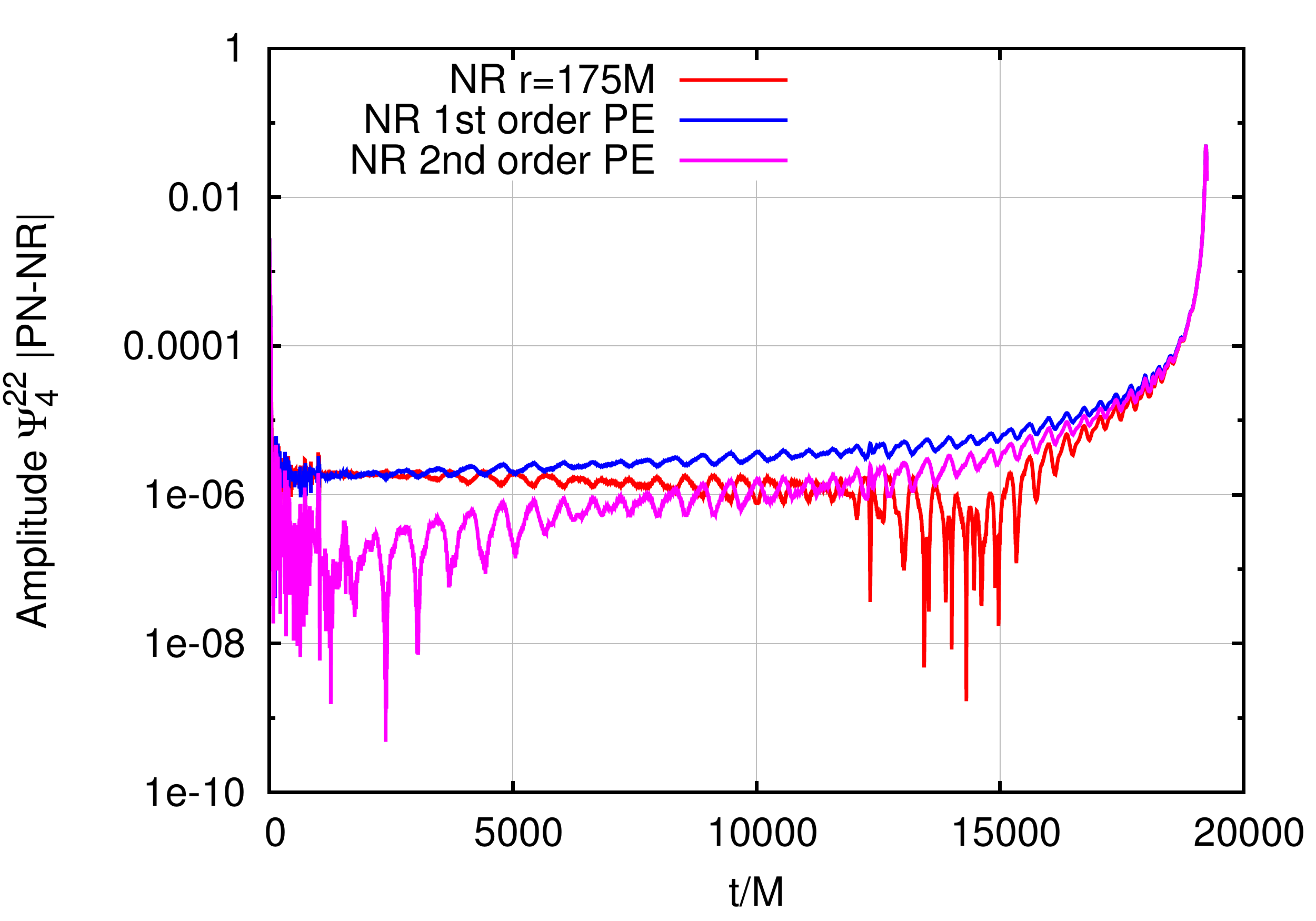}
\caption{The amplitude difference between the full numerically generated
$(\ell=2,\,m=2)$ waveform (NR) and 
the 3.5PN waveform. Differences are notably reduced
when the NR-waveform, extracted at $r=175M$, is extrapolated to
an infinite observer location by the perturbative formulae (PE) of
1st order, $(1/r)$, and second order, $(1/r^2)$.
\label{fig:amplitude}}
\end{figure}

Figures~\ref{fig:phase} and~\ref{fig:amplitude} display
the differences of the numerical and PN waveform for the $(\ell=2,\,m=2)$ mode phase
and amplitude, respectively. 
Since the full numerical waveforms are extracted at a finite
observer location (in this case at $r=175M$) we extrapolate
them to null infinity with a recent ${\cal O}(1/r^2)$ accurate
perturbative method~\cite{Nakano:2015rda,Nakano:2015pta}
to obtain further agreement, showing reasonable errors until near
the final merger.

\begin{figure}
\includegraphics[angle=270,width=\columnwidth]{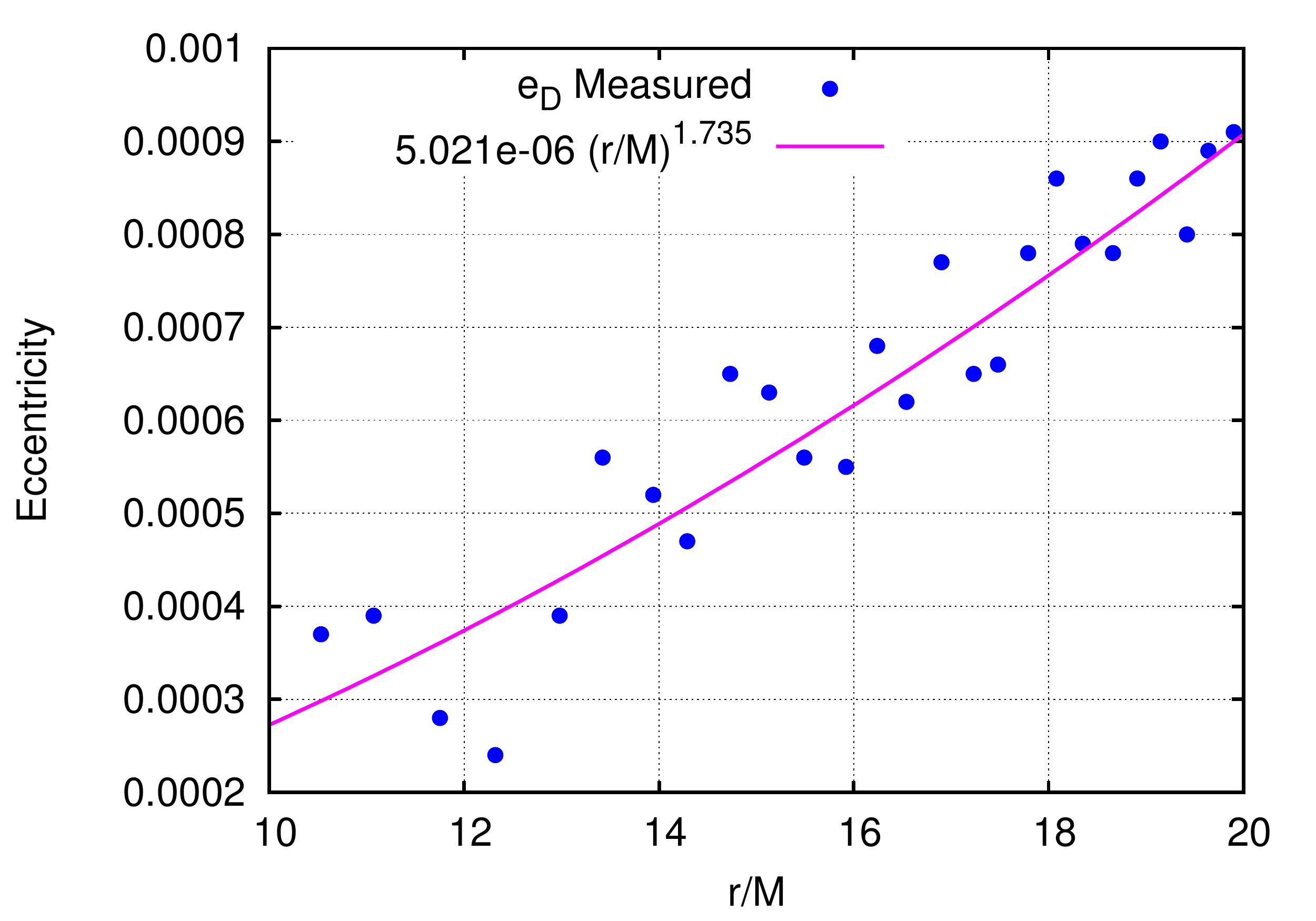}
\caption{Evolution of the eccentricity versus coordinate separation of
the BHB. Dots represent measurements of the eccentricity,
$e_D$, and the continuous curve a fit to its decay with $e\sim r^{1.73486\pm0.1495}$. In comparison, the theoretical prediction is $e\sim r^{1.5833}$.
\label{fig:ecc}}
\end{figure}

Since this BHB simulation is relatively long-term
it is well suited to study the evolution of the eccentricity as a function
of the binary's separation. Using the method of Ref.~\cite{Mroue:2010re}
to determine $e_D$,
we measure the eccentricity per orbit (we have not included the first
two orbits, contaminated by gauge settling, and last few orbits 
due to strong radiative effects). As the binary's separation
shrinks due to gravitational radiation the eccentricity reduces
as shown in Fig.~\ref{fig:ecc}.
We performed a fit to the measured eccentricity versus separation
of the form $a\,r^b$ obtaining an exponent $b=1.73486\pm0.1495$ to
compare with the lowest post-Newtonian prediction of Ref.~\cite{Peters:1964zz}
for radiation of eccentricity, $e\sim r^{19/12}$. With $19/12=1.5833$,
the post-Newtonian prediction is within the statistical error bars of 
the fit.

After merger the binary forms a single final remnant BH
with characteristics given in Table~\ref{tab:remnant}. Notably
the recoil velocity reaches $1500$~km/s in agreement with the
empirical predictions~\cite{Lousto:2012gt,Lousto:2013wta}.

\begin{table}
\caption{Remnant properties and recoil velocity.
The final mass and spin are measured from the horizon, and the
recoil velocity is calculated from the gravitational waveforms.
The error in the mass and spin is determined by the drift in 
those quantities after the remnant settles down.  The error in
the recoil velocity is estimated by the difference between first and second
order polynomial extrapolation to infinity of the waveforms.
\label{tab:remnant}}
\begin{ruledtabular}
\begin{tabular}{ccc}
$M_{rem}/m$ & $|\alpha_{rem}|$  & $V_{recoil}$~[km/s] \\
$0.94904 \pm 0.00000$ & $0.70377 \pm 0.00002$ & $1508.49 \pm 16.08$ \\
\hline
$\alpha_{rem}^x$ & $\alpha_{rem}^y$  & $\alpha_{rem}^z$ \\
$0.10815\pm0.00003$ & $-0.01986\pm0.00000$ & $0.69513\pm0.00002$ \\
\end{tabular}
\end{ruledtabular}
\end{table}

\begin{figure}
\includegraphics[angle=270,width=0.49\columnwidth]{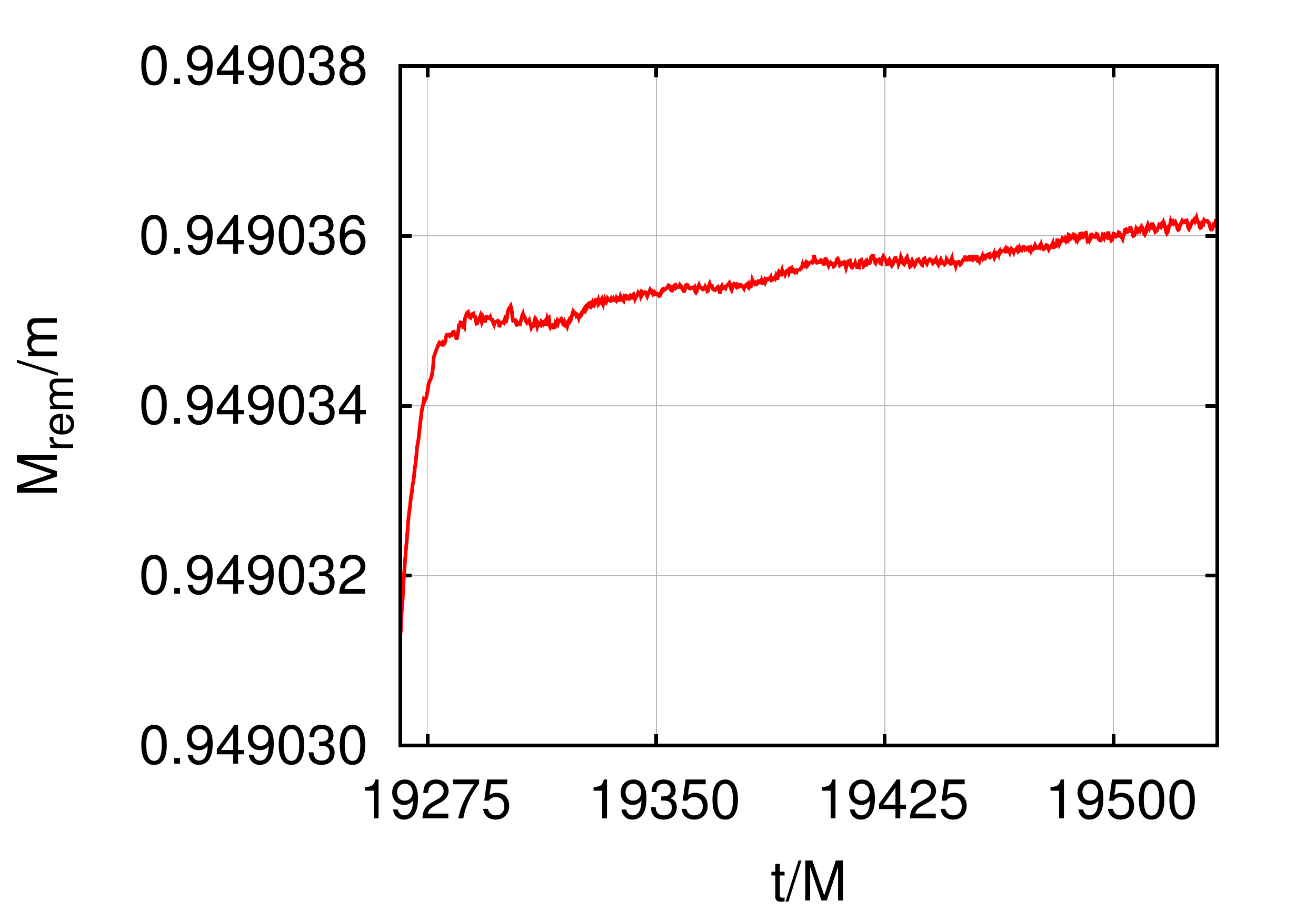}
\includegraphics[angle=270,width=0.49\columnwidth]{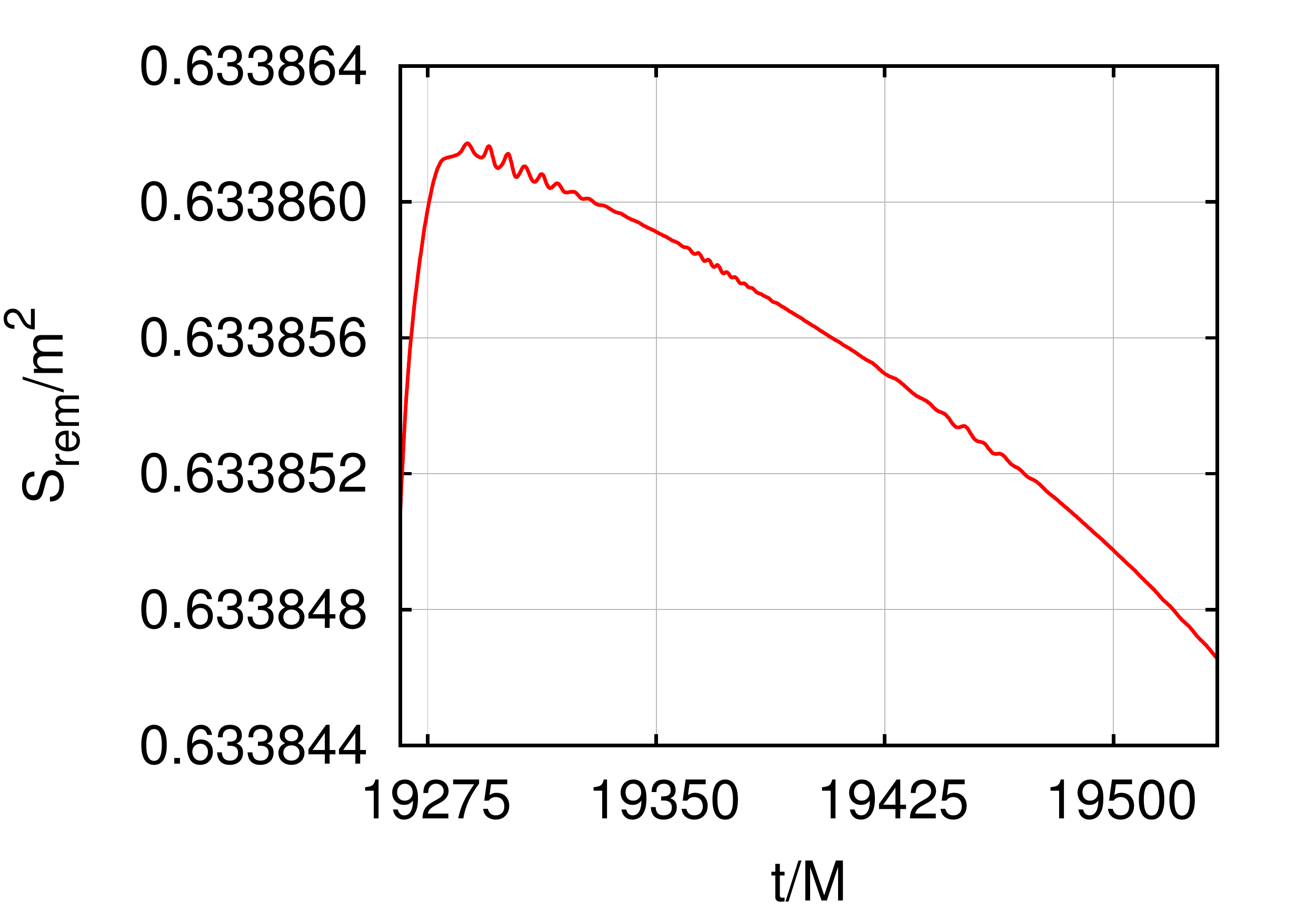}
\caption{The mass and spin of the final merged BH.
\label{fig:massspinfinal}}
\end{figure}

Figure~\ref{fig:massspinfinal} displays the settling of the
final mass and spin measures as the final BH radiates
away its last distortions and becomes a quiet Kerr
BH~\cite{Campanelli:2008dv}.

\subsection{Vector Analysis}\label{subsec:Vec}

\begin{figure}
\includegraphics[width=0.49\columnwidth]{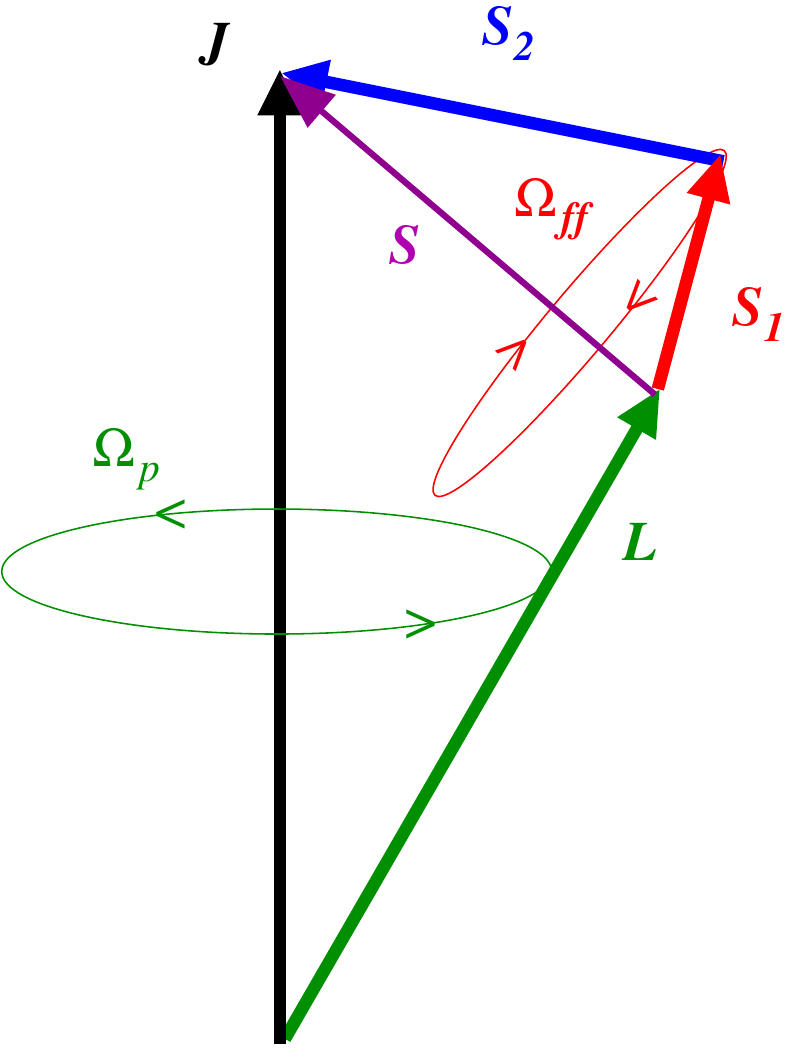}
\caption{Flip-flop of $\vec{S}_1$ and $\vec{S}_2$ as they precess with
$\vec{L}$.
\label{fig:JLS}}
\end{figure}

In order to visualize the origin of this {flip-flop} effect
we can use a simple vector addition model to represent the total
angular momentum of the system, $\vec{J}=\vec{L}+\vec{S}$, as
the sum of the orbital angular momentum $\vec{L}$ and the total
spin $\vec{S}=\vec{S}_1+\vec{S}_2$. This total spin being the
sum of the individual spin of the holes $\vec{S}_1$ and $\vec{S}_2$
as represented in Fig.~\ref{fig:JLS}.

In this picture, the six degrees of freedom of the spins 
$\vec{S}_1$ and $\vec{S}_2$ are shown as vectors in 3-space.
These degrees of freedom are constrained 
by the conservation of the magnitude of the individual spins
$S_1$, and $S_2$, the magnitude of the total spin $S$, and
its projection along the orbital angular momentum, $\vec{S}\cdot\hat{L}$.
The conservation of the magnitudes of the individual spins and
the magnitude of their sum $S$, can be seen in the 
2PN evolution of the spins in Eqs.~(\ref{spinevo}) of Sec.~\ref{subsec:q1PN}.
The conservation 
of $\vec{S}_0\cdot\hat{L}$ was shown in Ref.~\cite{Racine:2008qv} 
using the orbit averaged version of Eqs.~(\ref{spinevo}).
The conservation of these quantities in full numerical simulations
of BHBs has been observed to be approximately 
true in Refs.~\cite{Campanelli:2006fg} and~\cite{Lousto:2013vpa}. 
Note that in what follows to determine
the flip-flop angle, we do not use the PN (or full numerical) dynamics
beyond the referred conservation of the individual and total spin
magnitudes.

\begin{figure}
\includegraphics[width=0.66\columnwidth]{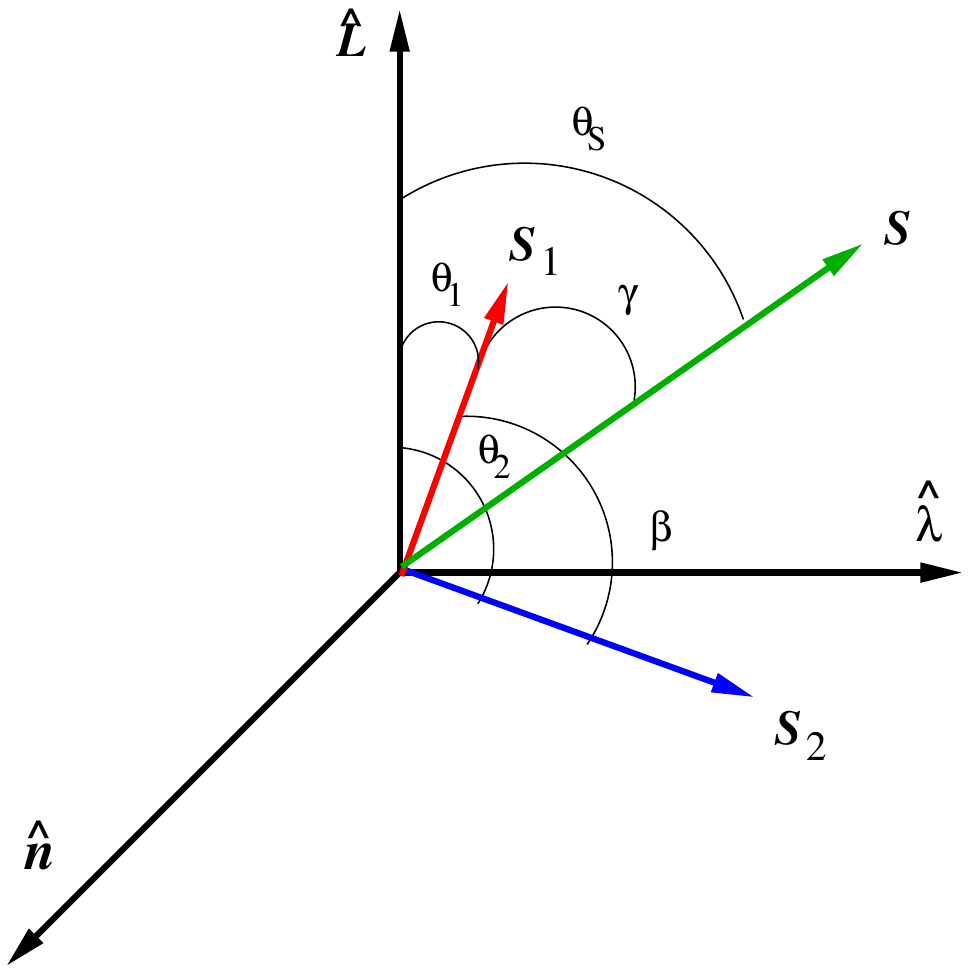}
\caption{The configuration of the spins in the orbiting frame.
Spin configurations $\vec{S}_1$ and $\vec{S}_2$
relative to the orbital angular momentum $\vec{L}$. 
Here $\vec{S}=\vec{S}_1+\vec{S}_2$.
\label{fig:frame}}
\end{figure}

It follows hence that the following quantities are conserved
(see Fig.~\ref{fig:frame})
\bea
\vec{S}\cdot\vec{S}&=&S^2=S_1^2+S_2^2+2S_1S_2\cos\beta=\text{constant} \,,
\cr
\vec{S}\cdot\vec{S}_1&=&SS_1\cos\gamma=S_1^2+S_2S_1\cos\beta=\text{constant} \,,
\cr
\vec{S}\cdot\vec{S}_2&=&SS_2\cos(\beta-\gamma)=S_2^2+S_2S_1\cos\beta
\cr &=&\text{constant} \,.
\eea
This leads to the conservation of $\beta$ and $\gamma$ during the evolution of
the binary.

In particular we find that $\vec{S}_1$ oscillates 
around $\vec{S}$ between
angles $\gamma$ and $-\gamma$ (when it is both coplanar to 
$\vec{S}$ and $\vec{L}$), where
\be\label{gamma}
\cos\gamma=\frac{S_1+S_2\cos\beta}{\sqrt{S_1^2+S_2^2+2S_1S_2\cos\beta}}
=\frac{S^2+S_1^2-S_2^2}{2SS_1} \,.
\ee
BH2 also oscillates at the same flip-flop
frequency  $\Omega_{ff}$, 
but with a smaller angle (since we consider $S_2>S_1$)
given by $\pm(\beta-\gamma)$ where
\be
\cos(\beta-\gamma)=\frac{S^2+S_2^2-S_1^2}{2SS_2} \,.
\ee
Thus both spins, $\vec{S}_1$ and $\vec{S}_2$, oscillate around $\vec{S}$ 
which in turn precesses around $\vec{L}$. 

The direction of oscillation of the spins 
can be seen from the leading precession
equation for $d\vec{L}/dt=(7/2r^3)(\vec{J}\times\vec{L})$ indicating
that $\vec{L}$ oscillates counterclockwise. Likewise the leading precession
for $d\vec{S}_1/dt = (1/r^3)((7/2) \vec{J}\times\vec{S}_1 -3 \vec{S}\times\vec{S}_1)$
indicates that $\vec{S}_1$ oscillates clockwise around $\vec{S}$
(see Eqs.~\eqref{spinevoave} for the precise expression). Thus  
Fig.~\ref{fig:JLS}, is represented as seen by an observer located up (north)
with respect to $\vec{J}$ direction.

We can define the {flip-flop} angle with respect to the polar coordinates
as measured from $\hat{L}$ (analogous formulae can be defined for $\hat{J}$)
\be
\Delta\theta_i^{ff}=\theta_i^{max}-\theta_i^{min} \,,
\ee
where $i=1,2$ labels either BH spin.

We have three cases to consider depending on
$\theta_S$, where $\cos(\theta_S)=S_L/S=(\vec{S}\cdot\hat{L})/S$,
\bea\label{eq:thetaff}
\Delta\theta_1^{ff} &=
\begin{cases}
\theta_S + \gamma   & \text{if  } 0 \leq \theta_S \leq \gamma \,, \\
2 \gamma            & \text{if  } \gamma \leq \theta_S \leq \pi-\gamma \,, \\
\pi - \theta_S + \gamma & \text{if  } \pi-\gamma \leq \theta_S \leq \pi \,,
\end{cases}
\eea
for $0\leq\gamma\leq\pi/2$,
and similar for BH2, replacing $\gamma$ by $\beta-\gamma$.
When $\pi/2\leq\gamma\leq\pi$, we replace $\gamma \leftrightarrow \pi-\gamma$
in the equation above.

Hence, for instance, in order to maximize the flip-flop angle, we can set
$\gamma=\pi/2$, which leads to the vanishing of $\cos\gamma$
on the left hand side of the above equation~(\ref{gamma}), leading
to the condition $S_1+S_2\cos\beta=0$, for $q=1$, used in our full numerical
configuration, as detailed in Sec.~\ref{subsec:FN}.

This oscillation of the spins represent a {\it genuine} spin-flip in the
sense that it is the same object that completely changes its spin orientation. 
This is different from the simple case where the final remnant spin has flipped
direction compared to the spin of one of the individual orbiting
BHs~\cite{Campanelli:2006fy}.
It is also different from a classic analog of the hyperfine transition
of an electron in the Hydrogen atom flipping its spin and
leading to the famous $21$\,cm radio emission. Here the flip also occurs in
a conservative set up, with no emission of gravitational waves, and it is
compensated by an opposite flop from the other BH.

\subsection{Post-Newtonian Spin Dynamics}\label{subsec:q1PN}

The vector analysis is helpful to visualize the angles of
flip-flop but in order to determine their rate of oscillation
one has to resort to a dynamical computation. In order to
provide simple, approximate, expressions we will perform a
2PN (conservative) study.

The precession equations for the spins 
$\vec{S}_1$ and $\vec{S}_2$ with a mass ratio $q=m_1/m_2$
to leading spin-orbit and spin-spin couplings in 
the (2PN) post-Newtonian expansion~\cite{Buonanno:2005xu}
take the form
\bea\label{spinevo}
\frac{d\vec{S}_1}{dt}&=&
\frac{1}{r^3}\left[\left(2+\frac{3}{2q}\right)\vec{L}-\vec{S}_2
+\frac{3( \vec{S}_0\cdot\hat{n})}{1+q}\hat{n}\right]\times\vec{S}_1 \,,\nonumber\\
\frac{d\vec{S}_2}{dt}&=&
\frac{1}{r^3}\left[\left(2+\frac{3q}{2}\right)\vec{L}-\vec{S}_1
+\frac{3q(\vec{S}_0\cdot\hat{n})}{1+q}\hat{n}\right]\times\vec{S}_2 \,,
\eea
where $\vec{n}=(\vec{r}_1-\vec{r}_2)/|\vec{r}_1-\vec{r}_2|$ and
\be\label{eq:S0}
\vec{S}_0=\left(1+\frac1q\right)\vec{S}_1+(1+q)\,\vec{S}_2 \,.
\ee
For more details see, for instance, the reviews in 
Refs.~\cite{Blanchet:2013haa,Schafer:2014cxa}.
The averaging of Eqs.~\eqref{spinevo} over the orbital period gives
\bea\label{spinevoave}
\langle \dot{\vec{S}}_1 \rangle &=&
\frac{1}{r^3}\left( \left( \frac{7}{2} 
- \frac{3}{2}\frac{{S}_{\hat{L}}}{\ell} \right) \vec{L} 
+ \frac{1}{2} \vec{S}_2
\right) \times\vec{S}_1
\cr 
&=& 
\frac{1}{r^3}\left( \left( \frac{7}{2} 
- \frac{3}{2}\frac{{S}_{\hat{L}}}{\ell} \right) \vec{J} 
-3 \vec{S}
\right) \times\vec{S}_1 \,,
\cr
\langle \dot{\vec{S}}_2 \rangle &=&
\frac{1}{r^3}\left( \left( \frac{7}{2} 
- \frac{3}{2}\frac{{S}_{\hat{L}}}{\ell} \right) \vec{L} 
+ \frac{1}{2} \vec{S}_1
\right) \times\vec{S}_2 
\cr
&=&
\frac{1}{r^3}\left( \left( \frac{7}{2} 
- \frac{3}{2}\frac{{S}_{\hat{L}}}{\ell} \right) \vec{J} 
- 3 \vec{S}
\right) \times\vec{S}_2 \,,
\eea
where $\ell=|\vec{L}|=q M^{3/2} r^{1/2} / (1+q)^2$
and we have ignored cubic terms of spins
in the most right hand side of each equation.

By decomposing the spins along
$\hat{L}$ and perpendicular to it with unit vectors $\hat{\lambda}$
and $\hat{n}$, as shown in Fig.~\ref{fig:frame},
in the fashion of Sec.~IV.A of Ref.~\cite{Lousto:2013wta},
we obtain the spin evolution equations for $q=1$ as
\bea
&&
\dot{S}_{1 \hat n} 
= {S}_{1 \hat \lambda} \left[
\frac{v_{\lambda}}{r}
- \frac{7\ell}{2r^3}  \left(1+\frac{ S_{1\hat L}}{\ell}\right)  + \frac{S_{2\hat L}}{r^3} \right]
- \frac{9 S_{1\hat L} S_{2 \hat \lambda}}{2r^3} \,,
\cr 
&&
\dot{S}_{1 \hat \lambda} 
=
- {S}_{1 \hat n} \left[
\frac{v_{\lambda}}{r}
- \frac{7\ell}{2r^3}  \left(1+\frac{ S_{1\hat L}}{\ell}\right)  + \frac{S_{2\hat L}}{r^3} \right]\cr
&&\quad\quad\quad
+ \frac{9 S_{1\hat L} S_{2 \hat n}}{2r^3}
- \frac{3 S_{1\hat L} S_{\hat n}}{r^3} \left(1+\frac{S_{\hat L}}{\ell}\right)  \,,
\cr
&&
\dot{S}_{1 \hat L}
= 
\frac{9 S_{1\hat n} S_{2 \hat \lambda}}{2r^3}
- \frac{9 S_{1\hat \lambda} S_{2 \hat n}}{2r^3}
+ \frac{3 S_{1\hat \lambda} S_{\hat n}}{r^3} \left(1+\frac{S_{\hat L}}{\ell}\right) \,,
\eea
for $\vec{S}_1$ and similar equations for $\vec{S}_2$ by 
exchanging labels 1 and 2 from Eqs.~(\ref{spinevo}).
In the above expression, 
$v_\lambda=r\Omega_{orb}$ is the tangential velocity of the binary
(in a quasicircular orbit, see Eq.~\eqref{eq:Omegaorb} for $\Omega_{orb}$),
${S}_{i \hat L}=\vec{S}_i\cdot\hat{L}$,
${S}_{i \hat n}=\vec{S}_i\cdot\hat{n}$,
${S}_{i \hat\lambda}=\vec{S}_i\cdot\hat{\lambda}$
and ${S}_{\hat L}=\vec{S}\cdot\hat{L}$.

We can then obtain equations of the form
$d^2(\vec{S}_i\cdot\hat{L})/dt^2=\Omega^2_{ff}\,\vec{S}_i\cdot\hat{L}
+\cdot\cdot\cdot$
for $i=1,\,2$ and analogously for the perpendicular component of $S_i$
giving $\Omega_p$.
From where we can read-off the orbit averaged polar and azimuthal oscillations 
frequencies of the spin $\vec{S}_i$ (see also Ref.~\cite{Racine:2008qv} for the
equations projected along $\hat{J}$)
\bea\label{Omegaff}
\Omega_{ff}&=&3\frac{S}{r^3}\left[1-\frac{2\,\vec{S}\cdot\hat{L}}{M^{3/2}r^{1/2}}\right]
+\cdot\cdot\cdot \,,\\
\Omega_{p}&=&\frac{7\ell}{2r^3}+\frac{2}{r^3}(\vec{S}\cdot\hat{L})
+\cdot\cdot\cdot \,,
\eea
that we identify with the flip-flop and precession frequencies respectively.
For the sake of completeness and further reference, the orbital frequency
of equal mass binaries in quasicircular orbits is given by Ref.~\cite{Kidder:1995zr}
\beq
\label{eq:Omegaorb}
M\Omega_{orb}=\left(\frac{M}{r} \right)^{3/2}-\frac{11}{8}\left(\frac{M}{r} \right)^{5/2}
+\cdot\cdot\cdot \,.
\eeq

\begin{figure}
\includegraphics[width=\columnwidth]{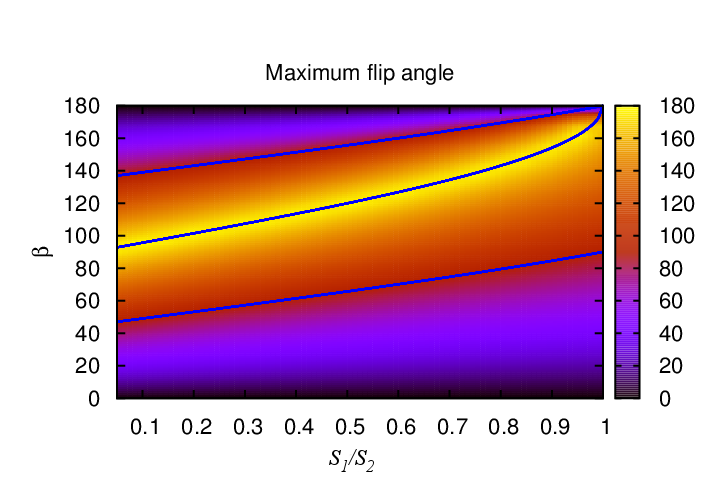}
\caption{$q=1$ binaries, inspiraling from $r=100M$ to $r=5M$. Initial
configurations having
$\phi_1^0=0=\phi_2^0$, $\theta_1^0=0$, and varying $\theta_2^0=\beta$ and
$S_1/S_2$. The central blue curve follows $\cos \beta=-S_2/S_1$ and agrees with the maximum (yellow) of the flip-flop angle of $\theta_2$.
The other two blue lines represent the $90$-degrees flips level.
\label{fig:maxAngle_ins_q1chi1}}
\end{figure}

\begin{figure}
\includegraphics[width=\columnwidth]{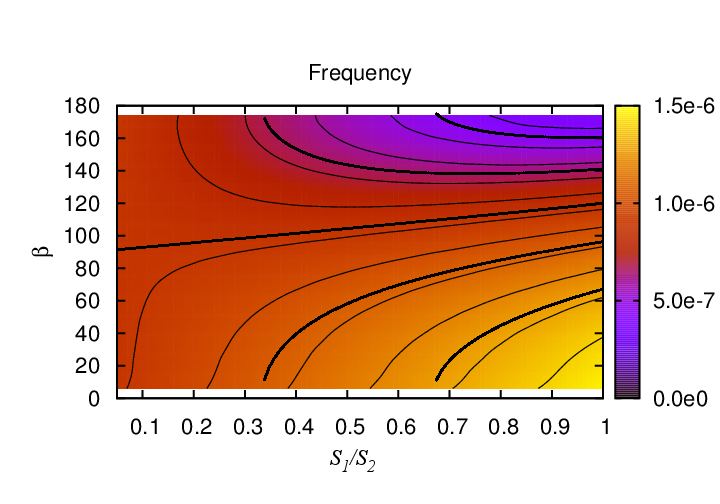}
\caption{Flip-flop frequency oscillation for
$q=1$ binaries, inspiraling from $r=100M$ to $r=5M$. Starting from
initial $\phi_1^0=0=\phi_2^0$, $\theta_1^0=0$, for different choices of
$\theta_2^0=\beta$ and $S_1/S_2$.
Different surface levels are represented in units of $10^{-7}$ for the
initial frequency at $r=100M$. Thin lines are the measured 
flip-flop frequency during evolution and thick lines their analytic
approximation for quasicircular orbits, Eq.~(\ref{Omegaff}).
\label{fig:freq_q1chi1_evo}}
\end{figure}

In order to verify the accuracy of the above expressions for
the flip-flop frequency and angle we performed a series of 
3.5PN numerical integrations representing the merger of BHBs
from $r=100M$ down to $r=5M$ and measured the flip-flop angle
and the frequency of oscillation. The results are presented
in Figs.~\ref{fig:maxAngle_ins_q1chi1} and~\ref{fig:freq_q1chi1_evo}
as a color map with the direct integration of the PN equations
of motion coupled to the spin evolutions and we superposed a
few curve levels based on the above analytic estimates. We
observe good agreement among them, in particular for the maximum
flip angle described by $\cos\beta=-S_1/S_2$.
Fig.~\ref{fig:freq_q1chi1_evo} displays as a color map
the frequency at the start of the evolution, $r=100M$ where the
term $(3S/r^3)$ in  Eq.~(\ref{Omegaff}) dominates.  
In this figure, the thin lines are the contours from the measured
flip-flop frequency, and the bold lines are the
analytic expression given by Eq.~(\ref{Omegaff}) for
quasicircular orbits at the corresponding $r$.

\subsection{Statistics}\label{sec:q1stats}

\begin{figure}
\includegraphics[angle=270,width=\columnwidth]{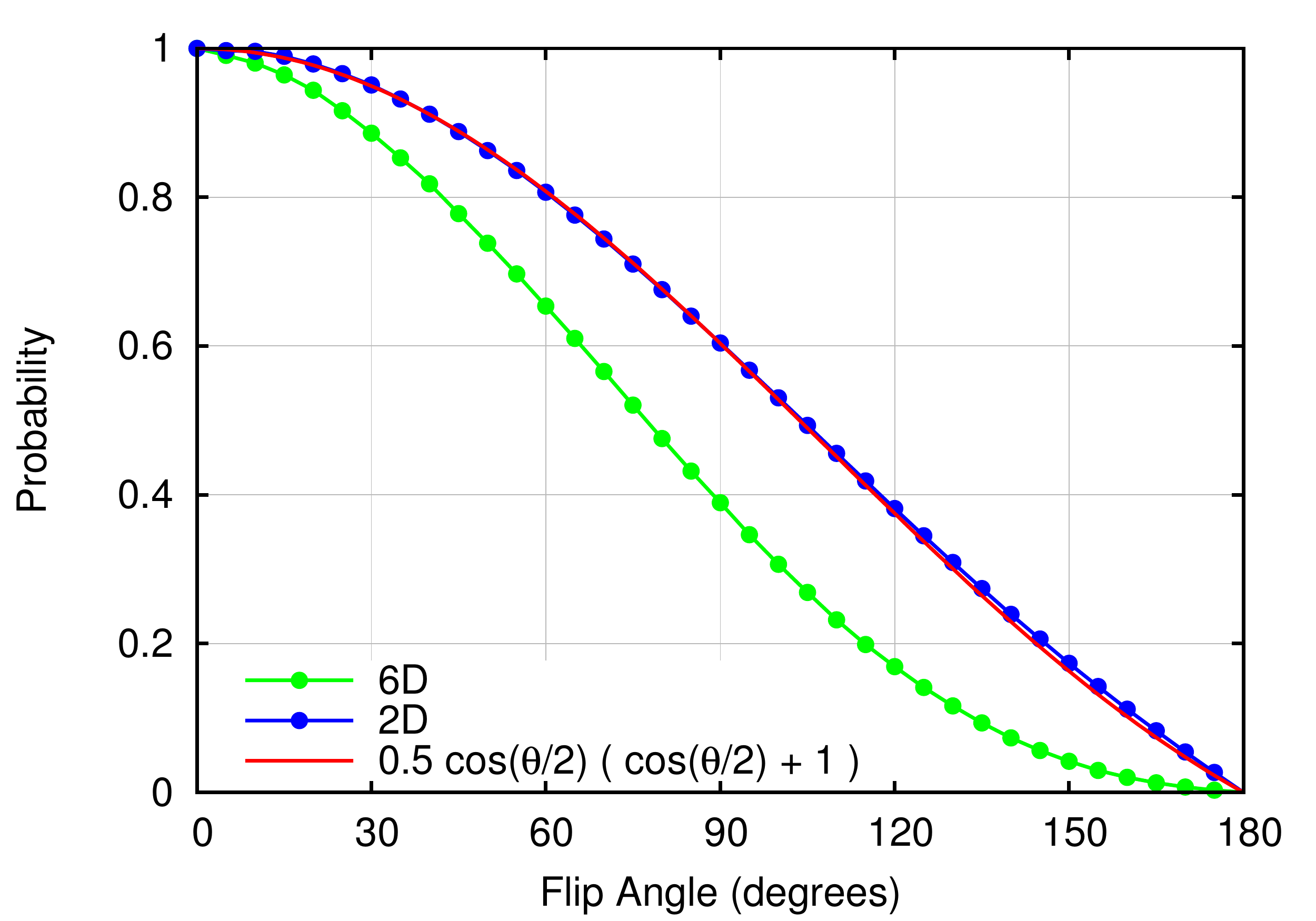}
\caption{The probability for the $q=1$ case of having a flip greater
than a certain angle. Distribution `2D' lets either $\theta_1^0$ or $\theta_2^0$ 
be aligned or counteraligned with $\hat{L}$ with a randomly varied 
$\cos$ of the other variable and spin magnitudes ratio.  
There are $209,450$ samples
that match the analytic distribution. Distribution `6D' shows
the probability of flip over a generic range of spins (6 parameters) for $q=1$, 
totaling 16 million evolutions.  
\label{fig:q1distribution}}
\end{figure}

Let us study first the simplest case of conservative precession 
for equal mass binaries.
In Fig.~\ref{fig:q1distribution}, we display the 
probability of a flip-flop angle $\Delta\theta_{ff}\geq x$
assuming equal masses with random spin orientations and magnitudes of
BH1 and 2. For the sake of simplicity, we 
used here spin evolution equations~(\ref{spinevo})
with no radiation reaction taken into account.
The line labeled `2D' considers random angular distribution
of one spin and random distribution of the ratio of spin magnitudes.
It could represent the scenario where accretion processes succeeded in
aligning the spin of one of the BHs, while the spin of the
other has still a random orientation due to a much larger 
time scale for alignment.  If we allow instead to vary all six 
parameters of the BH spins (magnitude and both spin directions) with
$q=1$, we observe the distribution labeled `6D'.

When either BH has its spin aligned or counteraligned
with the orbital angular momentum and the other spin is
chosen randomly we have a random angle $\beta$ between the
directions of the BH1 and 2 spins. In this case one
can obtain from Eq.~(\ref{gamma}) that the probability of an angle
$\gamma$ given a random distribution of $\cos\beta$ is approximated by
\be
P(\cos\gamma)=1+\left(\frac{S_1}{S_2}\right)
\left(2\cos\gamma-1\right) \,,
\ee
which upon assuming random distribution of spin magnitudes
$0\leq\left(S_1/S_2\right)\leq1$, leads to the probability for
a flip-flop angle larger than $x$
\be
P(\Delta\theta_{ff}>x)=
\frac12\cos\left(\frac{x}{2}\right)\left(\cos\left(\frac{x}{2}\right)+1\right) \,. 
\ee

This probability distribution, represented in Fig.~\ref{fig:q1distribution}
by the red continuous curve 
shows, for instance, that (for equal mass binaries)
there is nearly a $60\%$ probability of flip-flops
of more than $90^\circ$.

In the more general case, allowing for radiation reaction to drive
the BHBs towards merger and random variation of all
variables, we find the same qualitative behavior as above plus a detailed
$\Delta\phi$-dependence.
From the expressions for the flip-flop angle in Eq.~(\ref{eq:thetaff}),
we observe they depend entirely on the two angles $\gamma$ and $\theta_S$,
both being constants of motion (for equal mass binaries). Thus once
we express them in terms of initial data we can predict the flip-flop
angle. This can be achieved in terms of four of the six spin 
components: $\theta_1$, $\theta_2$, $\Delta\phi=\phi_2-\phi_1$, and $S_1/S_2$,
from Eq.~(\ref{gamma}) for $\gamma$ 
\beq
\cos\gamma=\frac{S_1/S_2+\cos\beta}{\sqrt{1+S_1^2/S_2^2+2S_1/S_2\cos\beta}} \,,
\eeq
and
\beq
\cos\theta_S=\frac{S_1/S_2\cos\theta_1+\cos\theta_2}
{\sqrt{1+S_1^2/S_2^2+2S_1/S_2\cos\beta}} \,,
\eeq
where
\beq
\cos \beta =\cos\theta_1 \cos\theta_2 + \sin\theta_1 \sin\theta_2 \cos\Delta\phi \,.
\eeq

\begin{figure}
\centerline{
\includegraphics[angle=270,width=\columnwidth]{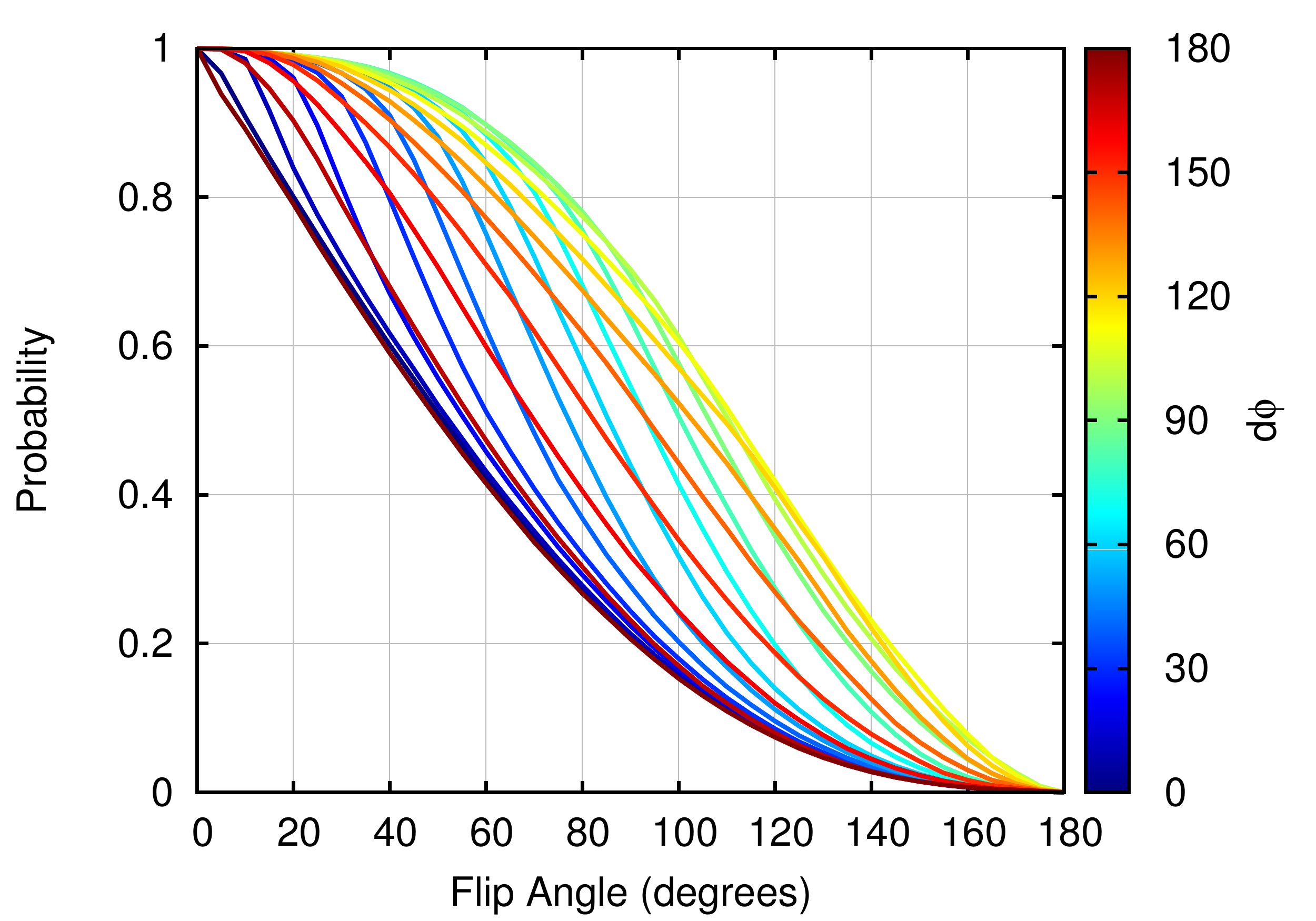}}
\caption{The probability of a flip-flop angle $\Delta\theta_{ff}\geq x$
assuming equal masses, $(q=1)$, and a random initial 
spin orientations and magnitudes of both, the primary and secondary spins.
Evolutions from initial separation $r=100M$ down to $r=5M$ are performed
using 3.5PN equations of motion. The dependence with the
azimuthal orientation $\Delta\phi=d\phi$ is displayed.
We observe the largest flip angles probabilities for configurations with
$\Delta\phi=\pi/2$ and smallest for $\Delta\phi=0$ and $\pi$. 
}\label{fig:q1probabilities}
\end{figure}

The largest flip-flop angles occurs when the two BH spins
started at (or pass through) some initial configuration with $\Delta\phi=\pi/2$
as is displayed in the Fig.~\ref{fig:q1probabilities}.
The curves in this figure represent the probability that an equal-mass
system with initial $\Delta\phi$ will have a flip-flop angle $\geq x$ given
random spin orientations and spin magnitudes of the primary and
secondary BHs (here labeled simply as 1 or 2).  
Each curve corresponds to a set of $203,401$ PN evolutions 
$(|\alpha_1|\times |\alpha_2|\times\cos(\theta_1^0)\times\cos(\theta_2^0)$ 
$= 11\times 11\times 41\times 41)$ starting from a separation of $r=100M$
and evolved with 3.5PN radiation reaction terms down to a separation of $5M$.
The color of the curve is determined by the initial value of $\Delta\phi$.

\section{Unequal Masses Binaries}\label{sec:Unequal}

The flip-flop of BH spins in a binary system is an interesting
effect since it singles out a conservative dynamical behavior
that may produce important astrophysical phenomena. In order to
further evaluate the relevance to astrophysics we have to study
BHBs with unequal masses and the likelihood of
a given flip-angle to occur as well as the timescale (or flip-flop
frequency) involved.

In the equal mass case, we have been able to discuss the geometry
of the flip-flop angles in Sec.~\ref{subsec:Vec} based on  
(to a good approximation) the
conservation of the spin magnitudes, $S_1$, $S_2$, and $S$ as well
as the component of the total spin along the orbital angular momentum
$S_{\hat{L}}$. These four conserved quantities allowed us to express the 
flip-flop angle in terms of the conserved quantities $\gamma$
and $\theta_S$, i.e., Eq.~(\ref{eq:thetaff}). The conservation of $\gamma$
and $\theta_S$ implies that we are able to determine the flip-flop
angle in terms of the initial configuration of the binary, and hence
by specifying $S_1/S_2$, $\theta_1$, $\theta_2$, and $\Delta\phi$.
Note that from the six variables that are needed to specify the
components of the two vectors $\vec{S}_1$ and $\vec{S}_2$ only
four are needed here due to the fact we compute a dimensionless 
flip-flop angle, hence the dependence on the ratio $S_1/S_2$ and
the dependence on $\Delta\phi=\phi_2-\phi_1$ appears due to the fact
that for large separations a rotation of the angle $\phi$ is
already provided by the precession of $\Vec{S}$ around $\vec{L}$,
spanning all orientations relative to the orbital linear momentum,
$\vec{p}_i$.

In retrospective,
while the flip-flop angle for equal masses can be determined
geometrically, the frequency at which spins flip-flop is a 
dynamical quantity that requires information about the evolution
of the spins as well as the orbital evolution of the BHBs.
At the lowest (2PN) post-Newtonian approximation we can
evaluate this frequency using the spin-evolution equations~(\ref{spinevo})
assuming that the binary is separated far enough
that the radial decay is negligible during a flip-flop cycle
(note that the gravitational radiation frequency scales as
$\sim r^{-4}$ compared to the flip-flop one scaling as 
$\sim r^{-3}$). The flip-flop frequency thus depends on the
orbital radius $r$, the total spin and its projection along
$\hat{L}$ as given by expression~(\ref{Omegaff}).

Here, in the unequal mass binary case we can assume to a certain
degree of accuracy the conservation of both spin magnitudes,
$S_1$ and $S_2$. $S_{\hat{L}}$ is not conserved, but
the projection of the vector $\vec{S}_0$ (as given in 
Eq.~(\ref{eq:S0})) along $\hat{L}$, denoted as $S_{0\hat{L}}$
is approximately conserved~\cite{Racine:2008qv}.
However, the total spin magnitude, $S$, is not conserved
since the angle between the two individual spins $\beta$ is
no longer conserved.
We will use a combination of analytic 
and 3.5PN numerical integrations to provide asymptotic
expressions for the flip-flop angle, flip-flop frequency,
and estimates of the likelihood for them to arise in
astrophysical scenarios.
Contemporary studies~\cite{Kesden:2014sla}
could also describe the spin flips in terms of the two extreme
values of the total spin.

Note that in generic binaries spin resonances of the
kind studied in Refs.~\cite{Schnittman:2004vq,Bogdanovic:2007hp}
bring the azimuthal angle differences towards $\Delta\phi=0$ and $\pi$
when the spin polar orientations are significantly different,
and to $\Delta\phi=\pi/2$ when they are
similar~\cite{Kesden:2010ji,Berti:2012zp,Gerosa:2013laa}; 
thus allowing a further statistical
reduction of the parameter space to explore.

\subsection{Post-Newtonian Analysis}\label{subsec:UPN}

In the unequal mass binaries case the flip-flop angle will
be a function of not only the intrinsic spins, 
$\vec{\alpha}_i=\vec{S}_i/m_i^2$,
but also the radius $r$ of the quasicircular orbit as well 
as the mass ratio $q$.

By studying the spin evolution equations~(\ref{spinevo})
projected along $\hat{L}$ (or along $\hat{J}$) we can
obtain the {\it maximum} flip-flop angle for a given
set of binary parameters, $\alpha_1$, $\alpha_2$, $q$, and $r$, 
\beq\label{eq:ffangle1L}
1-\cos(\Delta\theta_{1L}^{ff}) \approx \frac{2\alpha_2^2}{(1-q)^2}\left(\frac{M}{r}\right)+
\frac{4\alpha_1\alpha_2^2q}{(1-q)^3}\left(\frac{M}{r}\right)^{3/2} \,.
\eeq
This maximum flip-flop angle of the {\it smaller} BH with respect to $\hat{L}$
is achieved when the smaller hole starts (or pass during its oscillation)
through an anti-alignment with respect to $\hat{L}$, i.e., 
$\theta_{1L}^0=\pi$ and the larger BH spin
$\vec{S}_2$ forms initially an angle 
$\theta_{2L}^0$ with $\hat{L}$ given by
\bea\label{eq:beta2L}
\cos\theta_{2L}^0&&\approx
\frac{\alpha_2}{(1-q)}\sqrt{\frac{M}{r}}
+\frac{3q\alpha_1\alpha_2}{(1-q)^2}\left(\frac{M}{r}\right) \,.
\eea

On the other hand, if we seek to maximize the flip-flop angle of
the {\it larger} BH, we find
\beq\label{eq:ffangle2L}
1-\cos(\Delta\theta_{2L}^{ff}) \approx \frac{2\alpha_1^2 q^2}{(1-q)^2}\left(\frac{M}{r}\right)
+\frac{4\alpha_2\alpha_1^2q^2}{(1-q)^3}\left(\frac{M}{r}\right)^{3/2} \,.
\eeq
This maximum flip of the larger hole is achieved when
the spin of this hole goes through
alignment with the orbital angular momentum, i.e., 
$\theta_{2L}^0=0$, then the smaller BH spin forms
an angle $\beta_1^0$ with $\vec{S}_2$ given by
\bea\label{eq:beta1L}
\cos\theta^0_{1L}\approx
\frac{q\alpha_1}{(q-1)}\sqrt{\frac{M}{r}}
-\frac{3q\alpha_1\alpha_2}{(q-1)^2}\left(\frac{M}{r}\right) \,.
\eea

\begin{figure}
\includegraphics[angle=270,width=0.49\columnwidth]{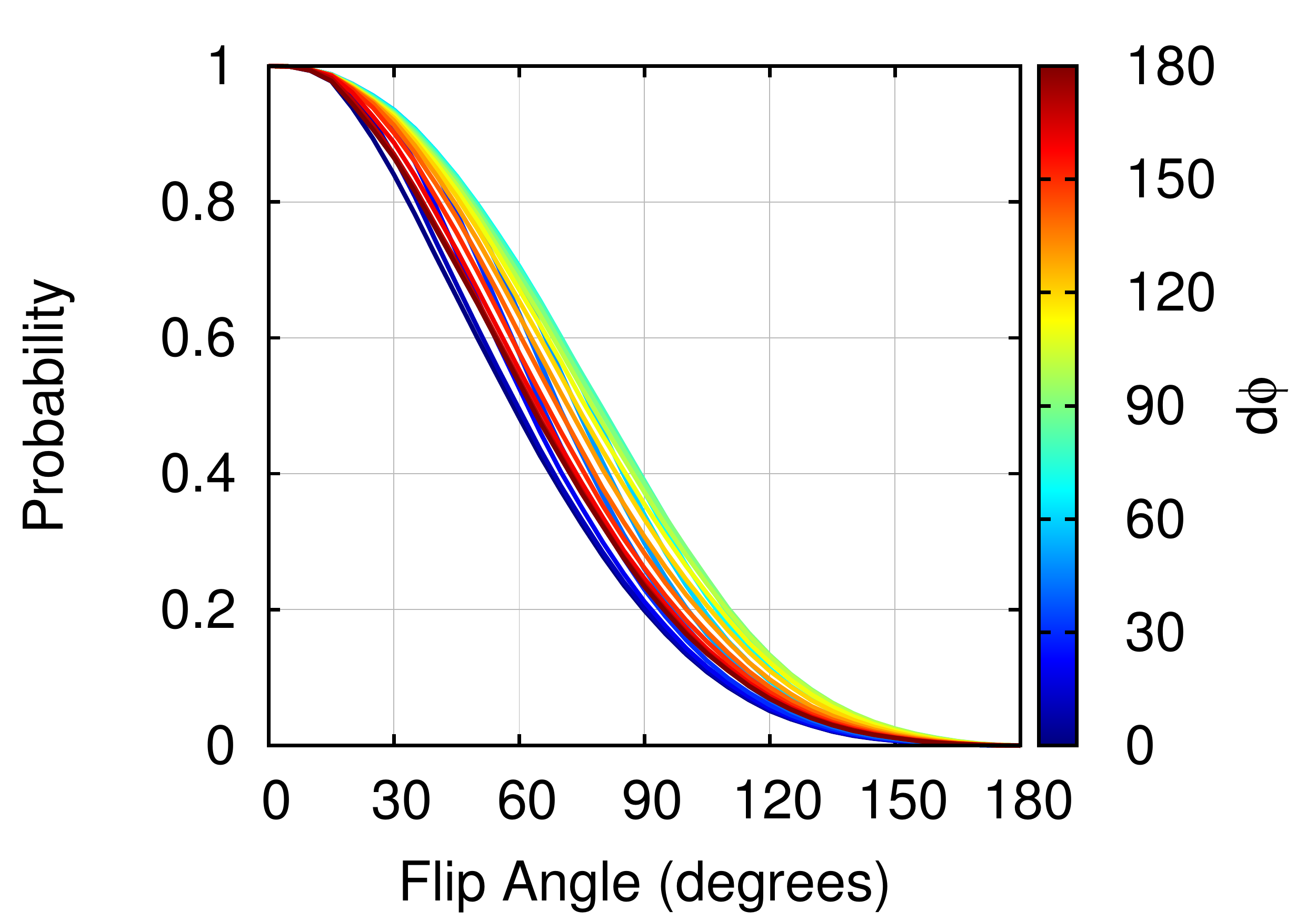}
\includegraphics[angle=270,width=0.49\columnwidth]{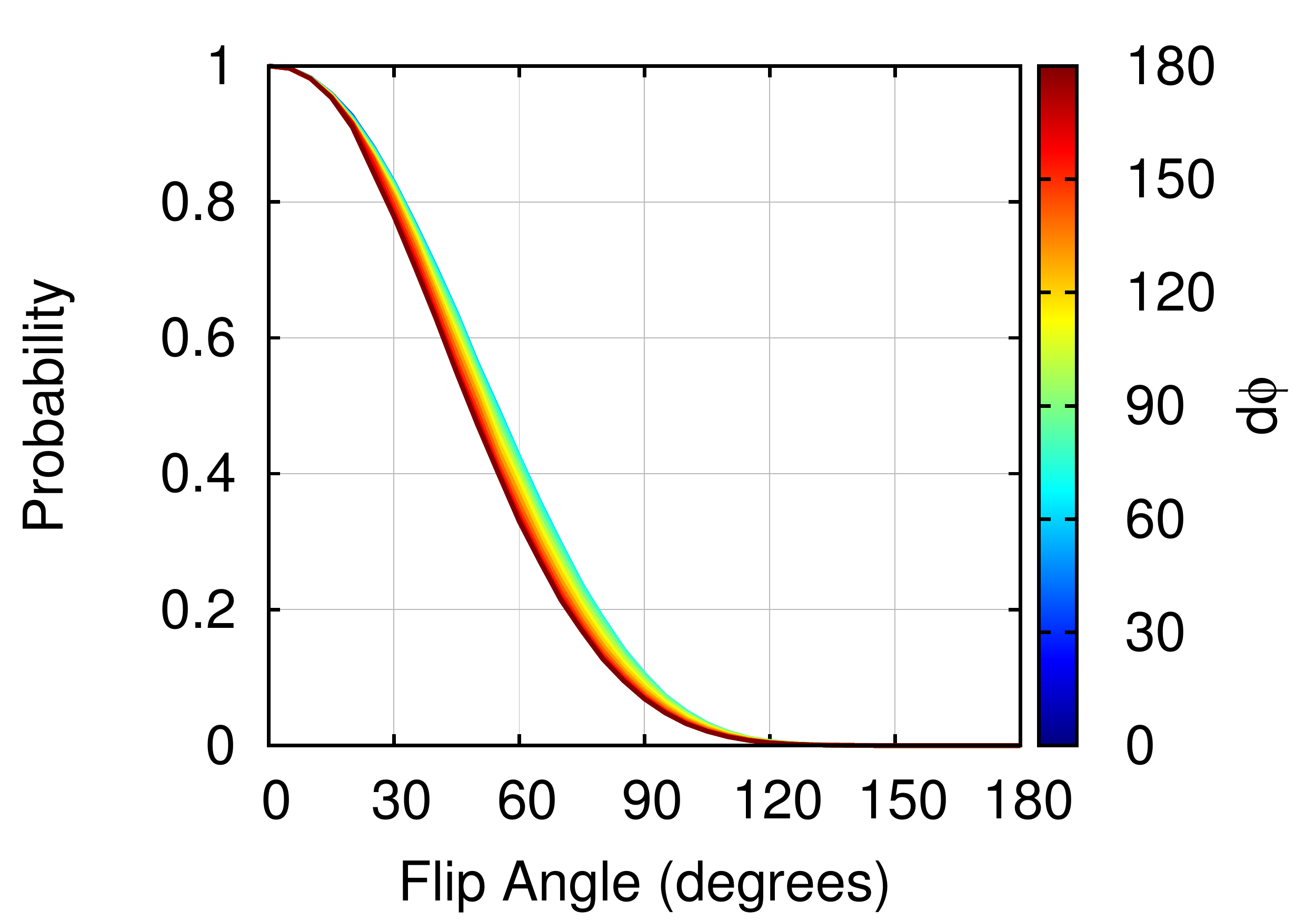}
\includegraphics[angle=270,width=0.49\columnwidth]{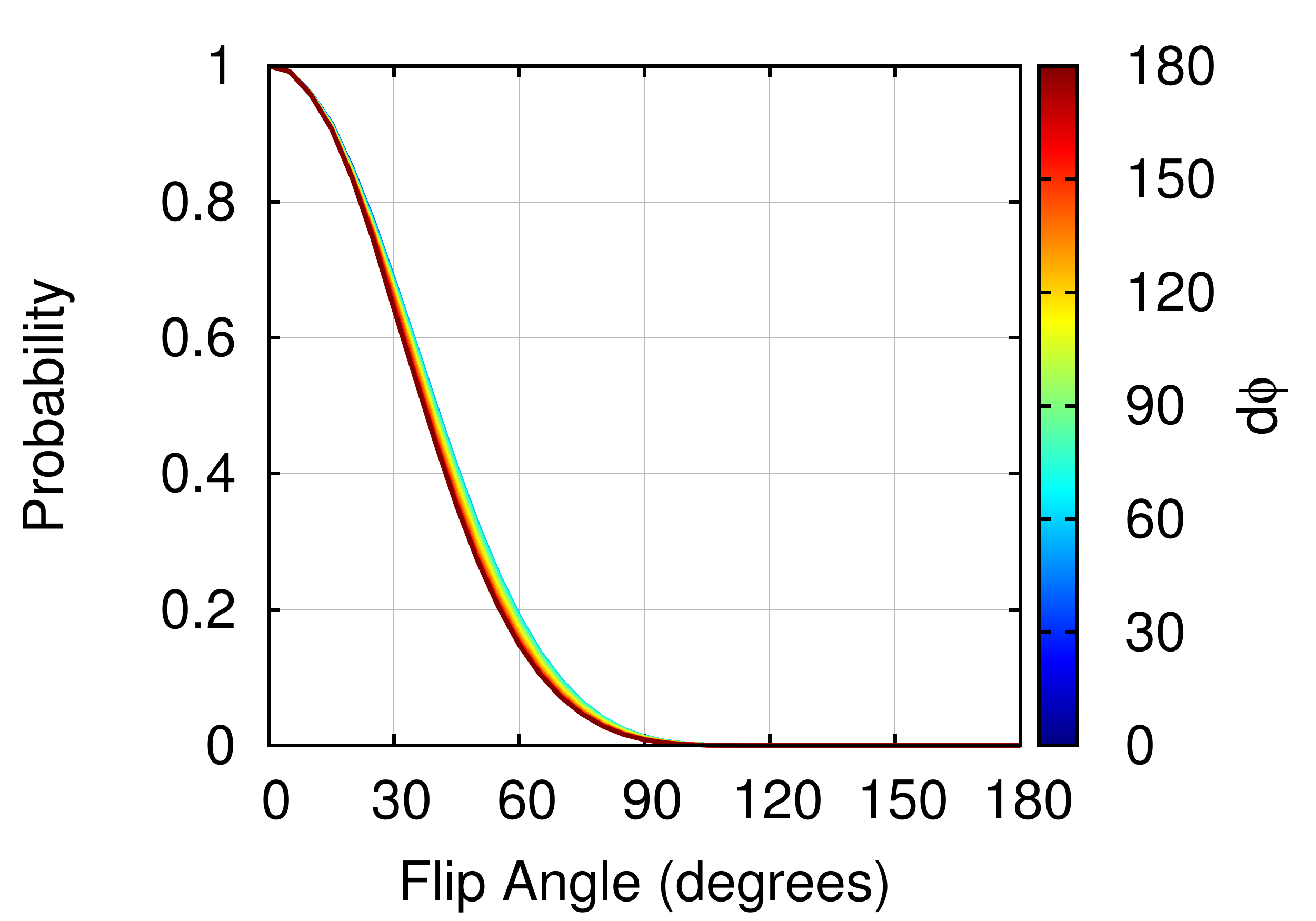}
\includegraphics[angle=270,width=0.49\columnwidth]{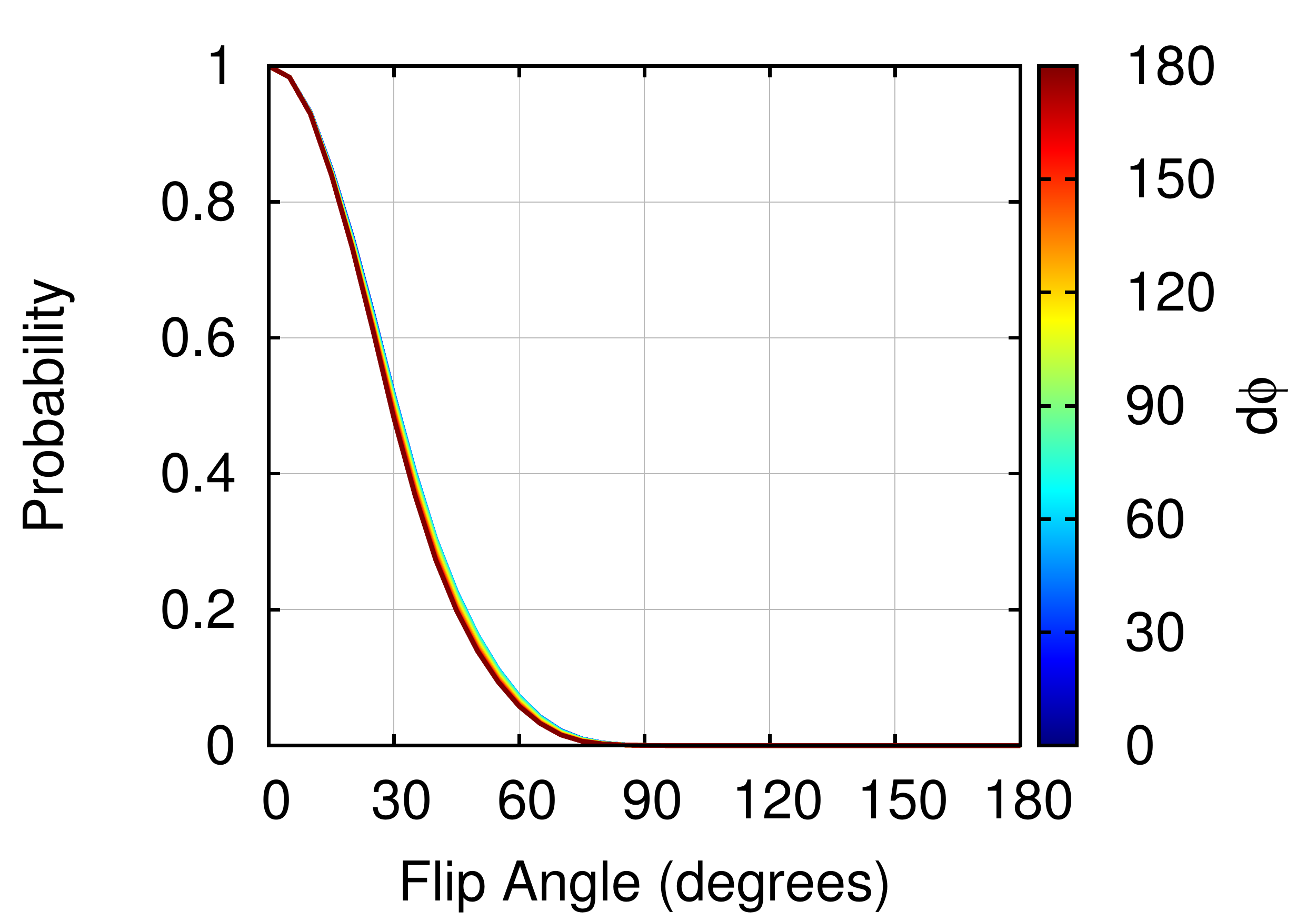}
\caption{The dependence of the probabilities of a flip-flop larger
than a given value as a function of the initial $\Delta\phi$
angle among spins in the $L$-frame. Evolutions from $r=100M$ down
to $r=5M$ are performed with 3.5PN including radiation reaction.
From left to right and top
to bottom for $q=0.9,\, 0.8,\, 0.7,\, 0.6$.
We observe an insensitivity on
$\Delta\phi$ for mass ratios of the binary below $q=0.7$. Similar
results are observed in the $J$-frame.
\label{fig:phiL}}
\end{figure}

In Fig.~\ref{fig:phiL} we display the results of 3.5PN evolutions
from $r=100M$ for 10 choices for $q$, 41 choices for $\cos \theta_1^0 $, 
41 choices for $\cos \theta_2^0 $, 11 choices for $|\alpha_1|$, 
11 choices for $|\alpha_2|$, and 37 choices
for $\Delta\phi$ by setting $\phi_1^0=0$ and choosing different $\phi_2^0$.
The systems are then evolved with radiation reaction down to $5M$.
The color-bar in the figures denote the initial $\Delta\phi$.
The highest values happen at $\Delta\phi=\pi/2$, but
the results are very insensitive on this dependence for
almost every mass ratio below $q\approx0.7$.
This same property is observed in both the $L$- and $J$-frames.

\begin{figure}
\includegraphics[angle=270,width=0.49\columnwidth]{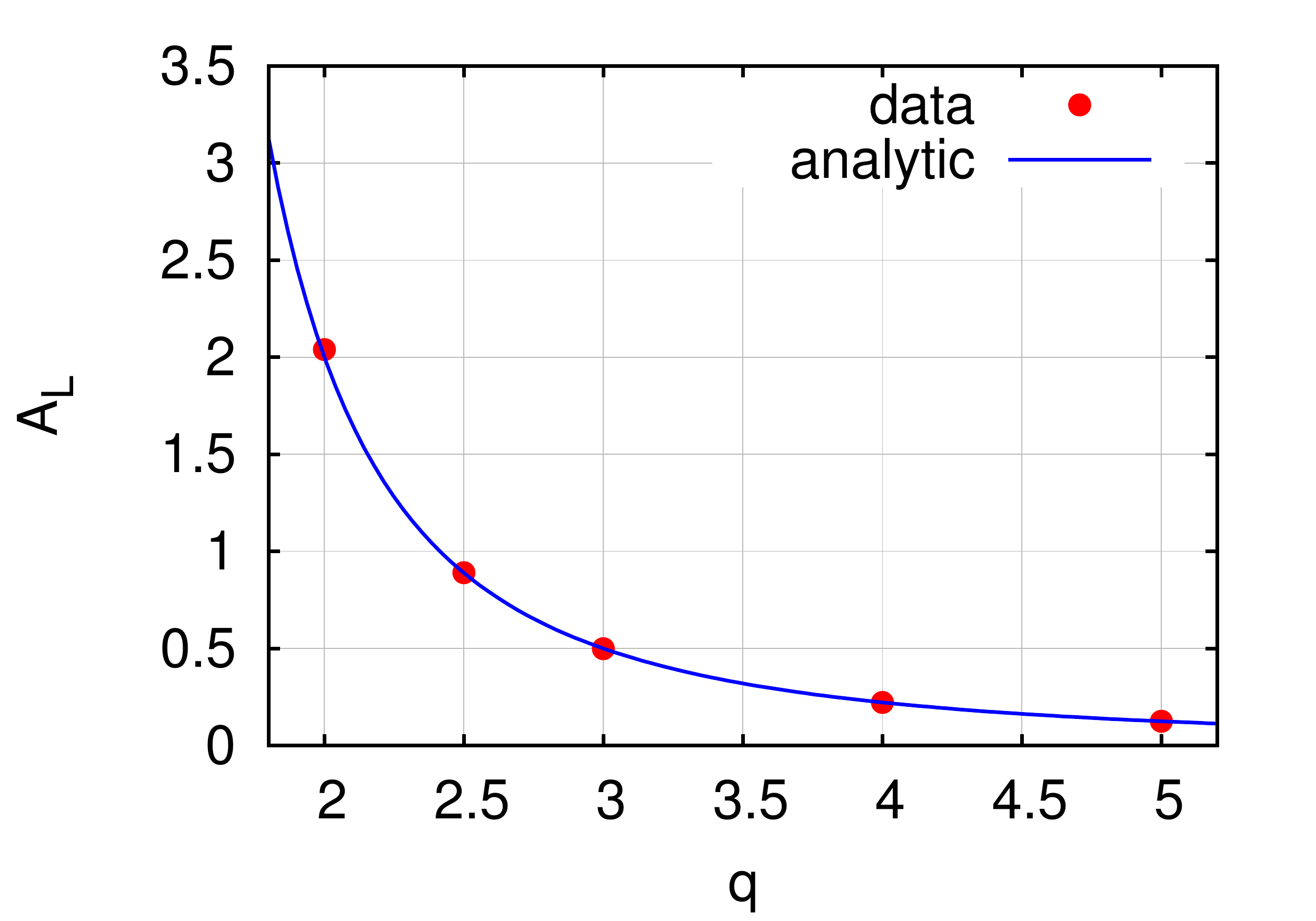}
\includegraphics[angle=270,width=0.49\columnwidth]{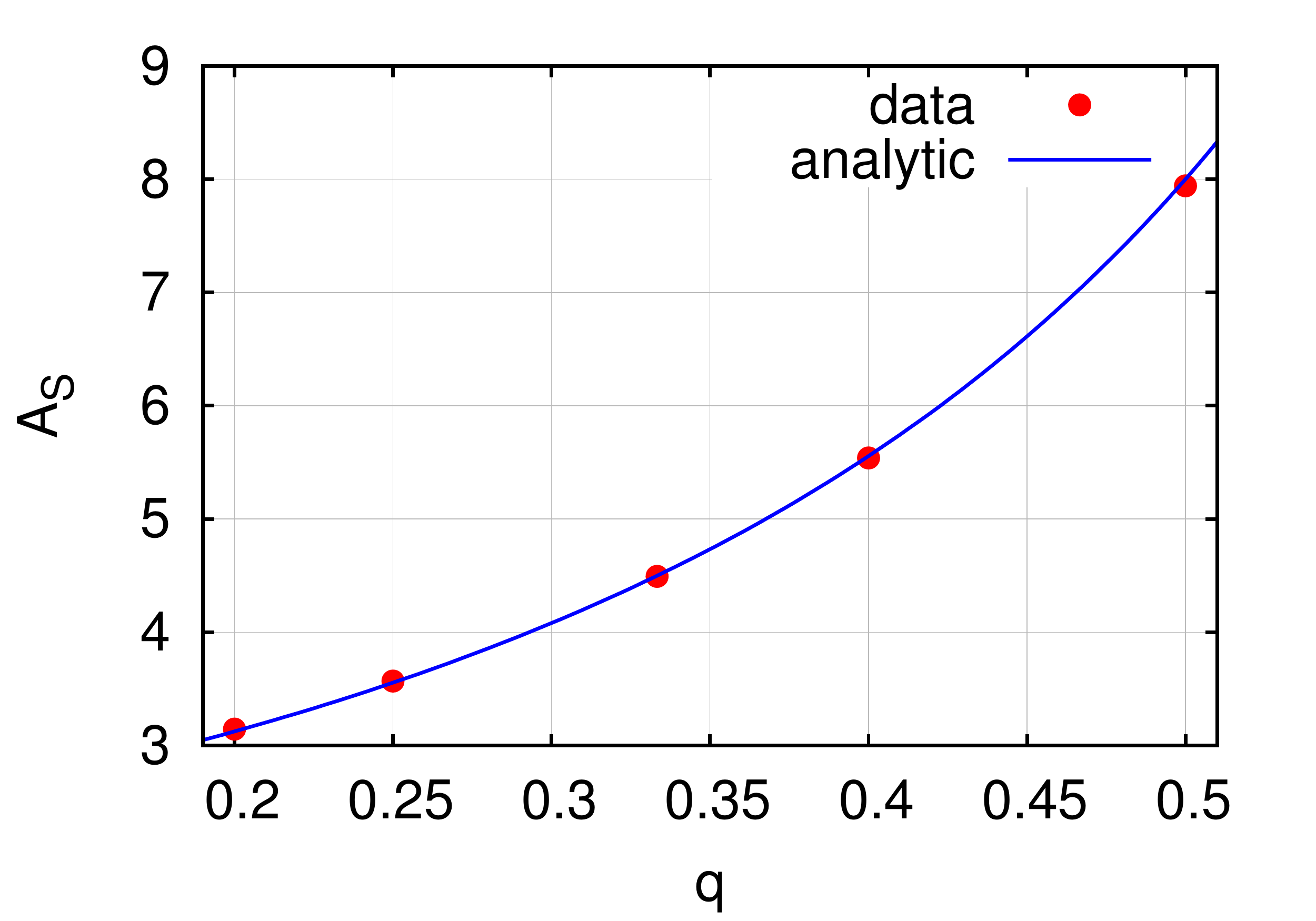}
\caption{Measured maximum flip-flop angle from PN evolutions
versus analytic prediction
(no fitting here). Here we display the leading coefficient of
Eqs.~(\ref{eq:ffangle1L}) and~(\ref{eq:ffangle2L}) for the larger BH
flip-flop angle (left) and smaller BH angle (right).
\label{fig:Angle}}
\end{figure}

\begin{figure}
\includegraphics[angle=270,width=0.49\columnwidth]{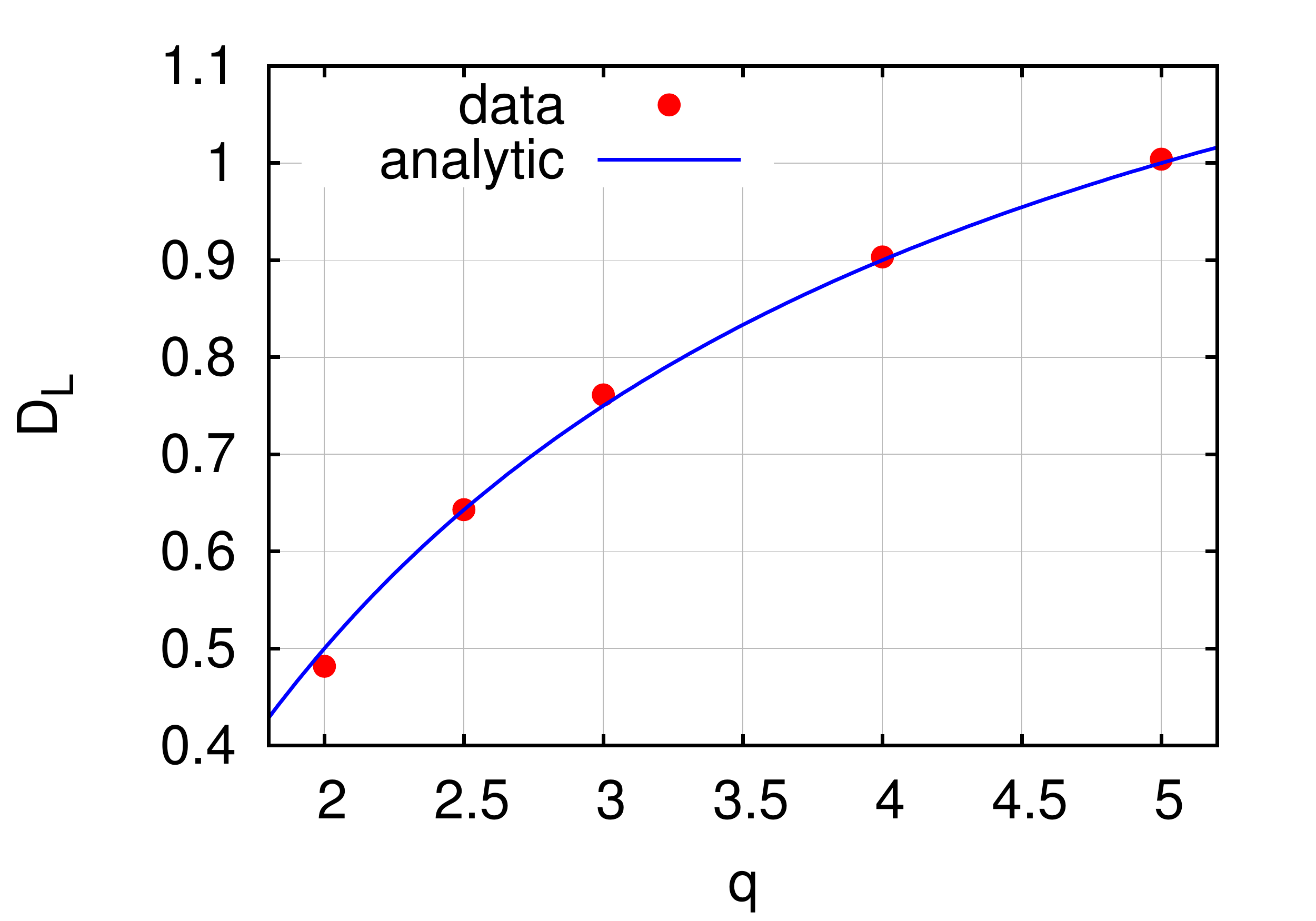}
\includegraphics[angle=270,width=0.49\columnwidth]{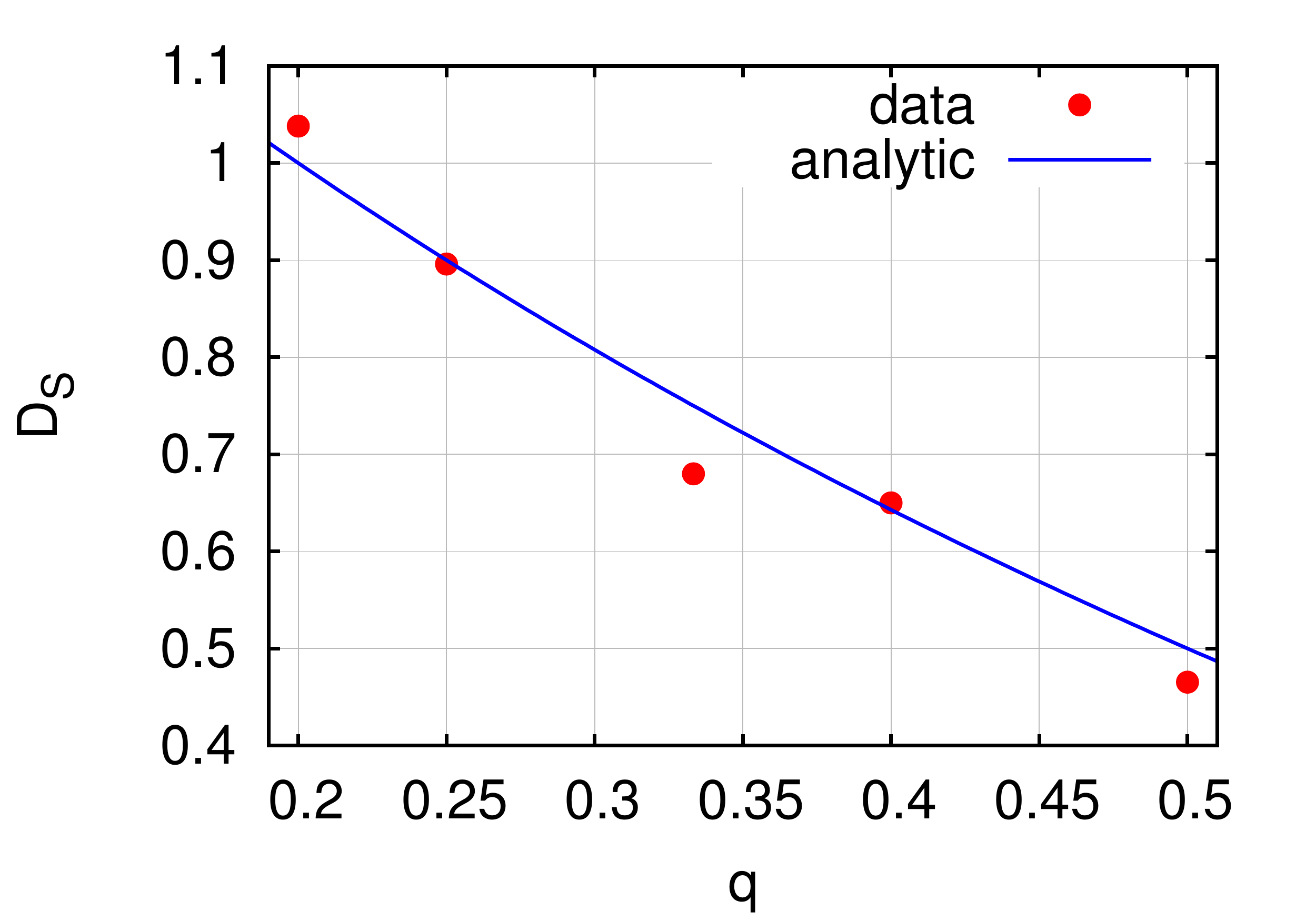}
\caption{Measured flip-flop frequency versus analytic prediction
(no fitting here). Here we display the leading coefficient of
Eq.~(\ref{eq:fffrequency}) for the larger BH
flip-flop angle (left) and smaller BH angle (right).
\label{fig:Frequency}}
\end{figure}

We verified these analytic dependencies by performing numerical
integrations of the spin evolution equations~(\ref{spinevo}) for different
radii, $r/M=30,\,50,\,100,\,150,\,200,\,250$, and
mass ratios $q=1/2,\,2/5,\,1/3,\,1/4,\,1/5$ and display a 
summary in Figs.~\ref{fig:Angle} and~\ref{fig:Frequency}. 
We compare in those plots (not fit) 
the analytic form of the leading flip angle
(on the right-hand side of Eqs.~(\ref{eq:ffangle1L}) and~(\ref{eq:ffangle2L}))
versus the measured values as obtained from assuming the forms
$1-\cos(\theta^{ff}_{1,2})=A_{L,S}/r+B_{L,S}/r^{1.5}+C_{L,S}/r^2$,
and $(\Omega^{ff}_{1,2})=D_{L,S}/r+E_{L,S}/r^{1.5}+F_{L,S}/r^2$.
In order to make explicit the $1\leftrightarrow2$, $q\leftrightarrow1/q$
symmetry of the $A$ and $D$ coefficients, we display the case of the
larger BH flip-flop on the left of Figs.~\ref{fig:Angle}
and \ref{fig:Frequency} in terms of $1/q>1$ variable.

The maximum angle of flip-flop with respect to 
${J}$-frames is different than with respect 
to the ${L}$-frame due to the fact that, even during conservative
evolution, the angle of precession of $\vec{L}$ around $\vec{J}$ is not
preserved. In fact, because for unequal masses, unlike the equal mass
case, the projection of the total spin $\vec{S}$ along $\vec{L}$,
$S_{\hat{L}}$, is not preserved,
\beq
\hat{J}\cdot\hat{L}=\frac{(\ell+S_{\hat{L}})}{j}\,,
\eeq
even if on average the magnitudes of the total angular momentum, $j$,
and the orbital angular momentum, $\ell$, are preserved.

This leads to the possibility of having larger fluctuations in
the angles as ``seen'' in the ${J}$-frame, given by
\bea\label{eq:ffangle1J}
1-\cos(\Delta\theta_{1J}^{ff}) & \approx & \frac{2\alpha_2^2}{q(1-q)^2}\left(\frac{M}{r}\right)
\cr &&
+\frac{2\alpha_1\alpha_2^2(q-3)}{(q-1)^3}\left(\frac{M}{r}\right)^{3/2} \,,
\eea
and
\bea\label{eq:ffangle2J}
1-\cos(\Delta\theta_{2J}^{ff}) & \approx & \frac{2\alpha_1^2 q^3}{(1-q)^2}\left(\frac{M}{r}\right)
\cr &&
-\frac{2\alpha_2\alpha_1^2q^2(3q-1)}{(q-1)^3}\left(\frac{M}{r}\right)^{3/2} \,,
\eea
for the maximum smaller and larger BH oscillations of their
spins respectively. 
The extreme values happen at a initial coordinate $\Delta\phi=\pi/2$
and seem to correspond to the $S_{max}$ (and $S_{min}$) configurations
in Fig.~2 of Ref.~\cite{Gerosa:2015tea}.

Note the $q \to 1/q$ symmetry of all the above expressions corresponding to
a $1\to 2$ exchange in the labels of the holes.
Also at each moment of time, the conservation of $S_0^L$ determines 
the angular location of the other BH given 
the spin of the BH with larger flip-flop angle.

The initial angle of the small hole (and
between the spins) that maximizes the flip-flop as seen in
the $J$-frame is approximated by 
\bea
\cos\theta^0_{1J} & \approx &
- \frac{1}{2}\,{\frac {\left( 1+q \right)q{\alpha_1} }
{ \left( 1-q \right) }} \sqrt{\frac{M}{r}}
\cr &&
-\frac{1}{2}\,{\frac {(q^2+4q+1){\alpha_1}{\alpha_2} }{(1-q)^2 }}
\left(\frac{M}{r}\right) \,,
\eea
for the larger BH spin initially pointing to
\bea
\cos\theta^0_{2J} & \approx &
1-\frac{1}{2}\,{{q}^{2}{\alpha_1}^{2}} \left(\frac{M}{r}\right)
+{q{\alpha_1}^{2}{\alpha_2}} \left(\frac{M}{r}\right)^{3/2}
\,.
\eea

While if we seek to maximize the flip-flop angle of the smaller
hole as seen in the $J$-frame, we find
\bea
\cos\theta_{1J}^0 & \approx &
- 1 + \frac{1}{2}\,{\frac {{\alpha_2}^{2}}{q^2 }}\left(\frac{M}{r}\right)
+{\frac {{\alpha_2}^{2}{\alpha_1}}{q}}\left(\frac{M}{r}\right)^{3/2}
\,.
\eea
In this case the spin of the larger hole points to
\bea
\cos\theta_{2J}^0 & \approx &
\frac{1}{2}\,{\frac {\left( 1+q \right){\alpha_2} }
{q \left( 1-q \right) }}\sqrt{\frac{M}{r}}
\cr &&
+ \frac{1}{2}\,{\frac { (q^2+4q+1){\alpha_2}{\alpha_1}}{(1-q)^2 }}\left(\frac{M}{r}\right) \,.
\eea
The detailed derivation is shown in Appendix~\ref{app:PNSpin}.

The frequency of the flip-flop oscillation in all these cases is given by
(see Appendix~\ref{app:PNSpin})
\bea\label{eq:fffrequency}
M\Omega_{1,2}^{ff} & \approx & 
\frac{3}{2}\frac{|1-q|}{(1+q)}\left(\frac{M}{r}\right)^{5/2}
\cr &&
+3\frac{S_1^L-S_2^L}{M^2}\left(\frac{M}{r}\right)^{3}{\it sign}(1-q) \,.
\eea

The origin of the additional term for unequal masses scaling with
$\sim r^{-5/2}$ is due to the non-conservation of the angle $\beta$
between the two spins (as opposed to its conservation in the $q=1$
case). These oscillations in $\beta$ are due to the differential
precessional angular velocity of $\vec{S}_1$ and $\vec{S}_2$ for
$q\not=1$ and hence provides the (precessional) scaling $r^{-5/2}$.


For the sake of completeness and comparison of the time-scales involved,
we give here the leading terms of the orbital frequency
in a PN expansion~\cite{Racine:2008kj}
\bea
M\Omega_{orb}&&\approx\left(\frac{M}{r}\right)^{3/2}
-\frac{3+5q+3q^2}{2(1+q)^2}\left(\frac{M}{r}\right)^{5/2} 
\cr &&
-\frac{2\alpha_2^L(1+4q)+q\alpha_1^L(3+2q+5q^2)}{2(1+q)^3}\left(\frac{M}{r}\right)^{3} \,.
\eea

The leading terms of the precession or azimuthal frequency
are~\cite{Racine:2008qv}
\bea
M\Omega_{pre}&&\approx\frac72 \frac{q}{(1+q)^2}\left(\frac{M}{r}\right)^{5/2}
\cr &&
+\left[\frac72\frac{(\alpha_1^Lq^2+\alpha_2^L)}{(1+q)^2}
-\frac34\frac{(\alpha_1^Lq+\alpha_2^L)}
{(1+q)}\right]\left(\frac{M}{r}\right)^3 \,. 
\eea


If we now request that the flip-flop angle be exactly $\pi$ in an
unequal mass configuration, this might only happen for quasicircular
orbits inside a critical separation $r_C$. In order to estimate this
critical separation we may use the analytic expression for the
maximum flip-flop angle of the smaller BH given in
Eq.~(\ref{eq:ffangle1L}) and request that $\Delta\theta^{ff}_{1L}=\pi$
for total flip-flop in the ${L}$-frame. This leads to the condition
\beq\label{eq:u}
1=u^2+2\,p\,u^3 \,,
\eeq
where
\beq
u=u_L=\frac{\alpha_2}{(1-q)}\sqrt{\frac{M}{r_C}} \,,
\quad
p=p^L=\frac{q\alpha_1}{\alpha_2} \,.
\eeq

The exact solution to Eq.~(\ref{eq:u}) is given by
\beq
u=\frac{1}{6p}\left(f^{1/3}+f^{-1/3}-1\right) \,,
\eeq
where
\beq
f=\left(\sqrt {108}\sqrt {27\,{p}^{2}-1}\,\,p+54\,{p}^{2}-1
 \right) \,.
\eeq

It is useful though to estimate the best scenario for large $r_C$,
which occurs in the small to intermediate $p$-regime where $u\sim1$.
Thus we obtain
\beq
r_C\sim\frac{M\alpha_2^2}{(1-q)^2} \,,
\eeq
with a best scenario for highly spinning BHs we get that $q>3/4$
is needed for $r_C>20M$.
Setting up a flip of $90$-degrees moves the critical radii outwards 
by approximately a factor two. This indicates that large flip-flop
angles, as measured in the orbital frame, 
are mostly expected for comparable mass binaries.

Similarly, to achieve the total flip-flop angle as seen in the
$J$-frame, the above analysis applies with
\beq
u=u_J=\frac{\alpha_2}{(1-q)}\sqrt{\frac{M}{qr_C}} \,,
\quad
p=p^J=\frac{\alpha_1(3-q)q^{3/2}}{2\alpha_2} \,.
\eeq
This leads to similar bounds as above for comparable masses,
but gets a notable increase for small mass ratios due to
large precessional effects (see the bottom panel of Fig.~\ref{fig:qFrequency}).

\subsection{Statistics}\label{subsec:stats}

In order to evaluate the probabilities of a flip-flop angle to occur,
we have performed numerical integrations of the 3.5PN equations of
motion~\cite{Damour:2007nc,Buonanno:2005xu} 
coupled with the spin evolution~(\ref{spinevo}) for randomly
varied spin magnitudes and orientations (thus representing unbiased
initial conditions). This may correspond to `dry' mergers of BHBs
or to the case of `chaotic accretion'~\cite{Sesana:2014bea}.
Other scenarios involving spin alignment due to coherent accretion
are possible~\cite{Krolik:2015jya} as well as the effect of
resonances~\cite{Schnittman:2004vq}, 
that lead to two `attractor' configurations, either $\Delta\phi=0$
or $\Delta\phi=\pi$. We will study those two cases below. 

We have thus performed a set of
$\alpha_1$ $\times$ $\cos(\theta_1^0)$ $\times$
$\alpha_2$ $\times$ $\cos(\theta_2^0)$ 
$= 20 \times 90 \times 20 \times 90 = 3,240,000$ simulations per $q$ 
(we have masked the points that have one of the spins $0$ 
or that have anti-aligned or aligned spins bringing the 
number for the statistics down to $2,859,120$).

\begin{figure}
\includegraphics[angle=270,width=\columnwidth]{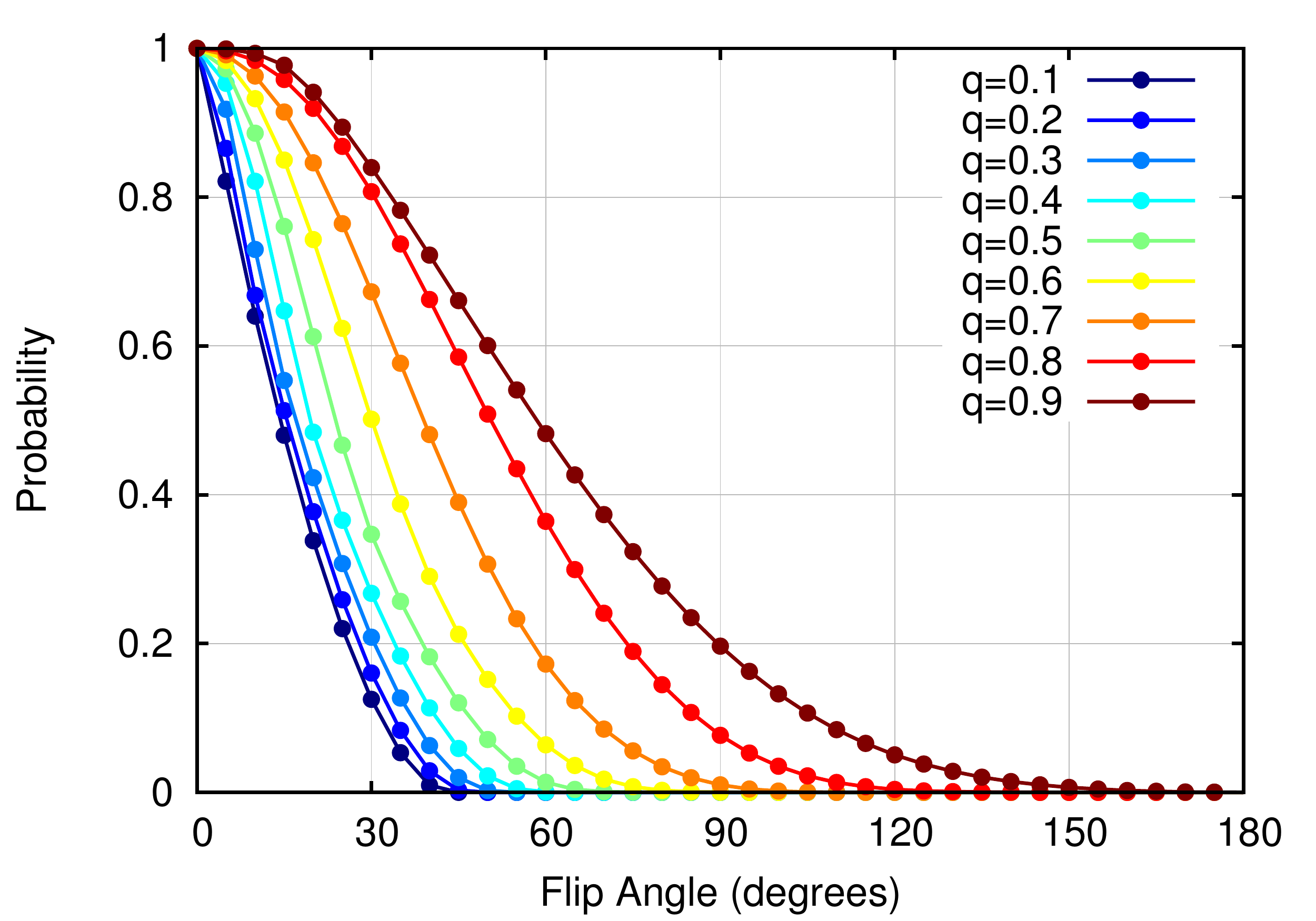}
\includegraphics[angle=270,width=\columnwidth]{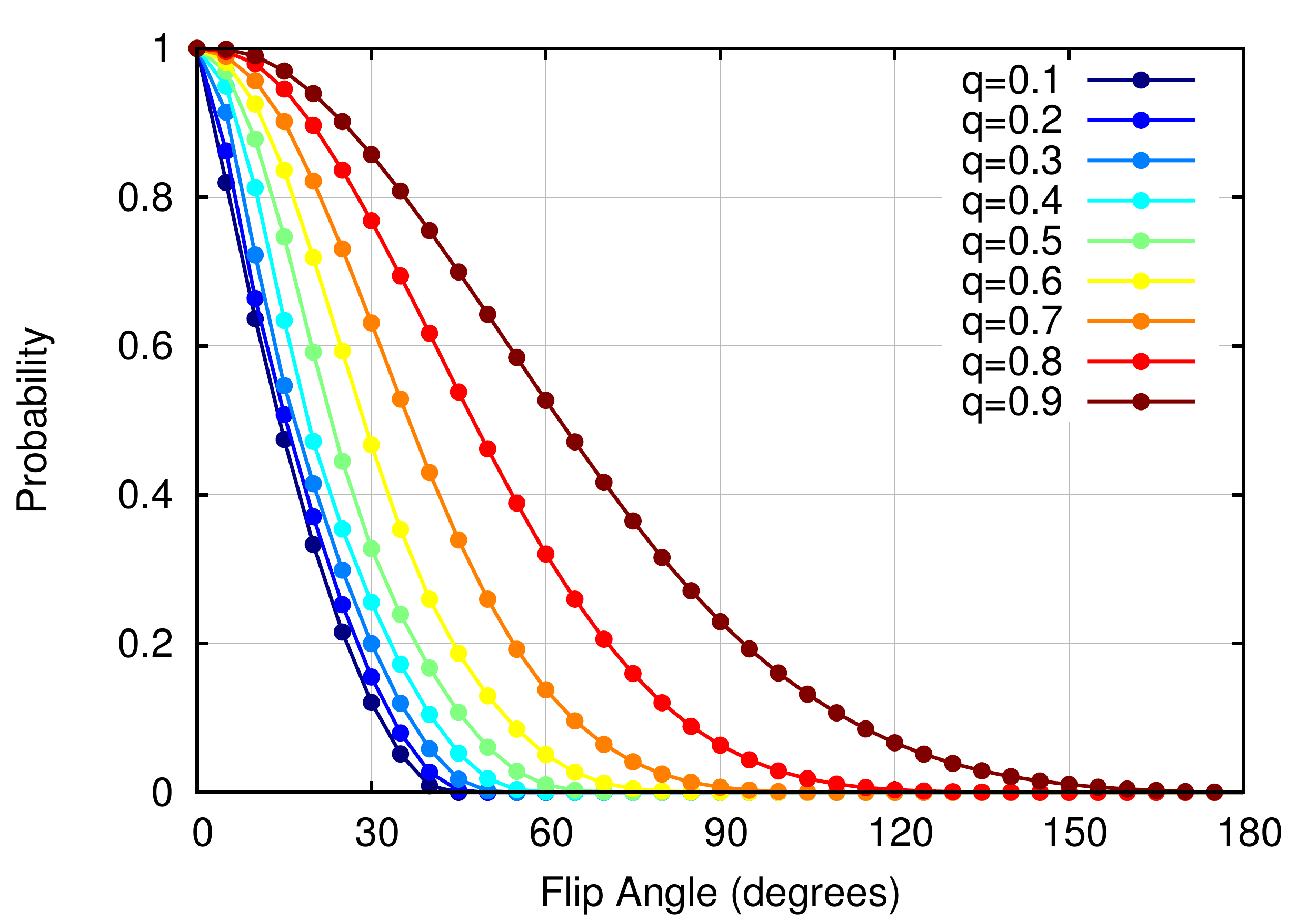}
\caption{The distributions of the flip angles larger than a given
(abscissa) value for a binary evolved
with 3.5PN from a separation
$r=100M$ to $r=5M$ for each $0.1\leq q\leq0.9$ in the $L$-frame.
The upper panel assumes an initial $\phi_1=\phi_2$ while
the lower panel assumes an initial $\phi_1=-\phi_2$ as
the resonant attractors at larger separations. 
\label{fig:probability_per_q}}
\end{figure}

Figure~\ref{fig:probability_per_q} displays the results of the
probabilities (as inferred from the occurring frequency of each value) of
a flip-flop angle larger than a given (by the abscissa) value to occur for each
mass ratio $q$ in the range $0.1\leq q\leq0.9$ (no resonances
exists for the $q=1$ case). We observe
that for some values of $q>0.7$ the probabilities for large angles
can be larger than for $q\approx1$ indicating that the leading
term in Eq.~(\ref{eq:ffangle1L}) is preponderant as the binary inspirals
down to smaller separations. This is a phenomena that we discussed
in the context of maximal flip-flop at the end of Sec.~\ref{subsec:UPN}.
We thus conclude that large flip-flop angles are frequent
in comparable masses binaries.

\begin{figure}
\includegraphics[angle=270,width=\columnwidth]{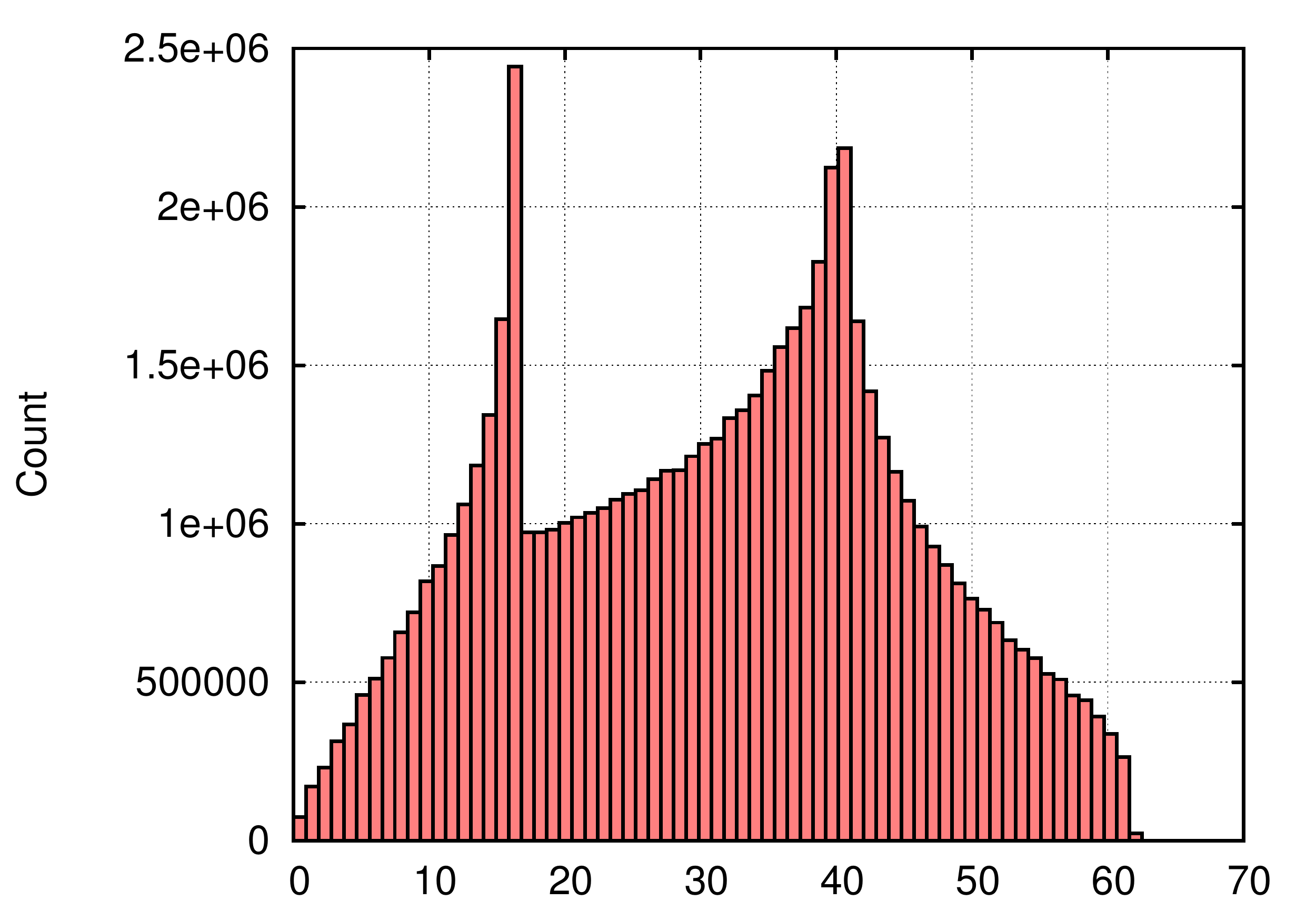}
\caption{The distributions of the flip angles for a binary at a 
constant separation
$r=20M$ and mass ratio $q=1/2$. Spin magnitudes and angles have been chosen
uniformly and the max(small, large) flip angles is evaluated.  
The peak at around $41^\circ$ corresponds to the smaller hole
spin flip. The larger hole gives an additional peak (at around $17^\circ$) 
towards low angles and bulks up the middle region
between peaks. 
\label{fig:Udistribution}}
\end{figure}

To further illustrate the probabilities of a given flip-flop
angle to occur, we consider a BHB system bearing
a mass ratio $q=1/2$ at a relatively short distance, $r=20M$.
Figure~\ref{fig:Udistribution} displays the results of a uniform 
initial distribution of spin magnitudes and angles.

We observe that the distribution of a given flip-flop angle
has a peak at about $41^\circ$ corresponding to the smaller hole
flip-flop and at around $17^\circ$ for the flip-flop angle 
of the larger hole. Flip-flop
angles larger than $63^\circ$ are very unlikely, in agreement
with the results displayed in Fig.~\ref{fig:probability_per_q}.

\begin{figure}
\includegraphics[angle=270,width=\columnwidth]{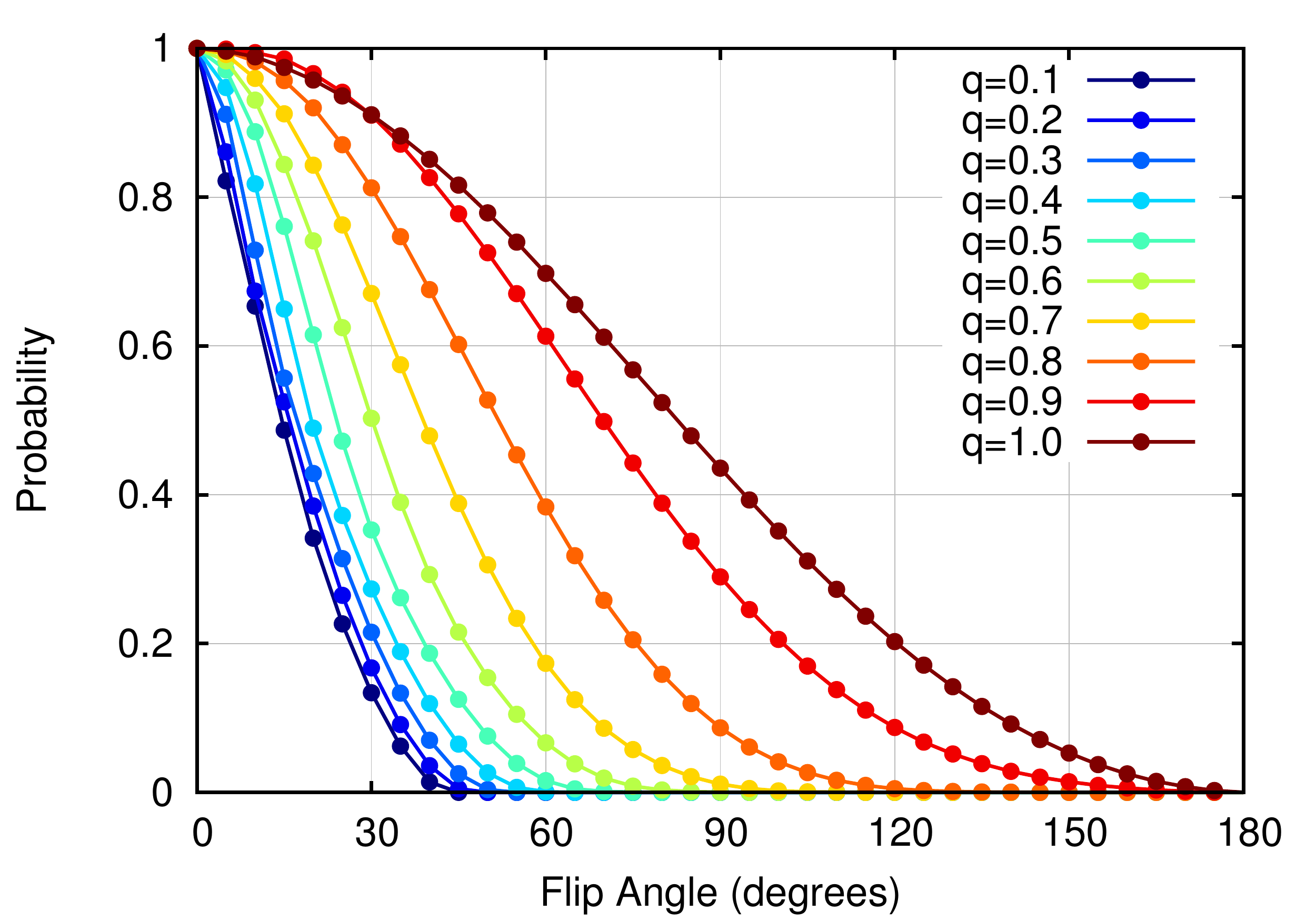}
\includegraphics[angle=270,width=\columnwidth]{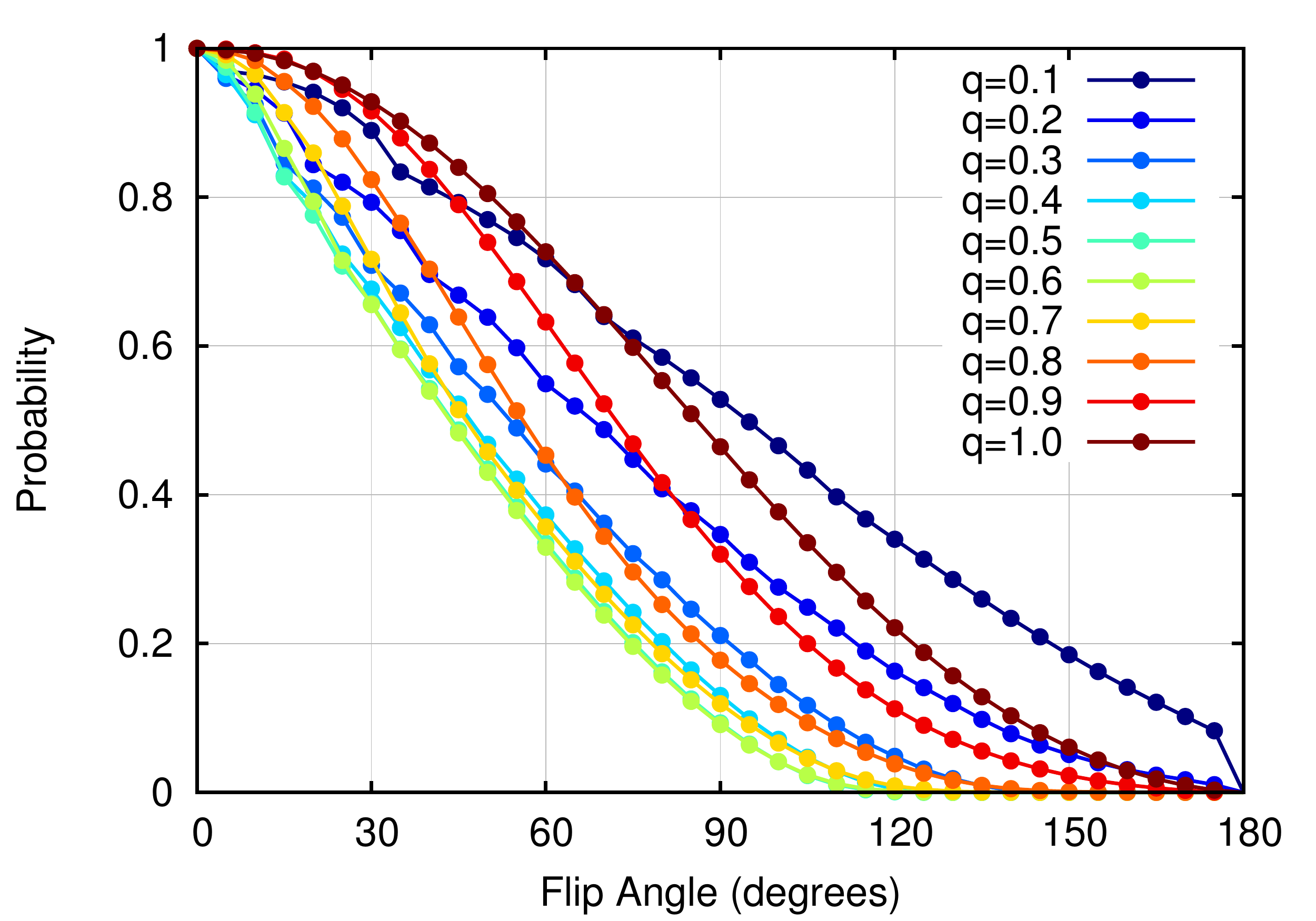}
\caption{Probability distribution of unequal mass flip-flops with respect to
${L}$-frame on top and with respect to ${J}$-frame below.
These results are for initial random spin distributions evolved with
3.5PN form $100M$ of initial separation down to $5M$ 
(approximating the merger).
\label{fig:qFrequency}}
\end{figure}

We mention that for unequal masses there is a difference in counting
flip-flop angles in the $J$-frame (total angular momentum) and the $L$-frame
(orbital angular momentum). In Fig.~\ref{fig:qFrequency} we display the 
differences for a few selected mass ratios. We observe in general that
larger angles are seen in the $J$-frame. Particularly for $q=0.1$,
due to the effects of transitional precession~\cite{Apostolatos94},
that may reverse the orientation of $\vec{J}$. But we also observe
larger angles for $q>1/4$, configurations that exclude total
transitional precession~\cite{Lousto:2013vpa,Lousto:2013wta}).

Both, the ${L}$- and ${J}$-frames are useful references to
account for different astrophysical accretion scenarios.

\section{Conclusions and Discussion}\label{sec:Discussion}

The unequal mass component of the flip-flop frequency, as given
by Eq.~(\ref{eq:fffrequency}), adds a leading term with a
dependence on the orbital separation as $\sim r^{-5/2}$. This term,
also proportional to $(1-q)$, eventually
becomes the dominant time-scale at large enough separations. 
We can compare this
time-scale with that for the realignment of BH spins via
the Bardeen-Peterson torque effect produced by
accreting matter. This alignment mechanism competes with
the spin flip-flop one.

The leading flip-flop period is now given by
\beq
T_{ff}\approx 2,000\, {\rm yr}\, \frac{(1+q)}{(1-q)}
\left(\frac{r}{10^3 M}\right)^{5/2}\left(\frac{M}{10^8M_\odot}\right) \,,
\eeq
which is much shorter than the gravitational radiation scale~\cite{Kidder:1995zr}
\beq
T_{GW}\approx 1.22 \times 10^6\, {\rm yr}\, \left(\frac{r}{10^3 M}\right)^{4}\left(\frac{M}{10^8M_\odot}\right) \,,
\eeq
and hence can take over the Bardeen-Peterson alignment 
mechanism~\cite{bardeen75}
at an order of magnitude further separations 
(i.e., $10,000M$) than the gravitational
radiation decay, used as reference time-scale in Ref.~\cite{Miller:2013gya}.
Thus, the flip-flop of spins shortens the time available for 
accretion to completely align
the spins of the BH with the orbital angular momentum.
However, according to Eq.~(\ref{eq:ffangle1L}) the angle of flip-flops
gets notably reduced at large distances, hence their statistical
relevance can be focused to the latest stages of merger, i.e.,
at closer separations or comparable mass binaries.

While in this paper we point out the relevance of studying
accretion and torque exchange of matter with BHs in
a {\it binary} system rather than on single BHs,
a definitive account of these competing mechanisms should be
provided by full BHB simulations in a matter filled
environment that incorporates all the needed elements of a
magnetohydrodynamical description of the accretion, like
in Ref.~\cite{Noble:2012xz} with spinning BHB backgrounds.

Since this flip-flop phenomena relies on the spin-spin coupling
(fed by the relative spin-orbit precession in the unequal mass case),
one can predict that not only BHs can see their spins flipped,
but material bodies with high spin, such as neutron stars, will as well,
since they have been observed to pair with comparable masses.
A simple evaluation of the flip-flop frequency and angles from
Eqs.~(\ref{eq:ffangle1L}) and~(\ref{eq:fffrequency}) applied 
to the double pulsar PSR J0737-3039~\cite{Burgay:2003jj} with
$M=2.587M_\odot$, $q=0.935$, $M/r=4.4\times10^{-6}$, 
leads to a flip-flop period of $1,230$ years, i.e., the spin direction
of the less massive neutron star B would change
by a degree every 3.4 years. However, the total change,
according to Eq.~(\ref{eq:ffangle1L}) will be less than a degree if
the highly spinning pulsar A has an intrinsic spin $\alpha_2<0.25$,
i.e., in order to produce observable effects nowadays its measured 
period should be below a millisecond instead of the actual 23\,milliseconds.
Notably, the spin precession frequency of pulsar B
has actually been observed to be around
$4.77^\circ$ per year~\cite{Breton:2008xy}. Unfortunately this precession 
of pulsar B meant that the highly beamed radio-emission is currently no 
longer pointing towards the Earth and we lost the detection of this one
pulsar in 2008~\cite{Burgay:2012xv}.
Yet the spin-orbit precession of another pulsar has been measured.
For the PSR B1534+12 (a $37.9$ milliseconds pulsar) 
a precession of $0.59^\circ$ per year has been detected~\cite{Fonseca:2014qla}.
This underlines that other potentially observable neutron star 
binary systems, at closer binary separations, 
may lead to larger observational 
flip-flop effects.

\acknowledgments 
The authors thank M. Campanelli, J. Krolik, H. Pfeiffer and Y. Zlochower for discussions on this work.
The authors gratefully acknowledge the NSF for financial support from Grants
PHY-1305730, PHY-1212426, PHY-1229173,
AST-1028087, PHY-0969855, OCI-0832606, and
DRL-1136221. Computational resources were provided by XSEDE allocation
TG-PHY060027N, and by NewHorizons and BlueSky Clusters 
at Rochester Institute of Technology, which were supported
by NSF grant No. PHY-0722703, DMS-0820923, AST-1028087, and PHY-1229173.
H.N. also acknowledges support by 
the Ministry of Education, Culture, Sports, Science and Technology (MEXT)
Grant-in-Aid for Scientific Research on Innovative Areas,
``New Developments in Astrophysics Through Multi-Messenger Observations
of Gravitational Wave Sources'', Grant No.~24103006.


\appendix

\section{Post-Newtonian Spin Evolution Equations in the ${L}$-
and ${J}$- frames}\label{app:PNSpin}

We discuss the spin evolutions in Eqs.~(\ref{spinevo})
by decomposing the spins 
along $\hat{L}$ and perpendicular to it, with unit vectors $\hat{\lambda}$
and $\hat{n}$, as shown in Fig.~\ref{fig:frame}. Here, we have that this frame
evolves as
\bea
 \dot{\hat{L}}&=&\frac{2 S_{{\rm eff} \hat{\lambda}}}{r^3} \hat{n}+
 \left(-\frac{2\, S_{{\rm eff} \hat{n}}}{r^3}+
 \frac{3\,\eta\, S_{0\hat{n}}\,S_{0\hat{L}}}{\ell r^3}
 \right)\hat{\lambda} \,, 
\nonumber\\
 \dot{\hat{n}}&=&\frac{v_\lambda}{r}\hat{\lambda}
 -\frac{2\, S_{{\rm eff} \hat{\lambda}}}{r^3} \hat{L} \,,  \\
\dot{\hat{\lambda}}&=&-\frac{v_\lambda}{r}\hat{n}-
\left(-\frac{2\, S_{{\rm eff} \hat{n}}}{r^3}+
\frac{3\,\eta\, S_{0\hat{n}}\,S_{0\hat{L}}}{\ell r^3}
\right)\hat{L} \,, \nonumber
\eea
where $\eta=q/(q+1)^2$,
$\vec{S}_{\rm eff}=[1+3/(4q)]\vec{S}_1+[1+3q/4]\vec{S}_2$, and
$\vec{S}_{0}=[1+1/q]\vec{S}_1+[1+q]\vec{S}_2$.
Then, we obtain the spin evolutions from Eqs.~(\ref{spinevo}) as
\bea
&&
\dot{S}_{1 \hat n} 
= \Omega_1 {S}_{1 \hat \lambda} 
- \frac{3\,S_{1\hat L} S_{2 \hat \lambda}}{r^3}\left(1+\frac{q}{2}\right) \,,
\cr
&&
\dot{S}_{1 \hat \lambda} 
=
- \Omega_1 {S}_{1 \hat n} 
+ \frac{3\,q\, S_{1\hat L} S_{2 \hat n}}{2r^3}
\left(1-\frac{2\,S_{0\hat L}}{(1+q)\ell}\right)
\cr 
&&\quad\quad\quad
- \frac{3 S_{1\hat L} S_{1\hat n}}{q\,r^3} 
\left(1+\frac{q\,S_{0\hat L}}{(1+q)\ell}\right)  \,,
\\
&&
\dot{S}_{1 \hat L} =
\frac{3\,S_{1\hat n}}{2\,q\,(1+q)r^3}\Bigg[
q(1+q)(2+q)S_{2\hat \lambda}
\cr
&&\quad\quad\quad
+2\left(1+q+q\frac{S_{0\hat L}}{\ell}\right)
\,S_{1\hat \lambda}
\Bigg]
\cr
&&\quad\quad\quad
+\frac{3\,q\,S_{2\hat n}\,S_{1\hat \lambda}}{2\,(1+q)r^3}\left[
-(1+q)+2\frac{S_{0\hat L}}{\ell}
\right]\,,\nonumber
\eea
where $S_{0\hat L}$ is a conserved quantity and
\beq
\Omega_1=
\frac{v_{\lambda}}{r}- \frac{\ell}{r^3}\left(2+\frac{3}{2q}\right)
\left(1+\frac{S_{1\hat L}}{\ell}\right)+\frac{S_{2\hat L}}{r^3}\,,
\eeq
for $\vec{S}_1$ and similar for $\vec{S}_2$ exchanging labels 1 and 2
and $q\to1/q$.
In the above expression ${S}_{i \hat L}=\vec{S}_i\cdot\hat{L}$,
${S}_{i \hat n}=\vec{S}_i\cdot\hat{n}$, and
${S}_{i \hat\lambda}=\vec{S}_i\cdot\hat{\lambda}$; 
$\ell=|\vec{L}|=q M^{3/2} r^{1/2} / (1+q)^2$, and 
${S}_{\hat L}=\vec{S}\cdot\hat{L}$.

\subsection{Flip-flop frequency}

The evolution of $\vec{S}_1$ and $\vec{S}_2$ along $\hat L$ becomes
\bea
\dot{S}_{1 \hat L} &=& 
\frac{1}{r^3} (\vec{S}_1 \times (\vec{S} + 2\vec{S}_{\rm eff})) \cdot \hat L
\cr &&
+ \frac{3}{r^3} \frac{S_{0\hat n}}{1+q} S_{1 \hat \lambda} 
\left(1 + \frac{q}{1+q}\frac{S_{0\hat L}}{\ell} \right) \,,
\cr
\dot{S}_{2 \hat L} &=& 
\frac{1}{r^3} (\vec{S}_2 \times (\vec{S} + 2\vec{S}_{\rm eff})) \cdot \hat L
\cr &&
+ \frac{3}{r^3} \frac{q S_{0\hat n}}{1+q} S_{2 \hat \lambda} 
\left(1 + \frac{1}{1+q}\frac{S_{0\hat L}}{\ell} \right) \,.
\eea
Then, the averaging over the orbital period (denoted by the brackets
$\langle\rangle$) gives
\bea
\langle \dot{S}_{1 \hat L} \rangle 
&=& \frac{3}{2r^3}(1+q) \frac{|\vec J|}{\ell} 
\left(1- \frac{q}{(1+q)^2} \frac{S_{0\hat L}}{\ell}\right)
\cr && \times
(\vec{S}_1 \times \vec{S}_2) \cdot \hat{J}
\,,
\cr
\langle \dot{S}_{2 \hat L} \rangle
&=& - \frac{3}{2r^3}\left(1+\frac{1}{q}\right) \frac{|\vec J|}{\ell}
\left(1- \frac{q}{(1+q)^2} \frac{S_{0\hat L}}{\ell}\right)
\cr && \times 
(\vec{S}_1 \times \vec{S}_2) \cdot \hat{J}
\,.
\label{eq:dS1LdS2L}
\eea
This shows the conservation of $\vec{S}_0\cdot\hat{L}$, as we have
$\langle \dot{S}_{1 \hat L} \rangle + q \langle \dot{S}_{2 \hat L} \rangle =0$.

On the other hand, the spin evolution projected along $\vec{J}$ can
be derived from Eqs.~(\ref{spinevo}) as
\bea
\dot{S}_{1 \hat J} &=& 
= \frac{1}{r^3 |\vec{J}|}
\left(
\left(3+\frac{3}{2q}\right)
(\vec{S}_1 \times \vec{S}_2) \cdot \vec{L}\right.
\cr
&&\left.+ \frac{3}{1+q} \ell {S}_{0\hat{n}}{S}_{1\hat{\lambda}} 
+ \frac{3}{1+q} {S}_{0\hat{n}} (\vec{S}_1 \times \vec{S}_2) \cdot \hat{n}
\right) \,,
\cr
\dot{S}_{2 \hat J} &=& 
= \frac{1}{r^3 |\vec{J}|}
\left(
\left(3+\frac{3q}{2}\right)
(\vec{S}_2 \times \vec{S}_1) \cdot \vec{L}\right.
\cr
&&\left.+ \frac{3q}{1+q} \ell {S}_{0\hat{n}}{S}_{2\hat{\lambda}} 
+ \frac{3q}{1+q} {S}_{0\hat{n}} (\vec{S}_2 \times \vec{S}_1) \cdot \hat{n}
\right) \,,
\nonumber \\
\eea
and then the averaging over the orbital period gives
\bea
\langle \dot{S}_{1 \hat J} \rangle
&=& \frac{3}{2 r^3}\left(1+\frac{1}{q} \right)
\left(
1
- \frac{q}{(1+q)^2}  \frac{{S}_{0\hat{L}}}{\ell}
\right)
(\vec{S}_1 \times \vec{S}_2) \cdot \hat{J} \,,
\cr
\langle \dot{S}_{2 \hat J} \rangle
&=& - \frac{3}{2 r^3} (1+q)
\left(
1
- \frac{q}{(1+q)^2}  \frac{{S}_{0\hat{L}}}{\ell}
\right)
(\vec{S}_1 \times \vec{S}_2) \cdot \hat{J} \,.
\nonumber \\
\label{eq:dS1JdS2J}
\eea
This shows the conservation of the quantity
$\vec{S}_N\cdot\hat{J}$, where, $\vec{S}_N=q\vec{S}_1+\vec{S}_2$, as we have
$q \langle \dot{S}_{1 \hat J} \rangle + \langle \dot{S}_{2 \hat J} \rangle =0$.

Note that the factor $\ell/(q|\vec{J}|)$ (or $q\ell/|\vec{J}|$) is different from
$\langle \dot{S}_{1 \hat L} \rangle$ (or $\langle \dot{S}_{2 \hat L} \rangle$).
This difference can be expressed as
\bea
\langle \dot{S}_{1 \hat J} \rangle
- \frac{\ell}{|\vec{J}|} \langle \dot{S}_{1 \hat L} \rangle
&=& \frac{1}{|\vec{J}|}
(\dot{\vec{S}}_1 \cdot \vec{S} - \vec{S}_1 \cdot \dot{\vec{L}})
\cr
&=& \frac{1}{|\vec{J}|}
(\dot{\vec{S}}_1 \cdot \vec{S}_2 - \vec{S}_1 \cdot 
(\dot{\vec{J}}-\dot{\vec{S}}_2))
\cr
&=& \frac{1}{|\vec{J}|} \left[ \vec{S}_1 \cdot \vec{S}_2 \right]^{\cdot} \,.
\eea
Therefore, the time evolution of the angle between
$\vec{S}_1$ and $\vec{S}_2$ creates the different $q$-behavior
into the two projection frames. This will affect the flip-flop angle,
but not its frequency as we see below.

To obtain the flip-flop frequency (measured along $\vec{L}$ or $\vec{J}$
will be the same) we first note that
\bea
\dot{S}_{0 \hat L} &=& 
\frac{3}{2r^3} \left( \frac{1}{q} -q \right)
(\vec{S}_1 \times \vec{S}_2) \cdot \hat{L}
\cr &&
+ \frac{3}{r^3} \frac{S_{0\hat n}}{q} S_{1 \hat \lambda} 
\left(1 + \frac{q}{1+q}\frac{S_{0\hat L}}{\ell} \right)
\cr &&
+ \frac{3}{r^3} q S_{0\hat n} S_{2 \hat \lambda} 
\left(1 + \frac{1}{1+q}\frac{S_{0\hat L}}{\ell} \right)
\,,
\cr
\langle \dot{S}_{0 \hat L} \rangle &=& 0 \,.
\eea
Therefore, the time dependent part of
the orbital-averaged first derivative of the spins,
$\langle \dot{S}_{1 \hat L} \rangle$, $\langle \dot{S}_{2 \hat L} \rangle$,
$\langle \dot{S}_{1 \hat J} \rangle$ and $\langle \dot{S}_{2 \hat J} \rangle$
is only $(\vec{S}_1 \times \vec{S}_2) \cdot \hat{J}$, on which we focus
in the following.

The first derivative of $(\vec{S}_1 \times \vec{S}_2) \cdot \hat{J}$
with respect to time can be written as
\bea
\frac{d}{dt}[(\vec{S}_1 \times \vec{S}_2) \cdot \hat J]
&=&
\frac{3}{2r^3}\frac{\ell}{|\vec{J}|}
\left(1- \frac{q}{(1+q)^2} \frac{S_{0\hat L}}{\ell}\right) 
\cr &&
\times \biggl[
\frac{(1-q^2)}{q}
((\vec{L} \times  \vec{S}_1) \times \vec{S}_2) \cdot \hat L
\cr &&
- \left( \vec{S}_0 \times
(\vec{S}_1 \times \vec{S}_2) \right) \cdot 
\hat{L}
\biggl] \,.
\label{eq:firstS1S2J}
\eea
Here, we are always using the averaged evolution equations.
Taking one more time derivative of the above equation, we have
\begin{widetext}
\bea
\frac{d^2}{dt^2}
[(\vec{S}_1 \times \vec{S}_2) \cdot \hat J]
&=&
- \biggl[ \frac{9}{4}\,
{\frac { (1-q)^2 (1+q)^2 \ell^2}{{q}^{2}{r}^{6}}}
+ {\frac { 9 (1-q)(1+q) {\ell S_{1\hat L}}}{{q}{r}^{6}}}
- {\frac { 9 (1-q)(1+q) {\ell S_{2\hat L}}}{q{r}^{6}}}
- \frac{9}{4}\,{\frac { \left( 3+5\,q \right)  
\left( 1-q \right) {S_{1\hat L}}^{2}}{{q}^{2}{r}^{6}}}
\cr &&
+\frac{9}{2}\,{
\frac { \left( 1-q \right) ^{2}{S_{1\hat L}}\,{S_{2\hat L}}}{q{r}^{6}}}
+\frac{9}{4}\,{
\frac { \left( 1-q \right)  \left( 5+3\,q \right) {S_{2\hat L}}^{2}}{{r}^{6}}}
+\frac{9}{4}\,{\frac { \left( 1+q \right) ^{2} 
\left( {S_1}^{2}+2\,q\,{\vec{S}_1 \cdot \vec{S}_2}+{q}^{2}{S_2}^{2} \right) }
{{q}^{2}{r}^{6}}}
+ \cdot\cdot\cdot
\biggr]
\cr && \times 
(\vec{S}_1 \times \vec{S}_2) \cdot \hat J \,,
\label{eq:ddS1cS2dJ}
\eea
where ($+\cdot\cdot\cdot$) means higher PN order terms.
This equation has the form of an harmonic oscillator (with slowly varying
frequency) and reaches the two extreme positions when $\vec{S}_1$,
$\vec{S}_2$, and $\vec{J}$ (or equivalently at those points $\vec{L}$)
are all coplanar (and hence the triple vector product vanishes).

Therefore, we can extract a frequency as 
\bea
\Omega_{ff}^2 &=&
\frac{9}{4}\,{\frac { \left( 1-q \right) ^{2} M^3}
  {\left( 1+q \right) ^{2}{r}^{5}}}\left(1+4\frac{M}{r}\right)
+ 9\,{\frac { \left( 1-q \right)(S_{1\hat L}-S_{2\hat L})M^{3/2}}
{(1+q) r^{11/2}}}
-\frac{9}{4}\,{\frac {  \left( 1-q \right) \left( 3+5\,q \right) 
{S_{1\hat L}}^{2}}{{q}^{2}{r}^{6}}}
\cr &&
+\frac{9}{2}\,{
\frac { \left( 1-q \right) ^{2}{S_{1\hat L}}\,{S_{2\hat L}}}{q{r}^{6}}}
+\frac{9}{4}\,{
\frac { \left( 1-q \right)  \left( 5+3\,q \right) 
{S_{2\hat L}}^{2}}{{r}^{6}}}
+\frac{9}{4}\,{\frac { S_0^2 }{{r}^{6}}}
+ \cdot\cdot\cdot
\,,
\label{eq:OmegaFF}
\eea
\end{widetext}
where 
we have used $\ell = (q/(1+q)^2) M^{3/2} r^{1/2}(1+2M/r)$ from
Eq.~(4.7) of Ref.~\cite{Kidder:1995zr}, and
($+ \cdot\cdot\cdot$) stands for $O(1/r^{13/2})$ which is higher PN order corrections.
Here, $\Omega_{ff}$ is constant at fixed $r$ for $q=1$.
In the unequal mass cases, this is not constant in time, but we may treat
it as a constant in the flip-flop time scale
because of $\dot{\Omega}_{ff}/\Omega_{ff} = O(1/r^{7/2})$.

\subsection{Maximum flip-flop angle}

The maximum flip-flop angle is calculated as follows.
First, we obtain from Eq.~(\ref{eq:ddS1cS2dJ}) as 
\bea
(\vec{S}_1 \times \vec{S}_2) \cdot \hat J
= A \sin(\Omega_{ff} t) \,,
\eea
where we have chosen the initial value
by $S_{1\hat L}(0)=-S_1$ 
($\alpha_{1\hat L}(0)=-\alpha_1$ where 
$\vec{\alpha}_{i}=\vec{S}_i/m_i^2$)
to obtain the maximum flip-flop angle for the smaller hole.
Using Eq.~(\ref{eq:firstS1S2J}),
the amplitude $A$ is determined from
\bea
\Omega_{ff} A
&=& \frac{3}{2r^3}\frac{\ell}{|\vec{J}|}
\left(1- \frac{q}{(1+q)^2} \frac{S_{0\hat L}}{\ell}\right) 
(1+q) S_1 
\cr && \times
({S_2}^2-{S_{2\hat L}(0)}^2) \,,
\eea
at $t=0$ as
\begin{widetext}
\bea
A &=& 
{\frac { \left( {\alpha_2}^2-{\alpha_{2\hat L}(0)}^2 \right)  
{\alpha_1}\,{q}^{2}{M}^{9/2}}{ 
\left( 1+q \right) ^{4} \left( 1-q \right) \sqrt {r}}}
+{\frac {\left( {\alpha_2}^2-{\alpha_{2\hat L}(0)}^2 \right) 
\left( -{\alpha_1}\,{q}^{3}+2\,{\alpha_1}\,{q}^{2}
+2\,{\alpha_{2\hat L}(0)}\,q-{\alpha_{2\hat L}(0)} \right)
{\alpha_1}\,q {M}^{5} }
{\left( 1+q \right) ^{4} \left( 1-q \right) ^{2} r}} 
\cr &&
+\frac{1}{2}\,{\frac {
\left( {\alpha_2}^2-{\alpha_{2\hat L}(0)}^2 \right)  
{\alpha_1} {M}^{11/2}}
{\left( 1+q \right) ^{4} \left( 1-q \right) ^{3}{r}^{3/2}} }
( 2\,{{\alpha_1}}^{2}{q}^{6}-6\,
{{\alpha_1}}^{2}{q}^{5}-6\,{\alpha_1}\,{q}^{4}{\alpha_{2\hat L}(0)}
+6\,{{\alpha_1}}^{2}{q}^{4}+20\,{\alpha_1}\,{q}^{3}{\alpha_{2\hat L}(0)}
+8\,{\alpha_{2\hat L}(0)}^{2}{q}^{2}
\cr &&
-2\,{\alpha_2}^{2}{q}^{2}
-6\,{\alpha_1}\,{q}^{2}{\alpha_{2\hat L}(0)}
-8\,q{\alpha_{2\hat L}(0)}^{2}
+2\,{\alpha_2}^{2}q
-{\alpha_2}^{2}+3\,{\alpha_{2\hat L}(0)}^{2} ) 
\,,
\eea
where ${\alpha_{2\hat L}(0)}$ is derived later.
Substituting this solution into Eqs.~(\ref{eq:dS1LdS2L})
and integrating it with respect to time, we obtain
\bea
\frac{\alpha_{1\hat L}(t)}{\alpha_1}
&=& 
-1+{\frac { \left( {\alpha_2}^2-{\alpha_{2\hat L}(0)}^2 \right)  M }
{r \left( 1-q \right) ^{2}}}
(1-\cos(\Omega_{ff} t))
+2\,{\frac { \left( {\alpha_2}^2-{\alpha_{2\hat L}(0)}^2 \right)    
\left( {\alpha_1}\,q+{\alpha_{2\hat L}(0)} \right) {M}^{3/2}}{ \left( 1-q \right) ^{3}{r}^{3/2}}}
(1-\cos(\Omega_{ff} t))
\cr &&
+{\frac { \left( {\alpha_2}^2-{\alpha_{2\hat L}(0)}^2 \right)  
\left( 3\,{{\alpha_1}}^{2}{q}^{2}
+10\,{\alpha_1}\,q{\alpha_{2\hat L}(0)}
-{{\alpha_2}}^{2}+4\,{{\alpha_{2\hat L}(0)}}^{2} \right) {M}^{2}}
{{r}^{2} \left( 1-q \right) ^{4}}}
(1-\cos(\Omega_{ff} t)) \,,
\eea
\end{widetext}
with $\alpha_{1\hat L}(0)=-\alpha_1$.
Therefore, we have the maximized flip-flop angle
at $\cos(\Omega_{ff} t)=-1$,
and $\alpha_{2\hat L}(0)$ is fixed
to maximize the angle as
\bea
\frac{\alpha_{2\hat L}(0)}{\alpha_2}
= \frac{\alpha_2 \sqrt{M}}{(1-q) \sqrt{r}}
+ \frac{3 q \alpha_2 \alpha_1\, M}{(1-q)^2 r} \,.
\eea
Then, the solution becomes
\bea
\frac{\alpha_{1\hat L}(t)}{\alpha_1}
&=& 
-1 + 
\left[ \frac{\alpha_2^2 M}{(1-q)^2 r}
+ \frac{ 2 q {\alpha_2}^2 \alpha_1 M^{3/2}}{(1-q)^3 r^{3/2}}
\right] 
\cr && \times (1-\cos(\Omega_{ff} t)) \,.
\eea
As the result, the maximum flip-flop angle becomes
\bea
\frac{\alpha_{1\hat L}(t_{\rm max})}{\alpha_1}+1
&=& 
2
\left[ \frac{{\alpha_2}^2 M}{(1-q)^2 r}
+ \frac{ 2 q {\alpha_2}^2 \alpha_1 M^{3/2}}
{(1-q)^3 r^{3/2}}
\right] \,.
\nonumber \\
\eea

Here, we need a special treatment for the $q=1$ case
because of the lack of the leading term in Eq.~(\ref{eq:OmegaFF}).
Furthermore, since $\Omega_{ff}$ has only the leading order for $q=1$,
we focus on the leading order calculation.
Using the same initial configuration, $\alpha_{1\hat L}(0)=-\alpha_1$,
the solution with the parameter $\alpha_{2\hat L}(0)$ is obtained as
\bea
\frac{\alpha_{1\hat L}(t)}{\alpha_1}
&=& 
- {\frac {\, \left( {\alpha_2}^2-{\alpha_{2\hat L}(0)}^2 \right)  }{{{\alpha_1}}^{2}
-2\,{\alpha_1}\,{\alpha_{2\hat L}(0)}+{{\alpha_2}}^{2}}}
\cos (\Omega_{ff} t)
\cr &&
-{\frac { \left( {\alpha_{2\hat L}(0)} -{\alpha_1} \right) ^{2}}
{{{\alpha_1}}^{2}-2\,{\alpha_1}\,{\alpha_{2\hat L}(0)}+{{\alpha_2}}^{2}}} \,.
\eea
The maximum flip-flop angle is found for $\alpha_{2\hat L}(0)=\alpha_1$
under an assumption $\alpha_1 < \alpha_2$,
and the solution becomes
\bea
\frac{\alpha_{1\hat L}(t)}{\alpha_1}
&=& -\cos (\Omega_{ff} t) \,.
\eea
Therefore, the maximum flip-flop angle is $\pi$.

On the other hand, in the case of $S_{2\hat L}(0)=S_2$ 
($\alpha_{2\hat L}(0)=\alpha_2$),
we can obtain the maximum flip-flop angle for the larger hole.
The amplitude $A$ is derived from Eq.~(\ref{eq:firstS1S2J}),
\bea
\Omega_{ff} A
&=& \frac{3}{2 q r^3}\frac{\ell}{|\vec{J}|}
\left(1- \frac{q}{(1+q)^2} \frac{S_{0\hat L}}{\ell}\right) 
(1+q) S_2 
\cr && \times
({S_1}^2-{S_{1\hat L}(0)}^2) \,.
\eea
at $t=0$ as
\begin{widetext}
\bea
A &=& 
{\frac {\left( {\alpha_1}^2-{\alpha_{1\hat L}(0)}^2 \right) 
{\alpha_2}\,{q}^{3}{M}^{9/2}}
{ \left( 1+q \right) ^{4} \left( 1-q \right) \sqrt {r}}}
+{\frac { \left( {\alpha_1}^2-{\alpha_{1\hat L}(0)}^2 \right)  
\left( {\alpha_{1\hat L}(0)}\,{q}^{3}
-2\,{\alpha_{1\hat L}(0)}\,{q}^{2}+2\,{\alpha_2}\,q
-{\alpha_2} \right) {\alpha_2}\,{q}^{2}{M}^{5} }
{\left( 1+q \right) ^{4} \left( 1-q \right) ^{2}r }}
\cr &&
+\frac{1}{2}
\,{\frac {\left( {\alpha_1}^2-{\alpha_{1\hat L}(0)}^2 \right) 
q{\alpha_2}\,{M}^{11/2} }
{\left( 1+q \right) ^{4} \left( 1-q \right) ^{3}{r}^{3/2} }}
( 3\,{\alpha_{1\hat L}(0)}^{2}{q}^{6}-{\alpha_1}^{2}{q}^{6}
-8\,{\alpha_{1\hat L}(0)}^{2}{q}^{5}+2\,{\alpha_1}^{2}{q}^{5}
+6\,{\alpha_{1\hat L}(0)}\,{q}^{4}{\alpha_2}
+8\,{\alpha_{1\hat L}(0)}^{2}{q}^{4}
\cr &&
-2\,{\alpha_1}^{2}{q}^{4}-20\,{\alpha_{1\hat L}(0)}\,{q}^{3}{\alpha_2}
+6\,{\alpha_{1\hat L}(0)}\,{q}^{2}{\alpha_2}+6\,{\alpha_2}^{2}{q}^{2}
-6\,{\alpha_2}^{2}q+2\,{\alpha_2}^{2} ) \,,
\eea
\end{widetext}
where ${\alpha_{1\hat L}(0)}$ is derived later. From a similar
analysis in the above, we have
\bea
\frac{\alpha_{2\hat L}(t)}{\alpha_2}
&=& 
1 - 
\left[ \frac{q^2 {\alpha_1}^2 M}{(1-q)^2 r}
+ \frac{ 2 q^2 {\alpha_1}^2 \alpha_2 M^{3/2}}
{(1-q)^3 r^{3/2}}
\right] 
\cr && \times
(1-\cos(\Omega_{ff} t)) \,,
\eea
and
\bea
\frac{\alpha_{1\hat L}(0)}{\alpha_1}
= - \frac{q \alpha_1 \sqrt{M}}{(1-q) \sqrt{r}}
- \frac{3 q \alpha_1 \alpha_2 M}{(1-q)^2 r} \,.
\eea
Therefore, the maximum flip-flop angle is
\bea
1 - \frac{\alpha_{2\hat L}(t_{\rm min})}{\alpha_2}
&=& 
2
\left[ \frac{q^2 {\alpha_1}^2 M}{(1-q)^2 r}
+ \frac{ 2 q^2 {\alpha_1}^2 \alpha_2 M^{3/2}}
{(1-q)^3 r^{3/2}}
\right] \,.
\nonumber \\
\eea

In the $J$-frame, we find the maximum flip-flop of the larger hole
with $\alpha_{2\hat L}(0)=\alpha_2$ and a parameter $\alpha_{1\hat L}(0)$.
The solution is
\bea
\frac{\alpha_{2\hat J}(t)}{\alpha_2}
&=&
1 + \left(
{\frac {{q}^{3}{{\alpha_1}}^{2}M}{\left( 1-q \right) ^{2}r }} 
- {\frac {{q}^{2} \left( 1 - 3\,q\right){{\alpha_1}}^{2}{\alpha_2}\,
{M}^{3/2} }{ \left( 1-q \right) ^{3}{r}^{3/2}}}
\right) 
\cr && \times 
\cos (\Omega_{ff} t)
-\frac{1}{2}\,{\frac {{q}^{2} \left( {q}^{2}+1 \right) {{\alpha_1}}^{2}M}
{\left( 1-q \right) ^{2}r }}
\cr &&
-{\frac { q \left( {q}^{3}+2\,q-1 \right) 
{{\alpha_1}}^{2} {\alpha_2}\,{M}^{3/2}}{ \left( 1-q \right) ^{3}{r}^{3/2}}} \,,
\eea
and the maximum flip-flop angle becomes
\bea
\frac{\alpha_{2\hat J}(t_{\rm max})}{\alpha_2}
&-&\frac{\alpha_{2\hat J}(t_{\rm min})}{\alpha_2}
= 
2\,
\biggl[ \frac{q^3 {\alpha_1}^2 M}{(1-q)^2 r}
\cr &&
- \frac{ q^2 (1-3q) {\alpha_1}^2 \alpha_2 M^{3/2}}
{(1-q)^3 r^{3/2}}
\biggr] \,.
\eea
To obtain this flip-flop angle, the initial configuration is
\bea
\frac{\alpha_{2\hat L}(0)}{\alpha_2}
&=&
1 \,,
\cr
\frac{\alpha_{2\hat J}(0)}{\alpha_2}
&=&
1-\frac{1}{2}\,{\frac {{q}^{2}{\alpha_1}^{2}M}{r}}
+{\frac {q{\alpha_1}^{2}{\alpha_2}\,{M}^{3/2}}{{r}^{3/2}}}
\,,
\cr
\frac{\alpha_{1\hat L}(0)}{\alpha_1}
&=&
-\frac{1}{2}\,{\frac {\left( 3-q \right)q{\alpha_1}\, \sqrt {M}}
{\left( 1-q \right) \sqrt {r}}}
\cr &&
+\frac{1}{2}\,{\frac {(q^2-8q+1){\alpha_1}{\alpha_2}\,M}{(1-q)^2 r}}
\,,
\cr
\frac{\alpha_{1\hat J}(0)}{\alpha_1}
&=&
- \frac{1}{2}\,{\frac {\left( 1+q \right)q{\alpha_1}\, \sqrt {M} }
{ \left( 1-q \right) \sqrt {r}}}
\cr &&
-\frac{1}{2}\,{\frac {(q^2+4q+1){\alpha_1}{\alpha_2} \,M}{(1-q)^2 r}} \,.
\eea

For the smaller hole case, when we choose the initial value
by $S_{1\hat L}(0)=-S_1$ and find ${\alpha_{2\hat L}(0)}$
to maximize the flip-flop angle that is the same ansatz as the $L$-frame analysis,
the solution is derived as
\bea
\frac{\alpha_{1\hat J}(t)}{\alpha_1}
&=&
- 1 + \left( 
-{\frac {{{\alpha_2}}^{2}M}{q \left( 1-q \right) ^{2} r}} 
- \frac{ (3-q) {\alpha_2}^2 \alpha_1 M^{3/2}}
{(1-q)^3 r^{3/2}}
\right)  
\cr && \times \cos (\Omega_{ff} t)
+\frac{1}{2}\,{\frac {\left( {q}^{2}+1 \right){{\alpha_2}}^{2} M}
{q^2 \left( 1-q \right) ^{2} r }} 
\cr &&
- {\frac { \left( {q}^{3}-2\,q-1 \right) {{\alpha_2}}^{2} 
{\alpha_1}\,{M}^{3/2} }{ q\left( 1-q \right) ^{3}{r}^{3/2}}} \,,
\eea
and the maximum flip-flop angle becomes
\bea
\frac{\alpha_{1\hat J}(t_{\rm max})}{\alpha_1}
&-&\frac{\alpha_{1\hat J}(t_{\rm min})}{\alpha_1}
=
2\,
\biggl[ \frac{{\alpha_2}^2 M}{q (1-q)^2 r}
\cr &&
+ \frac{ (3-q) {\alpha_2}^2 \alpha_1 M^{3/2}}
{(1-q)^3 r^{3/2}}
\biggr] \,.
\eea
The above solution is obtained for the initial configuration,
\bea
\frac{\alpha_{1\hat L}(0)}{\alpha_1}
&=&
-1 \,,
\cr
\frac{\alpha_{1\hat J}(0)}{\alpha_1}
&=&
- 1 + \frac{1}{2}\,{\frac {{\alpha_2}^{2}M}{q^2 r}}
+{\frac {{\alpha_2}^{2}{\alpha_1}\,{M}^{3/2}}{q{r}^{3/2}}}
\,,
\cr
\frac{\alpha_{2\hat L}(0)}{\alpha_2}
&=&
-\frac{1}{2}\,{\frac {\left( 1-3q \right){\alpha_2}\, \sqrt {M}}
{q \left( 1-q \right) \sqrt {r}}}
\cr &&
- \frac{1}{2}\,{\frac {(q^2-8q+1){\alpha_2}{\alpha_1} \,M}{(1-q)^2 r}}
\,,
\cr
\frac{\alpha_{2\hat J}(0)}{\alpha_2}
&=&
\frac{1}{2}\,{\frac {\left( 1+q \right){\alpha_2}\, \sqrt {M} }
{q \left( 1-q \right) \sqrt {r}}}
\cr &&
+ \frac{1}{2}\,{\frac { (q^2+4q+1){\alpha_2}{\alpha_1}\,M}{(1-q)^2 r}} \,.
\eea


\bibliographystyle{apsrev4-1}
\bibliography{../../../Bibtex/references}

\end{document}